\documentclass[a4paper,twocolumn,11pt, accepted=2024-06-14]{quantumarticle}

\pdfoutput=1
\usepackage[utf8]{inputenc}
\usepackage[english]{babel}
\usepackage[T1]{fontenc}
\usepackage{amsmath}
\usepackage{hyperref}
\usepackage{graphicx}
\usepackage[dvipsnames]{xcolor}
\usepackage{braket}
\usepackage{bbold}

\usepackage[numbers,sort&compress]{natbib}
\usepackage{amssymb}
\usepackage{placeins}
\usepackage{subcaption}

\usepackage[left=23mm,right=23mm,top=35mm,columnsep=15pt]{geometry}

\usepackage{pgfplotstable}
\usepackage{array}

\usepackage[qm]{qcircuit}

\title{A proposal to demonstrate non-abelian anyons on a NISQ device}
\author{Jovan Jovanovi\'c}
\affiliation{Rudolf Peierls Centre for Theoretical Physics, Parks Road, Oxford, OX1 3PU, UK}
\orcid{0000-0002-2508-3207}
\author{Carolin Wille}
\affiliation{Rudolf Peierls Centre for Theoretical Physics, Parks Road, Oxford, OX1 3PU, UK}
\orcid{0000-0002-9764-6937}
\author{Daan Timmers}
\affiliation{Cavendish Laboratory, University of Cambridge,
Cambridge CB3 0HE, UK}
\author{Steven H. Simon}
\affiliation{Rudolf Peierls Centre for Theoretical Physics, Parks Road, Oxford, OX1 3PU, UK}
\orcid{0000-0001-7757-5978}

\date{21.6.2023}

\begin{document}

\maketitle
\begin{abstract}
In this work we present a proposal for realising non-Abelian anyons on a NISQ device. In particular we explore the feasibility of implementing the quantum double model $D(D_4)$. We propose techniques to drastically simplify the circuits for the manipulation and measurements of anyons. Numerical simulations with realistic noise models suggest that current NISQ technology is capable of probing signatures of non-Abelian anyons far beyond elemental properties such as the non-commutativity of braids. In particular, we conclude that experimentally measuring the full modular data of the model is feasible.   
\end{abstract}
\tableofcontents

\section{Introduction}

In 1977 Leinaas and Myrheim~\cite{Leinaas1977OnTT} 
first proposed the idea of \emph{anyons}\footnote{The term was later invented by Frank Wilczek~\cite{Wilczek}.} -- particles in 2+1 dimensions with fractional statistics that are neither Bosons nor Fermions. Shortly later, Tsui, Stormer and Gossard~\cite{Tsui} discovered the fractional quantum Hall effect, and very rapidly~\cite{Halperin84,Arovas84} it was understood that such fractional quantum Hall systems harbour anyons. Since then, the investigation of topological order and its signature -- anyonic excitation -- has become a major topic in modern condensed matter physics.

Beyond the study of unconventional phases of matter, topological order has been explored and praised for its potential applications in quantum computation~\cite{Nayak} and simultaneously the study of its underlying, rather sophisticated, mathematical structure~\cite{Kitaev2006} has received a lot of interest from the mathematical community.

Today, almost half a century later, our theoretical understanding of anyons in 2+1 dimensions is slowly approaching completion. However, unambiguous experimental evidence of anyons in 'natural' physical systems is still scarce. Recent experiments have beautifully demonstrated the existence of quasiparticles outside the Boson-Fermion dichotomy in quantum Hall systems~\cite{Nakamura_2020,Bartolomei}. However, the more complex types of anyons, so called \emph{non-abelian anyons} for which  braiding two particles changes the (vector-valued) wave-function by a unitary rotation instead of just a phase factor, have not been unambiguously observed so far.

On that grounds, one may argue that most topological phases of matter are just too complicated to exist in nature and are thus more of a mathematical curiosity than an actual physical phenomenon. However, a series of striking experiments~\cite{iqbal2023creation,xu,andersen2022observation} performed recently on noisy intermediate scale quantum computers (NISQ) strongly refutes such  criticism. Two experiments performed on superconducting qubits~\cite{xu,andersen2022observation} demonstrated the non-abelian braiding of mobile lattice defects which behave like Ising anyons embedded into an abelian phase \cite{Bombin2010,Lensky2023}. Another experiment performed on trapped ions~\cite{iqbal2023creation} prepared the topologically ordered ground state of a non-abelian phase and detected an intrinsically non-abelian braiding processes (Borromean rings) via anyon interferometry.

All three experiments indicate that today's quantum computers are capable of simulating states of matter whose complexity exceeds that of abelian topological order which can be seen as a significant step towards topologically protected quantum computation. While open questions of the scalability and the improvement of noise levels are left for the future to decide, it is clear that by now non-abelian anyons have descended from the somewhat esoteric mathematical realm to the concrete and tangible.

Motivated by these findings, we propose an alternative scheme to realise non-abelian anyons on a NISQ device. This scheme has certain advantages and disadvantages compared to Refs.~\cite{iqbal2023creation,xu,andersen2022observation} which we will elaborate on in the next section. Our proposal focuses on Kitaev's quantum double models -- a discrete realisation of lattice gauge theory -- and in particular the topological phase $D(D_4)$. This phase is (Morita) equivalent to the phase realised in Ref.~\cite{iqbal2023creation}. However, its microscopic Hamiltonian and the protocols we propose to demonstrate non-abelian braiding are quite different from the ones in Ref.~\cite{iqbal2023creation}. 

We note, that the simulation of quantum double models and the more general string-net models \cite{Levin_2005}, which include quantum double models as a subset, has been investigated previously in several studies \cite{Cirac,2009Brennen,PRXQuantum.3.040315,goel2023unveiling}. The focus of our work is on the concrete implementation and the simplifications devised to achieve feasibility on a state of the art NISQ device.

In the next section we will present a non-technical summary of our main methods and results. All following sections are devoted to a more technical and in-depth discussion, starting with a review of Kitaev's quantum double models in Section \ref{sec:qm_double}, where we also discuss the implementation of ribbon operators, charge measurements, and the concrete example of $D(D_4)$. In Section \ref{sec:probing} we present a detailed description of the protocols to probe non-abelian anyons and their concrete implementation as quantum circuits of low depth. In Section \ref{sec:num} we show the results of numerical simulations. Section \ref{sec:other_gauge} discusses the feasibility of our protocols for other gauge groups, in particular $S_3$, which would be universal for quantum computation in contrast to $D_4$. In Section \ref{sec:outlook} we summarise our results and comment on future perspectives.

\section{Summary of results}\label{sec:summary_intro}
The main challenge that needs to be overcome in any experiment that realises non-abelian topological order on a NISQ device is an intrinsic and a profound one. By definition, a NISQ device is noisy meaning that beyond a certain circuit depth quantum information is scrambled beyond recognition. On the other hand, preparing a topologically ordered state without measurements and feed-forward protocols (which are prohibitive on certain modern architectures) requires a unitary circuit whose depth scales linearly with the system size. In addition to that, moving anyons on a topological background (again without measurements) requires operators whose circuit depth again scales with the length of the paths. Thus, realising non-abelian anyons on a NISQ device becomes a challenging game of finding ways to circumvent these rather daunting limitations. 

\subsection{A suitable topological phase}

There are three levels on which to tackle this problem. The first and most important is to identify a suitable type of topological order. It is reasonable to further refine our classification of anyons beyond the basic distinction of abelian versus non-abelian. Non-abelian anyons in particular, can be classified by their computational power which correlates with the difficulty of realising them to some extent. 

For some anyon theories, such as the Fibonacci anyons, braiding alone allows one to perform universal quantum computation~\cite{Freedman2002}. In contrast, all anyons obtained from quantum double models of finite groups are not universal for braiding alone. However, their computational power can be further divided and is determined by the complexity of the underlying group. In particular, for non-nilpotent groups like $S_3$, universal quantum computation can be performed with additional measurements~\cite{Mochon2004} while for nilpotent groups such a scheme does not exist. 
When it comes to the implementation of a quantum double model for a group $G$ on some quantum hardware, we note that the degrees of freedom take values in $G$. Thus, the order of the group needs to be small. For any hardware that relies on qubits, which is the case for most set-ups, $|G|=2^n$ immensely simplifies the design of circuits. Lastly, we note that the property of being solvable is beneficial for the reduction of the circuit depth for certain operations. This will be explored in more detail below in Section \ref{sec:probing}. 
With this in mind, we identify $D_4$, the dihedral group and $Q_8$, the quaternion group, both of order eight, which are both solvable and nilpotent groups as the most suitable candidates.
While $D(S_3)$ would be more desirable due to the fact that one can use it for universal topological quantum computation, we find that its order not being a power of two makes it significantly more difficult to implement on a qubit architecture. For an architecture with native qutrits, its implementation would require circuits of similar, if not lower, depths than the ones used for $D(D_4)$ or $D(Q_8)$. An example of an architecture supporting qutrit operations is the photonic simulator featured in Ref.~\cite{goel2023unveiling} on which a proof-of-principle simulation of the fusion rules for $D(S_3)$ on a single lattice site has recently been performed. 

For concreteness, we will focus on $D(D_4)$ in the following. In Ref.~\cite{iqbal2023creation} the topological phase chosen is $D_\alpha(\mathbb Z_2^3)$, i.e., a \emph{twisted} quantum double model of the abelian group $\mathbb Z_2^3$. This phase is (Morita) equivalent to $D(D_4)$~\cite{mapping, propitius1995topological}, which further indicates that $D(D_4)$ is just of the right complexity -- simple enough to be realised on a NISQ device, and complex enough to host non-abelian anyons.
\subsection{Ground state preparation}
Having identified a reasonable phase, i.e., $D(D_4)$ in our case, the next task is to find a suitable microscopic realisation of the model and to prepare its ground state. Here, our protocol differs drastically from that presented in Ref.~\cite{iqbal2023creation}. First of all, the Hamiltonian chosen in Ref.~\cite{iqbal2023creation} is that of $D_\alpha(\mathbb Z_2^3)$. This means, its degrees of freedom (dof) are valued in $G=\mathbb Z_2^3$, while for us the dof are $G=D_4$-valued. In fact, the Hamiltonian in Ref.~\cite{iqbal2023creation} is best understood as a gauged version of a symmetry protected topological phase and the ground state preparation reflects that. 

To be more precise, Ref.~\cite{iqbal2023creation} starts with the preparation of a $\mathbb Z_2^3$ symmetry protected topological (SPT) phase that can be prepared by a constant depth quantum circuit. The internal symmetry of the SPT is then gauged such that the system acquires intrinsic topological order. This gauging protocol is performed using a feed-forward protocol in which the system is entangled to an extensive number of ancillas which are then measured. The measurement outcomes correspond to successful ground state preparation or the preparation of a state with residual, but abelian anyons. The latter can be deterministically removed using error correction such that no post-selection is necessary. However, we emphasise that a feed-forward protocol in which the circuits to be executed depend on intermediate measurement outcomes, is not suitable for all machines, since with certain architectures measurement is expensive -- requiring as much depth as tens to hundreds of gates~\cite{weber}.

Therefore, in our protocol we refrain from using any feed-forward protocols and prepare the ground state directly via a unitary circuit. This has the disadvantage of limiting the achievable lattice size. However, we note, that in order to demonstrate signatures of non-abelian braiding, it is not necessary to use a lattice which is fully two-dimensional. In fact, it is sufficient to consider a quasi-one dimensional geometry, which we refer to as a \emph{braiding ladder} as shown in Fig.~\ref{fig:latticeGS}. For this geometry, we can prepare the ground state with a depth-two circuit. While this ground state does not feature long-range entanglement, this is not needed for the demonstration of non-abelian braiding as we will show explicitly in Section \ref{sec:num}. We also consider a small truly two-dimensional lattice, just for the sake of proving that a direct unitary circuit preparation of the ground state is feasible and feed-forward protocols are not mandatory for the preparation of non-abelian topological order.

The braiding protocols we propose in the following are independent of the specific method of ground state preparation and can be applied on any lattice.

\subsection{Manipulating anyons}
With the ground state preparation in place we lastly turn to the operators which allow us to create and move anyons and to the measurement. All of these operations need comparably short circuits. It is in this area where we think that our work contributes the most and provides results that can be generalised to other quantum double models. To elucidate our achievements we need to briefly review the basics of quantum double models. A full recap is deferred to the main text. The dof in a quantum double model are $G$-valued and group multiplication is an operation as elemental as a spin-flip in a spin-1/2 system. Unfortunately, a single group multiplication requires several Toffoli gates which are non-Clifford and quite costly on most architectures. To be concrete, a single Toffoli gate translates to a depth-6 circuit of elemental gates on Google's Sycamore chip, which we took as the benchmark for current state NISQ devices~\cite{weber}. However, a careful investigation reveals that full group multiplications can be entirely avoided for the creation, manipulation and measurements of anyons. This realisation is one of our main contributions.

To see this, we remind the reader that in a quantum double model for each anyon there is a so-called \emph{ribbon operator} which creates an anyon pair at its end-points. As the name suggest, a ribbon operator is a quasi-one dimensional operator that can be defined for any path and has a finite $\mathcal O(1)$-width. The ribbon operator corresponding to a non-abelian anyon is non-unitary and is most conveniently implemented with ancilla qubits which are measured at the end of the protocol~\cite{Cirac}. In applying the ribbon operators, a sequence of entangling operations between the ancillas and the dof on the lattice are performed giving rise to states that 'know' about the presence of anyons. These entangling operations depend on the anyon type. While they formally involve group multiplication and are, as such, costly, closer inspection reveals that for all anyon types drastic simplifications of the circuits can be performed once we tailor the circuits to the anyon type in question rather than applying a 'one size fits all' protocol. The key here is to make use of the structure of the excitations in the quantum double model. In particular each anyon corresponds to a pair $(\mathcal C,\chi)$, where $\mathcal C$ is a conjugacy class of $G$ and $\chi$ is an irreducible representation of the centraliser $Z_r$, $r \in \mathcal C$. The group multiplication involved in the ribbon operators needs to be performed only for elements of the respective conjugacy class. We show, that exploiting this property drastically reduces the circuit depth and removes all Toffoli gates. Such a complexity reduction for the ribbon operators generalises to other groups, in particular to solvable groups including $S_3$. 

\subsection{Charge measurements}
Finally, we aim to design a protocol which can detect and uniquely determine topological charges. This is a non-trivial task and has so far not been demonstrated. In the experiments performed recently, the existence of multiple fusion channels and the action of non-abelian braiding on the latter has been demonstrated indirectly by measuring abelian charges before and after the braid. However, no experiment directly probed a state where a superposition of several charges had been created from the fusion of non-abelian anyons and identified the anyon content.

This might be due to the inherent difficulty of uniquely determining non-abelian charge content. In the quantum double model the operators needed to measure general charges are explicitly known, however, they require full group multiplications on several lattice-dof and are therefore prohibitively costly. To circumvent this we propose a \emph{partial charge measurement}. To determine the total charge, one needs to evaluate how a given state transforms under the full group. However, one can instead measure how it transforms under a subgroup. Due to partial orthogonality of the characters of a subgroup with those of the group, a measurement outcome reveals partial information about the charge. In the case of $D(D_4)$ one finds that a certain outcome is only compatible with at most two different charges. Repeating the measurement for three different subgroups, we can unambiguously infer the charge. This procedure avoids costly multiplications with the full group and removes unfavourable Toffoli gates from the circuit at the cost of repeating the protocol three times.

\subsection{Probing non-abelian signatures}
With these simplification in place we argue that it is possible to demonstrate the properties of non-abelian anyons of $D(D_4)$ on a NISQ device. In particular we propose two elemental protocols, anyon fusion and anyon braiding, demonstrating the existence of multiple fusion outcomes and non-commutativity of exchange operations, respectively. We furthermore propose protocols for anyon interferometry that allow us to measure the entries of the S- and T-matrices, which fully characterises the anyon content of $D(D_4)$. For the protocols proposed we provide numerical simulations using Google's realistic noisy quantum circuit simulator. All protocols proposed are ready to be run on the actual Sycamore chip and the results obtained from the simulations are representative of the actual experiments, if they were performed on a chip with similar lay-out and noise levels. 

Our numerical findings indicate that current NISQ technology is ready to demonstrate the full signatures of non-abelian anyons in the $D(D_4)$ model. Similar results hold for $D(Q_8)$. We also investigate how our protocols need to be adapted for $D(S_3)$ which hosts non-abelian anyons that can be used for measurement assisted universal topological quantum computation. We find that on a device with native qutrits that support $\mathbb Z_3$ multiplication, the simplifications discussed for $D_4$ carry over to $S_3$. However, for a device with qubits, the circuits for all individual aspects of the protocols are considerably more complicated and involve several Toffoli gates where the equivalent operation for $D_4$ require just CNOT operations. When this is translated into device ready circuits of two-qubit gates, this yields to an increase of the depth by a factor ten or larger, rendering it unsuitable for execution on current NISQ devices.

\section{Quantum double models}\label{sec:qm_double}
In this section, we will discuss Kitaev's quantum double models~\cite{Kitaev_2003}. While we assume that our readers are largely familiar with quantum double models, we nevertheless include this review to set notations and conventions, which vary throughout the literature. In addition to that, the last two parts of this section include additional material crucial for our implementation. In particular, we discucss the concrete protocol for applying ribbon operators and a non-standard protocol to infer anyonic charge via a so-called \emph{partial} charge measurement. 

Kitaev's quantum double models are (2+1d) Hamiltonian formulations of lattice gauge theory for finite gauge groups. Gauss' law is enforced energetically at each vertex by a Hamiltonian term and the model is at the deconfinement fixed point, where there are no electric field terms.
The Hamiltonian, therefore, has two sets of terms -- the gauge-invariant (magnetic) plaquette terms and the Gauss' law vertex terms~\cite{cui2018topological, Kitaev_2003}. 

Quantum double models can also be understood as a subclass of the more general string-net models~\cite{Levin_2005}, which describe all non-chiral (2+1d) topological phases of matter, or as a generalisation of Kitaev's toric code~\cite{Kitaev_2003} for which the gauge group $\mathbb Z_2$ is generalised to an arbitrary discrete group $G$. While all quantum double models have anyonic excitations, the anyons for models with abelian gauge group are themselves abelian. In order to obtain non-abelian anyons, it is necessary to consider non-abelian gauge groups $G$. While the models for the latter are conceptually still very similar to the toric code, their definitions require slightly more care and notation, which we will introduce in the following.

\begin{figure*}
    \centering
    \begin{subfigure}[b]{0.45\textwidth}
        \centering
        \includegraphics[width= \linewidth]{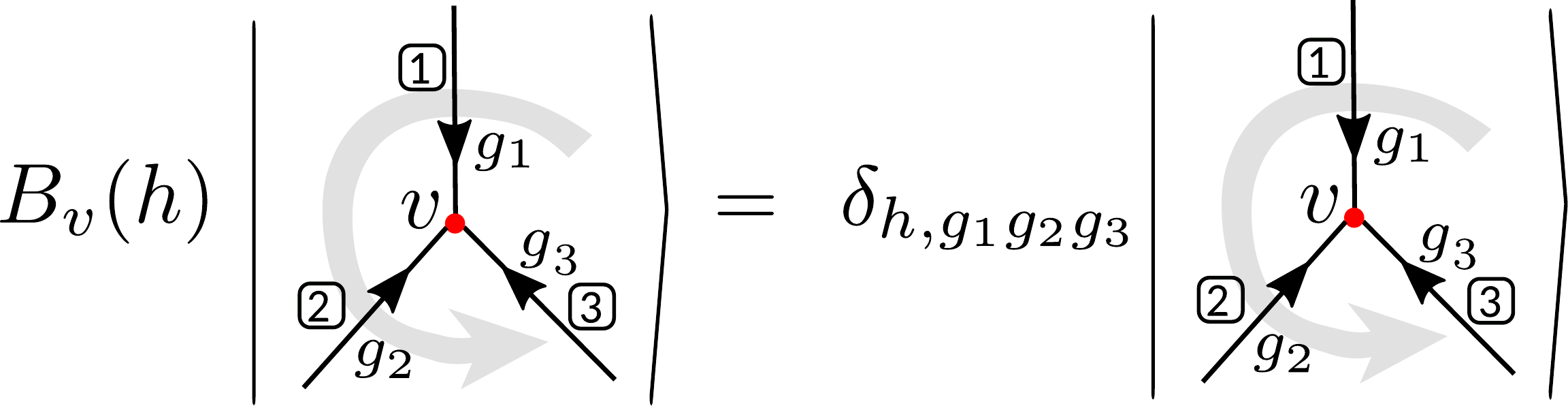}
        \caption{Vertex operator. The vertices are oriented in accordance with the plaquettes (counter-clockwise) and have a starting edge to make the group multiplication assignment unambiguous.}
        \label{eqn:Bs_def}
    \end{subfigure}\hfill
    \begin{subfigure}[b]{0.45\textwidth}
        \centering
        \includegraphics[width = \linewidth]{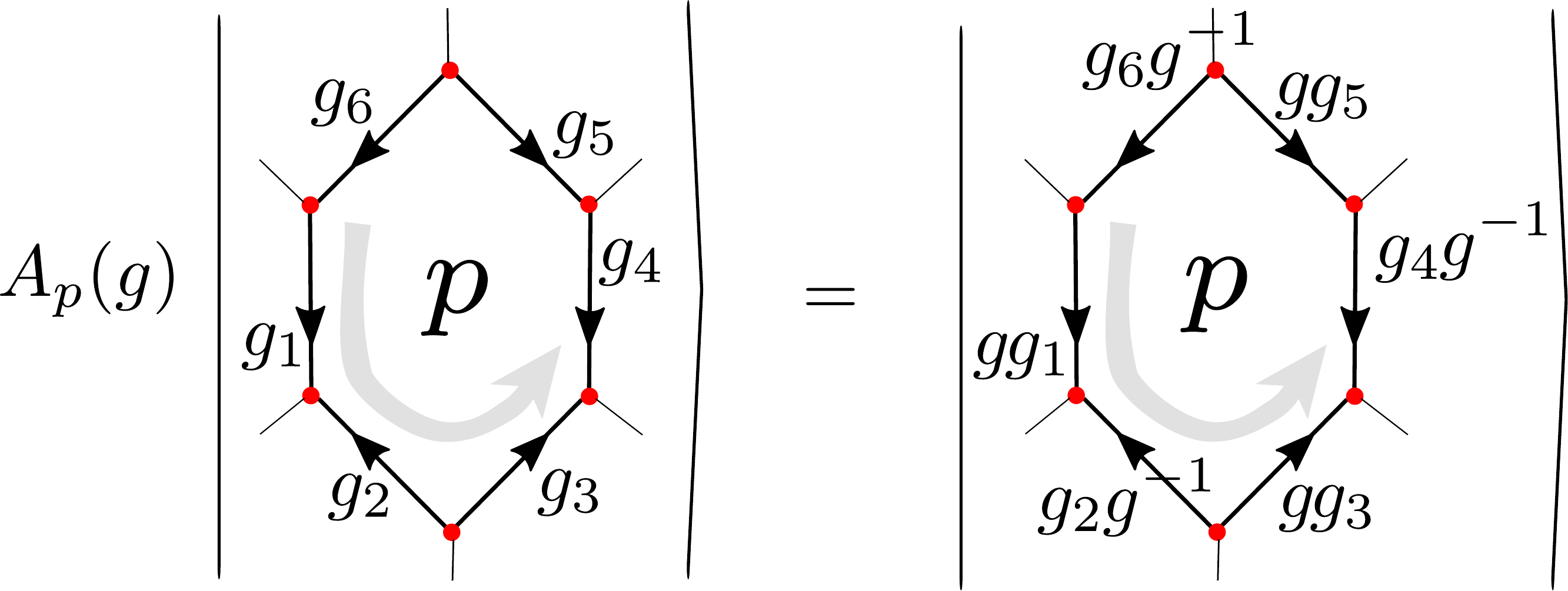}
        \caption{Plaquette operator. The orientation of the plaquettes determines pre or post multiplication with the inverse and is chosen in agreement with the orientation of the vertices (counter-clockwise).}
        \label{eqn:As_def}
    \end{subfigure}\hfill
    \caption{The vertex and plaquette operators.}
    \label{fig:vertex_ops}
\end{figure*}

\textbf{Hamiltonian.}
For a given group $G$ we can define its quantum double model on any arbitrary cellulation of a surface without boundary. To consistently define the model, the edges (1-cells) need to be oriented as will be explained in more detail below. The local degrees of freedom are $|G|$-dimensional and assigned to the edges. The basis of their local Hilbert space is labeled by the group elements, i.e., we think of edges as being labeled by elements $g\in G$. The Hamiltonian is given by a sum of mutually commuting terms that act on vertices (0-cells) $V$ and plaquettes (2-cells) $P$, respectively
\begin{equation}
    H = -  \sum_{v \in V} \mathbf B_v -  \sum_{p  \in  P} \mathbf A_p \label{eqn:ham} \;.
\end{equation}
Note, that we will here use a formulation of the theory on the \emph{dual} lattice compared to the lattice of the original work in Ref.~\cite{Kitaev_2003}.
We will now discuss these terms in more detail. As mentioned above, the vertex term enforces Gauss' law. To achieve this we first introduce a general vertex operator $B_v{(h)}$ for every vertex $v$. This operator projects onto all states for which the group elements assigned to the edges adjacent to the vertex multiply to $h$. To make the product unambiguous we need to order the edges. This ordering has to fulfil additional constraints to be specified momentarily. In addition, a group element $g$ assigned to an incoming (outgoing) edge enters as $g$ ($g^{-1}$). E.g., for the trivalent vertex depicted in Fig.~\ref{eqn:Bs_def} we have 
\begin{equation}
B_v{(h)} |g_1,g_2,g_3\rangle= \delta_{g_1 g_2 g_3,h} |g_1,g_2,g_3\rangle \;.	
\end{equation}
Gauss' law is then enforced by choosing $\mathbf B_v=B_v{(e)}$, where $e$ denotes the identity element of the group. To ensure that the vertex projector commutes with the plaquette projector introduced below, the ordering of the edges needs to be consistent with the orientation of the latter. This can be done by endowing both with a counter-clockwise orientation. The ordering is then obtained by additionally specifying a starting edge for each vertex.

The plaquette term 
	$\mathbf A_p=\frac{1}{|G|} \sum_{g \in G} A_p{(g)}$
is defined in terms of operators $A_p{(g)}$ which shift the labels of the edges forming the plaquette by $g$. As alluded to previously, the plaquettes have an \emph{orientation}. If the edge direction is aligned (anti-aligned) with this orientation, the shift acts as $g_i \rightarrow gg_i$ ($g_i \rightarrow g_ig^{-1}$). E.g., for the plaquette shown in Fig. \ref{eqn:As_def}, we have 
\begin{equation}
	A_p{(g)}\ket{g_1, g_2, \ldots} = \ket{gg_1, g_2g^{-1}, \ldots}.
\end{equation}

\textbf{Ground state.}
It is not hard to verify that all terms in the Hamiltonian commute. Hence we can diagonalise it term by term. One can show that the leftover degeneracy depends only on the genus of the surface the graph is embedded in~\cite{Kitaev_2003, cui2018topological}. We will work on a sphere-topology, for which the ground state is unique. For most quantum computing architectures, interactions need to be local, which limits the accessible topologies. However, a disk or sphere topology (a disk closed off by one large plaquette) is accessible.

All terms in the Hamiltonian are also projectors. Hence, one way to construct the ground state is to apply all projectors onto a state that has non-zero overlap with the ground state. In particular, we can start with the state $\ket{\{e\}}$, where every edge is labelled by the identity element.
This state trivially obeys all vertex projectors, so we just need to apply all plaquette projectors
\begin{equation}
    \ket{\psi} = \prod_{p \in P} \mathbf A_p \ket{\{e\}}.
\end{equation}
This state is the unique ground state and corresponds to the equal weight superposition of all states respecting Gauss' law.

\subsection{Anyon content}\label{sec:anyon}
In the following, we will discuss the anyonic excitations in quantum double models and the algebra describing them.

\textbf{The algebra $D(G)$.} The ground state is stabilised\footnote{Meaning, it is the $+1$ eigenstate.} by the projectors in Eq.~\eqref{eqn:ham}. Together with the property $A_p{(g_1)}A_p{(g_2)} = A_p{(g_1 g_2)}$ this implies the stronger condition
\begin{equation}
\begin{split}
    A_p{(g)} \ket{\psi} = \ket{\psi},\\
    B_v{(e)} \ket{\psi} = \ket{\psi},
\end{split}\label{eqn:ground_state}
\end{equation}
for all $v \in V$, $p \in P$ and $g \in G$. Therefore, the elementary excitation above the ground state will violate one or more of the equations above and are characterised by how the operators $B_v(h)$ and $A_p(g)$ act on them. More precisely, these operators form an algebra~\cite{cui2018topological, Kitaev_2003} (the quantum double algebra $D(G)$), which contains the full information about the anyonic excitations. In particular its irreducible representations (irreps) label the anyons.

For the toric code model, the $A$ and $B$ operators always commute and their algebras can be investigated independently. One finds the well-known $e$ (for electric) and $m$ (for magnetic) particles associated to vertex and plaquette violations, respectively, and their combination, the fermionic $(e,m)$-particle. However, for non-abelian gauge-groups the operators $A_p(g)$ and $B_v(g)$ no longer commute, if the vertex $v$ intersects the plaquette $p$. To discuss their joint algebra, we consider \emph{sites}
 $s_i = (v_i, p_i)$ of adjacent vertices and plaquettes.
 
On the same site we have the following algebraic relations 
\begin{equation}
    \begin{split}
        A_s{(g)}A_s{(h)} = A_s{(gh)}, \\
        B_s{(g)}B_s{(h)} = \delta_{g,h} B_s{(h)},\\
        A_s{(g)}B_s{(h)} = B_s{(ghg^{-1})}A_s{(g)}.
    \end{split}\label{eqn:alg}
\end{equation}
This is the on-site representation of the quantum double algebra $D(G)$~\cite{cui2018topological, Kitaev_2003}.

We will now discuss its irreducible representations. However, we will refrain from providing any derivations (see e.g. Ref.~\cite{Cui_2015}) and just state the results. 

The irreducible representations are labelled by two objects, a conjugacy class $C$ of the group $G$ and an irreducible representation $\chi$ of the centraliser $Z(r)$ of the class representative $r \in C$. The vector space on which $(C, \chi)$ acts is spanned by a basis $\ket{\mu} = \ket{c, i}$, where $c \in C$ and $i \in \{1, 2, \ldots, \text{dim}\chi\}$, i.e., the first index goes over the conjugacy class elements while the second goes over the vector indices of the irreducible representation $\chi$.

Note, that in the case of abelian groups, in particular the toric code, the conjugacy classes are trivial and identical to the group elements themselves. Their center is $G$, which has $|G|$ one-dimensional representations isomorphic to $G$ itself. Hence, the irreducible representations are given by $|G|^2$ tuples $(g,\rho_i)$, where any group element $g$ is paired with any irreducible representation $\rho_i$, $i=1,\ldots, |G|$.

In the general, non-abelian case the irreducible representations do not factorise as can be seen from the action of the algebra generators on the vector space spanned by $\ket{\mu}=\ket{c,i}$ %
\begin{equation}
    \begin{split}
        B_{\mu\nu}{(h)}=\bra{c, i} B{(h)} \ket{c', i'} = \delta_{c, h} \delta_{c, c'}\delta_{i, i'}, \\
        A_{\mu\nu}{(g)}= \bra{c, i} A{(g)} \ket{c', i'} = \delta_{c,gc'g^{-1}} \Gamma^\chi_{c,ii'}(g).
    \end{split}
\end{equation}
Here, $\Gamma^\chi_c$ is a map from the entire group, $G$, onto the $\chi$-representation matrices defined by composing the representation matrices $\Gamma^\chi$ themselves and a map $G\to Z(r)$ defined by $g \mapsto q_{c}^{-1}gq_{c'}$, where $q_c$ is a group element that satisfies $q_c c q_c^{-1} = r$ and $c'=g^{-1}cg$.

To get a better understanding of the meaning behind these expressions, we consider three simple examples. Let us start with the vacuum (or trivial) representation, labelled by $(\{e\}, \mathbb{1})$. This representation is one-dimensional and spanned by $\ket{e, 0}$
\begin{equation}
    \begin{split}
        B{(h)}\ket{e, 0} = \delta_{h,e}\ket{e, 0},\\
        A{(g)}\ket{e, 0} = \ket{e,0}.
    \end{split}
\end{equation}
Hence, Eq.~\eqref{eqn:ground_state} implies that for the ground state every site houses the trivial representation.

Other important examples are pure charges and pure fluxes. A pure flux is labelled by a conjugacy class and the trivial representation of its centre, $(C, \mathbb{1})$. Its basis vectors are $\ket{c, 0}$ for $c \in C$, with
\begin{equation}
    \begin{split}
        B_{\mu\nu}{(h)}=\bra{c, 0} B{(h)} \ket{c', 0} = \delta_{c, h} \delta_{c, c'}, \\
        A_{\mu\nu}{(g)}= \bra{c, 0} A{(g)} \ket{c', 0} = \delta_{c,gc'g^{-1}}.
    \end{split}
\end{equation}
Pure flux excitations only violate the vertex term, the $B$-term.

Pure charge excitations are labelled by the group identity and a representation of the group $G$ itself, $(\{e\}, \chi)$. Its basis vectors are $\ket{e, i}$ for $i \in \{1, 2, \ldots, \text{dim}\chi\}$, with
\begin{equation}
    \begin{split}
        B_{\mu\nu}{(h)}=\bra{e, i} B{(h)} \ket{e, i'} = \delta_{e, h},\\
        A_{\mu\nu}{(g)} = \bra{e, i} A{(g)} \ket{e, i'} = \Gamma^\chi_{ii'}(g).
    \end{split}
\end{equation}
Pure charge excitations only violate the plaquette term, the $A$-term.

In particular, if we have a gauge field state, $\ket{\chi, p; i}$, where each site houses a trivial representation except for one, $(v, p)$, which is occupied by a pure charge $(\{e\}, \chi)$\footnote{Note, that such a configuration is impossible on a sphere, but may occur on manifolds of genus $g>0$.}, this state satisfies all the constraints in Eq.~\eqref{eqn:ground_state} except for
\begin{equation}
    A_p{(g)}\ket{\chi, p; i} = \sum_{i'} \Gamma^\chi_{ii'}(g)\ket{\chi, p; i'},\label{eqn:tranfs}
\end{equation}
where $i$ and $i'$ are the internal degrees of freedom of the charge\footnote{Note that the charge can be vector valued for non-abelian symmetry groups.}.
This is the way a charged state transforms under gauge transformations in gauge field theory. Hence, we say that the plaquette terms generate gauge transformations.

All other excitations which are neither pure charge nor pure flux are called dyons. They violate vertex and plaquette terms simultaneously, meaning they have a flux component associated with a vertex of the site $(v,p)$ and a charge component associated with its plaquette, but unlike the toric code fermion they cannot generally be broken down to a combination of pure charge and pure flux sitting next to one another.

\textbf{Non-abelian anyons.} To understand the distinction between abelian and non-abelian anyons we will focus on the physical meaning of the dimension $d = \text{dim}(C, \chi) = |C|\text{dim}(\chi)$ of the irreducible representations.

If we have a gauge field state with an anyon of type $(C, \chi)$ at a site $s$, the plaquette and vertex terms of that site will transform this state in accordance with that $d$-dimensional algebra representation. This implies that specifying the type and location of this anyon does not uniquely fix the gauge field state. Instead, there is a $d$-dimensional subspace $\mathcal H_s(C,\chi)$ of the total Hilbert space associated with this anyon occupying this site. This $d$-fold degeneracy can be interpreted as a spin-like internal degree of freedom of the anyon.

Generalising this, we find that for a state with specified charge content $\{(C_s, \chi_s)\}_s$ on all (non-overlapping) sites $s$ the subspace associated to this configuration is
\begin{equation}
	\mathcal{H}_{\{s\}} = \bigotimes_s \mathcal H_s (C_s, \chi_s).
\end{equation}

A more powerful alternative to this local description can be derived, if we notice that
there is an algebra associated with the tensor product of representations, analogous to the Clebsch-Gordan (CG) decomposition of tensor products of linear representations of a group into the direct sum of irreducible representations. 

In particular, if we have two charges $a$ and $b$ the associated Hilbert space can be written as a direct sum of the Hilbert space associated to charges $c$. We write this as 
\begin{equation}
	a \otimes b = \bigoplus_{c}N^c_{ab} c,\label{eqn:fuse} 
\end{equation}
with $a$, $b$ and $c$ going over a set of anyon labels $(C, \chi)$ and $N_{ab}^c$ being integer coefficients.

How this manifests physically is that if we have two anyons $a$ and $b$ in some region and measure the topological charge associated to that region we may get any label $c$ for which $N_{ab}^c \neq 0$. This process is referred to as anyon fusion.

The general expression for $N_{ab}^c$ is cumbersome. For pure charge anyons $(\{e\}, \chi_i)$ it readily reduces to the well-known  decomposition of a tensor product of group irreps into the direct sums of irreps $\chi_i\otimes\chi_j = \bigoplus_k n^k_{ij} \chi_k$.

If the gauge group is abelian, all algebra representations are one-dimensional and there is no degeneracy once the charge content of a gauge field is specified. The fusion is unambiguous. 
We can see that by looking at the dimensions of the LHS and RHS of Eq.~\eqref{eqn:fuse}. For every $a$ and $b$ there is only one $c$ for which $N_{ab}^c=1$ and it is zero for all other $c$.

The Hilbert space associated with the presence of multiple non-abelian anyons is the stage on which all striking phenomena of non-abelianess are played out. Besides the possibility of multiple fusion outcomes discussed above,  also moving anyons around one another (braiding) acts non-trivially on this space and corresponds to a unitary operation. In such a braiding process the order of exchanges matter as in general the unitary matrices associated to the individual exchanges do not commute. The full theory that describes the braiding and fusion of anyons is a so-called unitary modular tensor category (UMTC)~\cite{Kitaev_2003}.
\subsection{Ribbon operators}\label{sec:ribbon_ops}

\begin{figure}
    \centering
    \includegraphics[width= \linewidth]{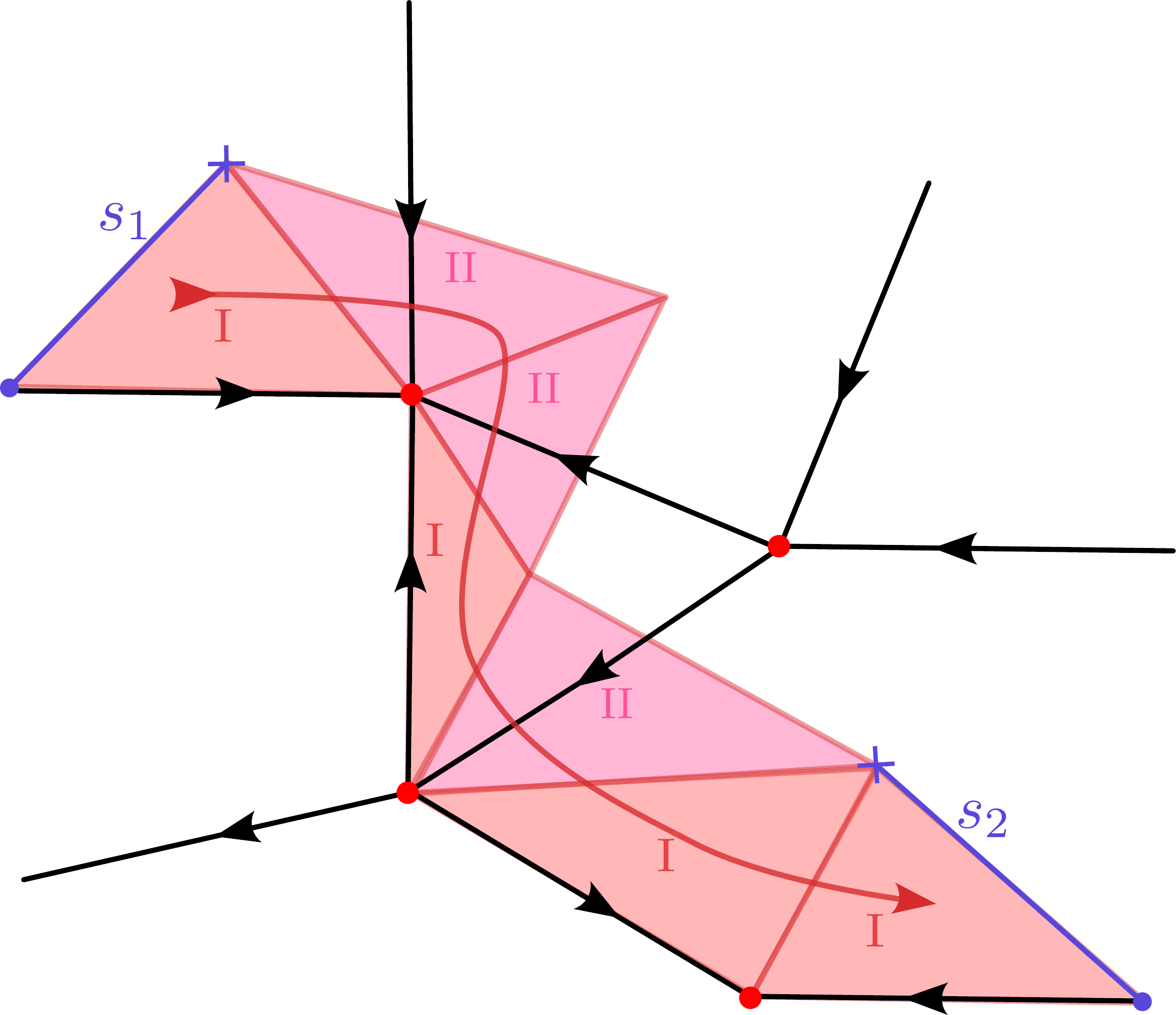}
    \caption{An example of a ribbon, $R = \{t_1, t_2, \ldots, t_7\}$, between site $s_1$ and $s_2$. The black lines are the edges of the graph. The ribbon is made up from four type I elementary triangles and three type II elementary triangles.}
    \label{fig:rib_exampl}
\end{figure}

In the following, we will explain how to create and move anyons. Anyons can always be created in pairs from the vacuum and for any anyon type, $(C, \chi)$, and any path on the graph between two sites, see Figure \ref{fig:rib_exampl}, there is a \emph{ribbon operator} that creates a pair of said anyons at the end sites.

If the ribbons are closed and contractable, the associated ribbon operators leave the ground state unchanged. Moreover, they span the loop operator algebra that leave the ground state invariant. It is in that way that the quantum double ground state knows about the anyon spectrum.

We will not explain the derivation of the ribbon operators themselves, see Ref.~\cite{Kitaev_2003,cui2018topological}, just how to apply them to a state.
The ribbon operators are not unitary in general. If the anyons have a dimension larger than one, the operators that create and manipulate them are non-local projectors. To simulate them on a digital quantum computer requires ancillas and measurements~\cite{Cirac}. 

In Ref.~\cite{andersen2022observation} the group overcame this issue by means of unitary lattice deformations, unitarily transforming from a state with a set of non-abelian defects on one graph to another state with the same defect content but defined on a different graph, hence they were able to move the non-abelian Majorana defects unitarily. However, these are extrinsic and static lattice defects on top of a theory that is an abelian $\mathbb Z_2$ gauge theory, hence, the nature of their non-abelian anyons is different from intrinsic gauge field excitations.

As we mentioned, a pair of $d$-dimensional non-abelian anyons defines a degenerate subspace of the gauge field Hilbert space that encodes the outcomes of their fusion. This encoding is increasingly non-local as we separate the anyons. However, we require accessing this information locally to move the anyons. To this end, we keep one $d$-dimensional ancilla qudit per end of the ribbon.

Every ribbon (cf. Fig.~\ref{fig:rib_exampl}) is made up from two types of elementary triangles
\begin{itemize}
    \item[I)] A triangle consisting of two vertices and one plaquette center of the underlying graph. One side of the triangle coincides with an edge of the graph.
        \item[II)] a triangle consisting of two plaquette centers and one vertex of the underlying graph. One of the triangle's sides crosses an edge of the graph.
\end{itemize}
Moving anyons means appending elementary triangles onto a ribbon. Each triangle type corresponds to a specific operator, the details of which also depend on the orientations of the edges and whether the triangle is attached to the back or front end of the ribbon. A detailed list is provided in Fig.~\ref{fig:al_trigs} in Appendix \ref{app:ribs}.

\textbf{The algorithm} that creates two anyons of type $(C,\chi)$ along a ribbon path in the internal state $\ket{\alpha; \beta} = \ket{c', i'; c, i}$ is the following
\begin{enumerate}
    \item Initialise two ($|C| \times \operatorname{dim} \chi$-dimensional) ancilla qudits in the states $\ket{\alpha; \beta} = \ket{c', i'; c, i}$.  We will think of the first qudit as the ribbon's back end and the second as its front end.
    \item For each triangle in the ribbon path, depending on the triangle \textit{type} (and the edge orientations), sequentially apply one of the following unitaries acting on ancilla qudits $\ket{c, i}$ and qudits encoding the group element $g$ associated to the edge of the lattice $\ket{g}_\text{phys}$  
    \begin{itemize}
        \item[I)] Multiply on the coinciding edge\\  $\ket{c, i}\ket{g}_\text{phys} \rightarrow \ket{c, i}\ket{cg}_\text{phys}$.
       \item[II)] Generalised-conjugate by the crossed edge\\ $\ket{c,i}\ket{h}_\text{phys} \rightarrow \ket{hch^{-1}} \Gamma^\chi_c(h) \ket{i} \ket{h}_\text{phys}$.
    \end{itemize}
    \item To complete the application of the ribbon operator project the ancilla qudits to the Bell state $\bra{\Phi^+} = \frac{1}{\sqrt{d}}\sum_\nu \bra{\nu; \nu}$. The projection is done by measurement and post-selection.
\end{enumerate}
For the other variants of Step 2 (for different edge orientations etc.) consult Figure \ref{fig:al_trigs} in Appendix \ref{app:ribs}. 

In the way presented above, we have started building up the ribbon from start-to-finish using only forward-type elementary triangles. Similarly, we could have started in the middle and extended in parallel both forward and backwards with appropriate types of elementary triangles, see Appendix \ref{app:ribs}. Backwards-type elementary triangles, analogously to Step \textbf{2.} stated above, act on the backwards ancilla qudit.

As we can see, the operation is sequential so at best the depth of a circuit implementing this is $\mathcal{O}(|R|)$, with the best depth achieved by starting at the middle and growing it both ways.

We can make this constant depth by separating the main ribbon into $|R|$ smaller ribbons which however requires $|R|$ pairs of qudits. To merge these ribbons, we also need to work with $2R$ ancillas (one pair for each cut of the circuit) and $|R|$ Bell-pair projections. As these are done via measurement and post-selection this requires an exponential in $|R|$ number of repetitions for a $\mathcal{O}(1)$ number of successful projections. Note, that this procedure is independent of quantum double models and can be applied to any sequential circuit.

In the variant of the protocol we proposed, the post-selection will necessarily raise concerns about the scaling of the number of total runs needed to build up an adequate number of successful runs. In fact, the protocol for applying an open ribbon operator onto the ground state succeeds with a probability of $1/d^2$ (neglecting the effect of noise). However, if the ribbon is closed, due to the no-flux condition of the ground state, the protocol succeeds with certainty. The number of runs needed scales exponentially with the number of open ribbon operators in the braiding protocol, which, in the cases we consider, is either one or two. 

To interpret what we have here, let us look at the quantum resources. We have qudits representing the matter $\mathcal{H}_{\text{matter}}$ and ancilla qubits representing the internal state of the anyons, or the fusion space, $\mathcal{H}_{\text{ancillas}} \equiv \mathcal{H}_{\text{fusion}}$ which is already embedded non-locally in  $\mathcal{H}_{\text{matter}}$. The two types of triangles couple the two spaces in two different ways. For type I the $\mathcal{H}_{\text{ancillas}}$ controls an action on $\mathcal{H}_{\text{matter}}$ and for type II it is the other way around. After the measurement that disentangles the redundant copy of $\mathcal{H}_{\text{fusion}}$, $\mathcal{H}_{\text{matter}}$ is left in a state with the ribbon operator imprinted on it as signalled by the topological charge, braiding amplitude and phase.

\subsection{Charge measurements}\label{sec:redchmmt}

In this section, we will explain how we measure the topological charge, i.e., the anyon label. This is the last element of our toolkit for probing topological order. An ideal charge measurement would differentiate any type of excitation on any site, such measurement is associated with the following set of projectors ~\cite{cui2018topological}
\begin{equation}
    P_s^{(C, \chi)} = \frac{\chi(e)}{|Z(r)|}\sum_{c \in C}\sum_{z \in Z(r)}\chi^*(z)B_s^{(c)}A_s^{(q_c z \bar{q}_c)},
\end{equation}
for each irreducible representation of $D(G)$.

Since we work in a basis where projectors onto a given $G$-valued flux through a vertex $v$, $B_v{(h)}$, are diagonal, we can just do a full projective measurement of the gauge field degrees of freedom in that basis. 
The result of such a measurement is a labelling of all edges by group elements from which we can compute the flux through any vertex. 

However, in order to measure the charge component, one needs to implement a controlled $A_s{(g)}$ operator which needs a controlled group multiplication applied to all edges of $s$'s plaquette, i.e. $\ket{g}\ket{g_i} \rightarrow \ket{g}\ket{gg_i}$ for all $\ket{g_i}$ around the plaquette.
This requires circuits which, in the case of non-abelian groups such as $D_4$ which we will study in more detail, are prohibitively expensive. 

The main idea to circumvent this problem is to instead use partial charge measurements that have a substantially reduced circuit depth. Such partial measurements do not determine the charge completely. However, when we combine a set of different partial measurements we are able to deduce the full charge content from the measurement outcomes.\footnote{Note, that these measurements are destructive, meaning we need to prepare the state again for the next measurement. However, if we only need to use a small number of different subgroups (in the case of $D(D_4)$ considered here at most three different subgroups are sufficient), this trade-off is acceptable.}

The key idea behind the partial measurement protocol is that controlled $A_s^{(g)}$ becomes significantly simpler to perform once we restrict ourselves to a proper subgroup of the full group, i.e. $g \in H \subset G$.

With Eq.~\eqref{eqn:tranfs} in mind, we propose the following algorithm for the $H$-partial charge measurement on a plaquette $p$
\begin{enumerate}
    \item Prepare an ancilla qudit, $a$, encoding the elements of $H\subset G$, in an equal superposition over all elements, so that the joint total state of the system is $$ \frac{1}{\sqrt{|H|}} \sum_{h \in H} \ket{h}_a\ket{\psi}_\text{phys}, $$where $\ket{\Psi}_\text{phys}$ is the physical system.
    \item Apply an $a$-controlled $A$-multiplication onto the edges of the plaquette $p$$$ \frac{1}{\sqrt{|H|}} \sum_{h \in H} \ket{h}_a\ket{\psi}_\text{phys} \rightarrow  $$ $$ \frac{1}{\sqrt{|H|}}\sum_{h \in H} \ket{h}_a A_p{(h)} \ket{\psi}_\text{phys}. $$
    \item Apply a unitary $$ U_a = \sum_{\chi_H}\sum_{i, j}\sum_{h\in H} \sqrt{\frac{d_\chi}{|G|}}  \bar \Gamma^{\chi_H}_{ij}(h)  \ket{\chi_H; i, j}_a\bra{h}_a $$ onto the ancilla qudit $a$. Here $\chi_H$ labels the irreducible representations of $H \subset G$ and $\Gamma^{\chi_H}(h')$ are the representation matrices, with $i$ and $j$ being the vector indices for a given representation $\chi_H$. The state $\ket{\chi;i,j}$ should be understood as one of the basis states of a $\sum_{\chi} d_\chi^2$-dimensional vector space which simply enumerate the irreps and their corresponding vector spaces. When $H$ is Abelian, this reduces to a $\sum_\chi=|G|$ dimensional space, labeled by the irreps alone. A concrete qubit encoding for $H=\mathbb Z_4$ and $H=\mathbb Z_2 \times \mathbb Z_2$ is presented in Eqs.~\eqref{eq:enc_irr1} and \eqref{eq:enc_irr2}.
        \item Measure the ancilla qudit $a$.
\end{enumerate}

To see how this protocol works, let us examine the case of a  gauge field with well-defined pure charge on a plaquette  $\ket{\psi}_\text{phys} = \ket{\chi; i}$. Using  Eq.~\eqref{eqn:tranfs} we can write the joint state after Step~2 as
\begin{equation}
   \frac{1}{\sqrt{|H|}} \sum_{h \in H,j} \ket{h}_a \Gamma^{\chi}_{ij}(h) \ket{\chi; j}.
\end{equation}
If $H = G$, the state after Step~3 becomes\begin{equation}
    \begin{split}
       \frac{1}{\sqrt{|G|}}  \sum_{h \in G,j} U_a\ket{h}_a \Gamma^{\chi}_{ij}(h) \ket{\chi; j} = \\
        \frac{1}{|G|}  \sum_{h,\chi',i', j',j} \sqrt{d_\chi} \bar \Gamma^{\chi'}_{i'j'}(h)\Gamma^{\chi}_{ij}(h) \ket{\chi'; i', j'}_a \ket{\chi; j} =\\
        \sum_j \frac{1}{\sqrt{d_\chi}} \ket{\chi; i, j}_a \ket{\chi; j}
    \end{split}
\end{equation}
The decoupling we see in the last line after summing over $h$ is guaranteed by Schur's orthogonality lemma.

In the case above, the result of the measurement in Step~4 is a label $(\chi, i, j)$, representing the charge, the internal state before the measurement and the internal state after the measurement.

If we, however, take $H \subset G$, then the charge information is partial.
By partial charge information, we mean that the result of the measurement in the last step, the label $(\chi_H, i, j)$ is no longer compatible with only one charge but a set of charges, i.e., the charge is not fully determined.

We may repeat the procedure using different subgroups $H \in G$ to gather further partial information on the charge in the hope that we will be able to deduce the charge fully. In the considered examples, choosing different subgroups proves to be sufficient. This relies on the partial orthogonality of character tables of a group and its subgroup, and is demonstrated for the example of the group $D_4$ below.

\textbf{Beyond pure charges.} The partial charge measurement scheme is exact, i.e., unambiguous, when the chosen subgroup coincides with the centre of the conjugacy class of an anyon $(C, \chi)$, i.e., $H = Z(r)$ for $r \in C$.
The label $\chi_H$ in fact is the label of the irreducible representation of the $Z(r)$ labelling the dyon.

The full measurement protocol, in general, then consist of performing a $H$-partial charge measurement and then reading-off the flux $f$ on the same site. We then compute the center of the measured flux $Z(f)$ and consider the following three cases.

If $H = Z(f)$, the protocol is complete. The measurement outcome corresponds to $(C_f, \chi_H)$.

If $H \subset Z(f)$, we need to perform partial charge measurements for other subgroups of $G$ that are also subgroups of $Z(f)$. We then combine these results to determine the charge label uniquely. This requires partial orthogonality of the character tables, see Section \ref{sec:D4_double} for an example.

If $H \not\subset Z(f)$ the measurement is discarded in post-selection and we switch to a different subgroup $H$.

\subsection{Quantum double of $D_4$}\label{sec:D4_double}

Throughout the rest of the paper (except for Section \ref{sec:other_gauge}) we will focus on the group $D_4$ and its lattice gauge theory.
Hence in this section, we will describe the group structure of $D_4$, its quantum double algebra, as well as the representation theory of both.

The dihedral group of order 8 is the symmetry group of a square. It is generated by a $\pi/2$-rotation $r$ and a reflection  $m$ along a diagonal.
The group law is defined by the following identities
\begin{equation}
	\begin{split}
		r^4 = e,\;
		m^2 = e,	\;	mr = r^3m. \label{eqn:group}
	\end{split}
\end{equation}

This group is solvable and all of its proper subgroups are abelian. It can be decomposed as $D_4 = \mathbb{Z}^m_2 \ltimes \mathbb{Z}^r_4$.
This decomposition is not unique, $D_4 = \mathbb{Z}_2^m\ltimes(\mathbb{Z}_2^{r^2}\times\mathbb{Z}^{mr}_2)$ is also a valid decomposition.\footnote{The superscripts in $\mathbb{Z}_n^x$ label the group generator.}

The conjugacy classes of $D_4$ alongside their centres are listed in Table \ref{tab:conjs}.
\begin{table}[h]
\centering    \begin{tabular}{|c | c|}\hline
         $C$ & $Z(r) \text{, } r\in C$\\ \hline $\mathcal C_e = \{e\}$ & $D_4$\\ $\mathcal C_{r^2}= \{r^2\}$ & $D_4$\\ $\mathcal C_{r}=\{r, r^3\}$ & $\mathbb{Z}_4^r \equiv H_r$\\ $\mathcal C_m=\{m, mr^2\}$& $\mathbb{Z}_2^m \times \mathbb{Z}_2^{r^2}\equiv H_m$ \\ $\mathcal C_{mr}=\{mr, mr^3\}$& $\mathbb{Z}_2^{mr} \times \mathbb{Z}_2^{r^2}\equiv H_{mr}$\\
         \hline 
    \end{tabular}
    \caption{The conjugacy classes of $D_4$ alongside their centres.}\label{tab:conjs}
\end{table}
There are five conjugacy classes, which implies that there are also five irreducible representations. Their dimensions are $(1,1,1,1,2)$. The characters, i.e., the traces of the representation matrices, are given in Table \ref{tab:char}.
\begin{table}[h]
\centering    \begin{tabular}{|c|c c c c c |}
         \hline
         $D_4$ & $\mathcal C_e$ & $\mathcal C_{r^2}$ & $\mathcal C_{r}$ & $\mathcal C_m$ & $\mathcal C_{mr}$ \\
         \hline
         $1$ & 1 & 1 & 1 & 1 & 1 \\
         $\alpha_r $& 1 & 1 & 1 & -1 & -1  \\
         $\alpha_{m}$ & 1 & 1 & -1 & 1 & -1  \\
         $\alpha_{mr}$ & 1 & 1 & -1 & -1 & 1  \\
         $\epsilon$ & 2 & -2 & 0 & 0 & 0  \\
         \hline 
    \end{tabular}
    \caption{Character table of $D_4$.}\label{tab:char}
\end{table}

When labelling the representations of the algebra $D(D_4)$, we will also need the irreducible representations of the centres of the conjugacy classes. The centres of the three non-trivial conjugacy classes are abelian and have four elements. Hence, they have four one-dimensional irreducible representations. Their character tables are shown in Table~\ref{tab:char_sub}.
\begin{table}[h]
\centering
\begin{tabular}{|r|rrrr|}\hline
  $H_r$ & $e$ & $r$ & $r^2$ & $r^3$ \\ \hline
$1$ & $1$   & $1$            & $1$             & $1$                                  \\ 
$i$ & $1$   & $i$            & $-1$             & $-i$                                  \\ 
$-1$ & $1$   & $-1$            & $1$             & $-1$                                  \\ 
$-i$ & $1$   & $-i$            & $-1$             & $i$                                  \\ \hline
\end{tabular}

\vspace{0.3cm}

\begin{tabular}{|r|rrrr|}\hline
  $H_m (H_{mr})$ & $e$ & $m(mr)$ & $r^2$ & $mr^2(mr^3)$ \\ \hline
$(1,1)$ & $1$   & $1$            & $1$             & $1$                                  \\ 
$(1,-1)$ & $1$   & $-1$            & $1$             & $-1$                                  \\ 
$(-1,1)$ & $1$   & $1$            & $-1$             & $-1$                                  \\ 
$(-1,-1)$ & $1$   & $-1$            & $-1$             & $1$                                  \\ \hline
\end{tabular}
\caption{Character tables of relevant subgroups of $D_4$. The groups $H_m$ and $H_{mr}$ are isomorphic.}
\label{tab:char_sub}
\end{table}

\textbf{Anyon content of $D(D_4)$.}
The task of listing all of the irreducible representations of $D(D_4)$ and their dimensions from this data is straightforward. There are $2 \times 5  + 3 \times 4 = 22$ irreducible representations, i.e., types of anyons in the $D_4$ quantum double model.

Eight of them are one-dimensional, i.e. abelian, while the rest are two-dimensional. Other than the vacuum, four of them are pure charges and four are pure fluxes. For dyons with the $\mathcal{C}_{r^2}$ flux component we say that they have a trivial flux even though it is not the vacuum flux. The flux can be factored out, just like in the case of the toric code fermion. This comes from the fact that $r^2$ commutes with all group elements, just like the identity.

 We label the vacuum and the trivial flux as $0 \equiv (\mathcal{C}_e, 1)$ and $\tilde 0 \equiv (\mathcal{C}_{r^2}, 1)$, respectively. Likewise, the pure charges and dyons with vacuum and trivial flux are labelled as $\Sigma_{\chi} \equiv (\mathcal{C}_e, \chi)$ and $\tilde{\Sigma}_{\chi} \equiv (\mathcal{C}_{r^2}, \chi)$, with $\chi$ being one of the nontrivial irreducible representations of $D_4$.
Nontrivial pure fluxes are labelled $\Psi_x = (\mathcal{C}_x, 1)$, with $C_x$ being one of the nontrivial conjugacy classes of $D_4$. The rest of the dyons have less informative labels such as $\{\tilde{\Psi}_x, \Phi_x, \tilde{\Phi}_x\}$ and can be found in Appendix \ref{app:reps}, where we also provide the quantum dimensions and the topological twists together with the fusion rules. 

\textbf{Charge measurements reprise.}
Let us review the partial charge measurements for $D(D_4)$.
We choose the subgroups $\{H_r, H_m, H_{mr}\}$ for the measurement protocol. All the irreducible representations of these subgroups, $\chi_{H_x}$, are one dimensional. This means that the measurement outcome of the partial charge measurement is just the irrep label $(\chi_{H_x})$. Looking at their character tables, we find a partial orthogonality with the characters of the irreps of $G$. Concretely, we compute $ \braket{\chi_{H_x}, \chi} = \sum_{h\in H_x} \chi^*_{H_x}(h)\chi(h)$ for the five different charges and list the results in Table \ref{tab:red_ch}.

\begin{table}[h]
\centering
\begin{tabular}{|c|lllll|}\hline
  $\braket{\chi_{H_r}, \chi}$ & $1$ & $\alpha_{r}$ & $\alpha_{m}$ & $\alpha_{mr}$ & $\alpha_{\epsilon}$ \\ \hline
$1$ & 4   & 4            & 0             & 0               & 0                   \\ 
$-1$ & 0   & 0            & 4             & 4               & 0                   \\ 
$i$ & 0   & 0            & 0             & 0               & 4                   \\ 
$-i$ & 0   & 0            & 0             & 0               & 4                   \\ \hline
\end{tabular}

\vspace{0.3cm}

\begin{tabular}{|c|lllll|}\hline
  $\braket{\chi_{H_m}, \chi}$ & $1$ & $\alpha_{r}$ & $\alpha_{m}$ & $\alpha_{mr}$ & $\alpha_{\epsilon}$ \\ \hline
$(1,1)$ & 4   & 0            & 4             & 0               & 0                   \\ 
$ (1,-1)$ & 0   & 4            & 0             & 4               & 0                   \\ 
$(-1,1)$ & 0   & 0            & 0             & 0               & 4                   \\ 
$(-1,-1)$ & 0   & 0            & 0             & 0               & 4                   \\ \hline
\end{tabular}

\vspace{0.3cm}

\begin{tabular}{|c|lllll|}\hline
  $\braket{\chi_{H_{mr}}, \chi}$ & $1$ & $\alpha_{r}$ & $\alpha_{m}$ & $\alpha_{mr}$ & $\alpha_{\epsilon}$ \\ \hline
$(1,1)$ & 4   & 0            & 0             & 4               & 0                   \\ 
$ (1,-1)$ & 0   & 4            & 4             & 0               & 0                   \\ 
$(-1,1)$ & 0   & 0            & 0             & 0               & 4                   \\ 
$(-1,-1)$ & 0   & 0            & 0             & 0               & 4                   \\ \hline
\end{tabular}
\caption{Partial orthogonality of $D_4$ with respect to its three four-element subgroups.}
\label{tab:red_ch}
\end{table}

For example, imagine we performed a $H_m$-partial charge measurement on a plaquette $p$ and obtained the measurement outcome $(1,1)$. This label has a non-vanishing overlap with the trivial charge $1$ and charge $\alpha_m$. Hence, both $0$ and $\Sigma_{\alpha_m}$ can be anyons present on the plaquette.

Now imagine we have performed all three measurement and obtained the set of labels $\{-1, (1, 1), (1, -1)\}$. This set is only compatible with the charge $\alpha_m$ and hence $\Sigma_m$ must be on a plaquette $p$.

\textbf{Beyond pure charges.} The four main subgroups we consider in the partial charge measurement are also the centralisers of the respective conjugacy classes, $H_x = Z(x)$. Hence, as mentioned in the last section, if we read-off the flux whose centre is the subgroup we used in the partial charge measurement, the result uniquely determines the dyon label.

It is only for dyons of trivial flux, $\{e, r^2\}$, that we need to use the repeated partial charge measurements with different subgroups alongside partial orthogonality of the character tables of $D_4$ and its four-element subgroups to uniquely determine the topological charge.

\section{Probing non-abelian anyons} \label{sec:probing}

In this section, we will present a set of protocols that allow us to demonstrate the non-abelian character of the anyons of $D(D_4)$ on realistic to-date quantum hardware. The main obstacle to overcome here is the noise which limits the depth of the circuits. We will show how low circuit depths can be achieved for all protocols. A benchmark of the proposed experiments is presented in the next section showing numerical simulation results for a realistic noise model.

The main experiments showcasing the non-abelian nature of the excitations we propose are the following.
\begin{itemize}

\item[i)] Anyon fusion. Here, the non-abelian nature is signalled by non-unique fusion outcomes.

\item[ii)] Non-abelian braiding. The order of the braids does not commute.

\item[iii)] S- and T-matrix measurements. This data describes the amplitudes of links and twists and almost uniquely\footnote{There are some exceptions, where the anyon theory, i.e., the UMTC is not determined uniquely by the $S$-and $T$-matrices alone~\cite{notdetermined2021}. The smallest known example where this is the case has 49 particle types.} characterises the full anyon theory.
\end{itemize}

\subsection{Achieving low circuit depth }\label{subsec:enc}

In this section, we will discuss how to achieve low circuit depths for creating and moving the anyons of $D(D_4)$. We will present the encoding of group elements into qubits, and show how short circuits for the elementary triangles of the ribbon operators discussed in Section~\ref{sec:ribbon_ops} can be obtained.

\textbf{Encoding.} 
The order of $D_4$ is 8, hence we need three qubits to encode a group element.
We chose the following map
\begin{equation}
    \ket{g} \equiv \ket{a}_m\ket{b}_r\ket{c}_{r^2} \iff g = m^a r^b (r^2)^c,
\end{equation}
where $a,b,c \in \{0,1\}$, $r$ is the $90^{\text{o}}$ rotation and $m$ is the reflection.

When encoding the internal space of the anyon, which we need for the ribbon operator protocol, we note that the basis $\ket{\nu} = \ket{c, i}$, contains a group element $c\in C$ restricted to a conjugacy class. All the non-trivial conjugacy classes contain just two elements and can be encoded with only one qubit. Since the centralisers of these conjugacy classes are abelian and their representations are one-dimensional, we can encode the full internal space of the anyon with just one qubit. For anyons with $m$-flux, $(C_m,\chi)$,
we have
\begin{equation}\label{eq:enc_C}
	\ket{c, 0} \equiv \ket{a}_{a} \iff c = m^1 r^0(r^2)^a \in \{m , mr^2\},
	\end{equation}
for anyons with $mr$-flux, $(C_{mr},\chi)$, we have	
\begin{equation}\label{eq:enc_C2}
	\ket{c, 0} \equiv \ket{a}_{a} \iff c = m^1 r^1(r^2)^a \in \{mr , mr^3\},
	\end{equation}
	and for anyons with $r$-flux, $(C_r,\chi)$, we have
\begin{equation}\label{eq:enc_C3}
	\ket{c, 0} \equiv \ket{a}_{a} \iff c = m^0 r^1(r^2)^a \in \{r , r^3\}.	
\end{equation}

Similarly the subgroups are encoded with two qubits. For example for $h\in H_r$ we use\begin{equation}
    \ket{h} \equiv \ket{a}_r\ket{b}_{r^2} \iff h = m^0r^a(r^2)^b.
\end{equation}

To encode the four different representations of the subgroups $\chi_H$ for $H_m$ and $H_{mr}$ we use
\begin{equation}
    \ket{(-1)^a,(-1)^b} \equiv \ket{a}_a\ket{b}_b, \label{eq:enc_irr1}
\end{equation}
where $(a,b)$ corresponds to the representation label used in Table \ref{tab:char_sub}, 
and for $H_{r}$ we use
\begin{equation}
    \ket{i^{2a+b}} \equiv \ket{a}_a\ket{b}_b, \label{eq:enc_irr2}
\end{equation}
where $i^{2a+b}$ corresponds to the representation label used in Table \ref{tab:char_sub}.

\textbf{Circuits.} To implement the anyon protocols we need circuits for the following operations.
\begin{enumerate}
    \item Controlled group multiplication $\ket{g, h} \rightarrow \ket{g, gh}$  \begin{enumerate}
        \item full domain variant ($g \in G$), used for ground state preparation,
        \item restricted domain variant ($g \in C$), used in ribbon operators,
        \item restricted domain variant ($g \in H$), used for partial charge measurements. 
    \end{enumerate}
    \item Controlled generalised conjugation\\ $\ket{c} \ket{i}\ket{g} \rightarrow \ket{gcg^{-1}} \Gamma^\chi_{c} \ket{i}\ket{g}$.
    \item Decoupling unitary of the partial charge measurement $$ U_a = \sum_{\chi_H,i', j',h'\in H} \sqrt{\frac{d_\chi}{|G|}} \bar \Gamma^{\chi_H}_{i'j'}(h')  \ket{\chi_H; i', j'}_a\bra{h'}_a. $$
\end{enumerate}

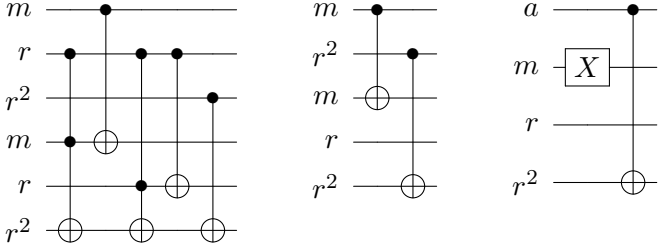
\begin{figure}
\begin{equation*}
\Qcircuit @C=0.4em @R=0.7em @!R{
\lstick{m} & \qw & \ctrl{3} & \qw & \qw & \qw & \qw\\
\lstick{r} & \ctrl{2} & \qw & \ctrl{3} & \ctrl{3} & \qw & \qw\\
\lstick{r^{2}} & \qw  & \qw & \qw & \qw & \ctrl{3} & \qw
\\
\lstick{\quad m} &  \ctrl{2} & \targ & \qw & \qw & \qw & \qw\\
\lstick{r} & \qw & \qw & \ctrl{1} & \targ & \qw & \qw\\
\lstick{r^{2}} & \targ & \qw & \targ & \qw & \targ & \qw
}\qquad\qquad
\Qcircuit @C=0.4em @R=0.7em @!R{
\lstick{m} & \ctrl{2} &  \qw & \qw\\
\lstick{r^{2}} & \qw  & \ctrl{3} & \qw
\\
\lstick{\quad m} &  \targ & \qw & \qw \\
\lstick{r} & \qw & \qw & \qw\\
\lstick{r^{2}} & \qw & \targ & \qw
}\qquad\qquad
\Qcircuit @C=0.4em @R=0.7em @!R{
\lstick{a} & \qw  & \ctrl{3} & \qw
\\
\lstick{\quad m} &  \gate{X} & \qw & \qw \\
\lstick{r} & \qw & \qw & \qw\\
\lstick{r^{2}} & \qw & \targ & \qw
}
\end{equation*}

    \caption{Circuits for controlled group multiplication $\ket{g,h} \rightarrow \ket{g,gh}$. Left: Both $g$ (first three qubits) and $h$ (last three qubits) are unrestricted, i.e., $g,h\in G$. Center: $g \in H_{m}$ is encoded by just the first two qubits. Right: $g \in C_m$ is encoded by just the first qubit $a$ (cf. Eq.~\eqref{eq:enc_C}).}
    \label{fig:restMult}
\end{figure}

We will focus on a set of examples illustrating the expected circuit depths. The full domain group multiplication (see Fig. \ref{fig:restMult}) consists of three Toffoli and 2 CNOT gates. Note, that on most hardwares a 3-qubit Toffoli gate must be decomposed into 2-qubit gates. Hence, each Toffoli gate should be seen as a depth-6 circuit in itself\footnote{If the coupling gates are restricted to act on neighbouring qubits, additional swaps may be needed which increase the depth to up to 12.}. With this in mind, the circuit depth for a single group multiplication is 22. Current noise levels limit the circuit depth to about 100 gates. Therefore, a naive protocol based on full domain group multiplication is doomed to fail. 

We now contrast this with the reduced domain multiplication, where the control qubits are restricted to a certain conjugacy class. A representative example is shown in Fig.~\ref{fig:restMult}. The circuit is reduced to a depth-1 circuit with only one CNOT gate. We can appreciate a dramatic reduction of circuit depth achieved by restricting the domain of the controlled multiplication circuit.

This is one of the main facts that allows us to greatly simplify many circuits related to ribbon operators and partial charge measurements. However, the direct ground state preparation does not benefit from these simplifications. We will discuss the ground state preparation separately in the next section.

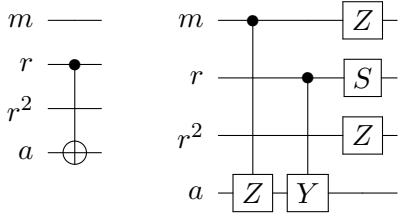
\begin{figure}
\begin{equation*}
\begin{split}
\Qcircuit @C=0.5em @R=0.7em @!R{
\lstick{m} & \qw & \qw \\
\lstick{r} & \ctrl{2}  & \qw \\
\lstick{r^{2}} & \qw  & \qw \\
\lstick{a} &  \targ & \qw
}\qquad\qquad
\Qcircuit @C=0.5em @R=0.7em @!R{
\lstick{m} & \ctrl{3} & \qw & \gate{Z} & \qw\\
\lstick{r} & \qw & \ctrl{2} & \gate{S} & \qw\\
\lstick{r^{2}} & \qw  & \qw & \gate{Z} & \qw\\
\lstick{a} & \gate{Z}  & \gate{Y} & \qw & \qw 
}
\end{split}
\end{equation*}
 
    \caption{Circuits for generalised conjugation. The first three qubits encode the physical, group valued, gauge field $\ket{g}$. The last qubit encodes the ancilla qubit representing the internal state of the (two-dimensional) anyon $\ket{c}$. Left: The conjugation unitary for a pure flux $\Psi_m$ $\ket{c}\ket{g} \rightarrow \ket{gcg^{-1}}\ket{g}$. Right: The generalised conjugation unitary for the dyon $\tilde{\Phi}_m$: $\ket{c}\ket{g} \rightarrow \Gamma(g)\ket{gcg^{-1}}\ket{g}$, where the representation 'matrix' $\Gamma(g) \in U(1)$ and  $S = \text{diag}(1, i)$.}

    \label{fig:genConj}
\end{figure}

Let us move on to the generalised conjugation. The circuit depth for this heavily depends on the type of anyon. For a pure flux the representation matrix $\Gamma$ is trivial and the generalised conjugation simplifies to a conventional conjugation which can be implemented by a circuit of depth one or two (see Fig.~\ref{fig:genConj} for an example). In contrast, for a dyon with a one-dimensional but non-trivial irrep of the center, the circuit is more involved. Since the irrep is one dimensional, it corresponds to a phase factor and does not need to be encoded in an additional vector space. I.e., we can still use $\ket{c}$ instead of $\ket{c,i}$. However, the conjugation of $c$ has to be accompanied by appropriate phase factors implementing the action of $\Gamma^\chi_c(g)$. A circuit showing such a generalised conjugation for the $m$-dyon $\tilde{\Phi}_m \equiv (C_m, (-1, -1))$ is shown in Fig.~\ref{fig:genConj}. We note, that there is still no need for Toffoli gates, however, the circuit is more complex than for pure fluxes. The full list of circuits for generalised conjugations is given in Appendix~\ref{app:reps}.

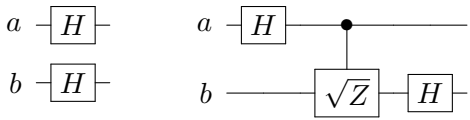
\begin{figure}
\begin{equation*}
\begin{split}
\Qcircuit @C=0.5em @R=0.7em @!R{
\lstick{a} & \gate{H} & \qw \\
\lstick{b} & \gate{H}  & \qw
}\qquad\qquad
\Qcircuit @C=0.5em @R=0.7em @!R{
\lstick{a} & \gate{H} & \qw &\ctrl{1} &\qw &\qw & \qw\\
\lstick{b} & \qw & \qw & \gate{\sqrt{Z}}& \qw&\gate{H} & \qw 
}
\end{split}
\end{equation*}

    \caption{The decoupling unitary map used in partial  charge measurements for subgroups of $D_4$. Left: $H_m,H_{mr} \simeq \mathbb{Z}_2 \times \mathbb{Z}_2$. Right: $H_r \simeq \mathbb{Z}_4$. In the circuits above $H$ denotes the Hadamard gate.}
    \label{fig:decopU}
\end{figure}

We note that there are adaptive constant-depth circuits (measurement-based schemes) for applying ribbon operators that have better scaling for solvable gauge groups~\cite{bravyi2022adaptive} (such as $D_4$), but the measurement overhead makes them less preferable for small systems. 

Lastly, we look at the decoupling unitaries for partial charge measurements. The unitaries for subgroups $H\subset G$ are shown in Figure \ref{fig:decopU}.

\textbf{Geometry and ground state preparation.}
Our proposal for ground state preparation does not use any feed-forward protocols. Therefore, it is suitable for an experimental set-up, where measurements terminate the circuit. We also note, that so far known feed-forward protocols for ground state preparation only cover quantum double models for solvable groups excluding more complex quantum double models and the more general string-net models based on fusion categories beyond groups.

Not using a feed-forward protocol, however, comes at the cost of limiting the lattice geometry to either small two-dimensional graphs or quasi one-dimensional graphs in order to keep the circuit depth short.

\begin{figure*}
    \begin{subfigure}{0.7\textwidth}\hfill
    \includegraphics[width=\linewidth]{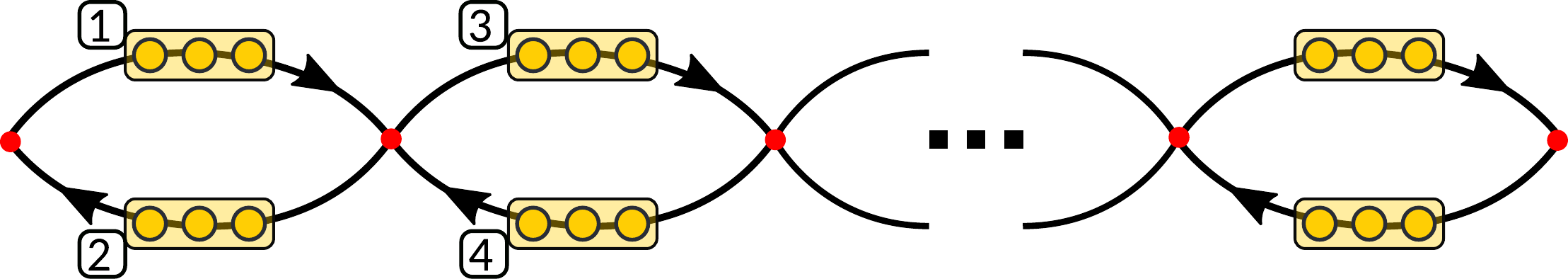}
    \vspace{0.05cm}
    \end{subfigure} %
    \hfill
    \begin{subfigure}{0.25\textwidth}
    \begin{equation*}
    \Qcircuit @C=0.2em @R=0.2em {
\lstick{m} & \gate{H} &       \ctrl{3} & \qw & \qw &\qw \\
\lstick{r} & \gate{H} &   \qw & \ctrl{3} & \qw & \qw \\
\lstick{r^{2}} & \gate{H} &  \qw & \qw & \ctrl{3} & \qw 
\\
\lstick{\quad m} &  \qw &   \targ & \qw & \qw & \qw \\
\lstick{r} & \qw&   \qw &  \targ & \qw & \qw \\
\lstick{r^{2}} & \qw & \qw&   \qw &  \targ & \qw 
}
\end{equation*}\vfill
\end{subfigure}
    \caption{Quasi one-dimensional lattice allowing for shallow ground state preparation of the quantum double model $D(D_4)$. Left: Yellow bars denote individual spins associated to edges, which are composed of three qubits each. Edge orientations are needed to define the vertex- and plaquette operators of the corresponding Hamiltonian. The lattice is embedded into a sphere, meaning in addition to the 2-gons, there is one large 'outer' plaquette. Right: Circuit for groundstate preparation per loop.}
    \label{fig:latticeGS}
\end{figure*}

For the majority of our result, we will focus on a quasi one-dimensional graph, that we call \emph{braiding ladder}. 
This geometry is shown in Figure~\ref{fig:latticeGS}. 

The ground state on this geometry can be prepared with a constant-depth circuit, i.e., it does not scale with the system size. The ground state on an $n$-segment ladder is given by
\begin{equation}
    \ket{GS} = \sum_{g_1,g_2,\ldots,g_n} \ket{g_1,g_1,g_2,g_2,\ldots,g_n,g_n} \;.
\end{equation}
This state has no long-range entanglement and factorises into a sequence of qudit Bell-pairs. 
However, entanglement is built up once one introduces anyons via ribbon operators.
In fact, this geometry is sufficient to correctly show the braiding statistics of the anyons. The only requirement for correctly reproducing the braiding statistics is that the ribbon paths only \emph{touch}, but do not intersect and is fulfilled here.

The shallow circuit preparing this state is shown in Figure \ref{fig:latticeGS}.

\begin{figure}
    \centering
    \includegraphics[width=\linewidth]{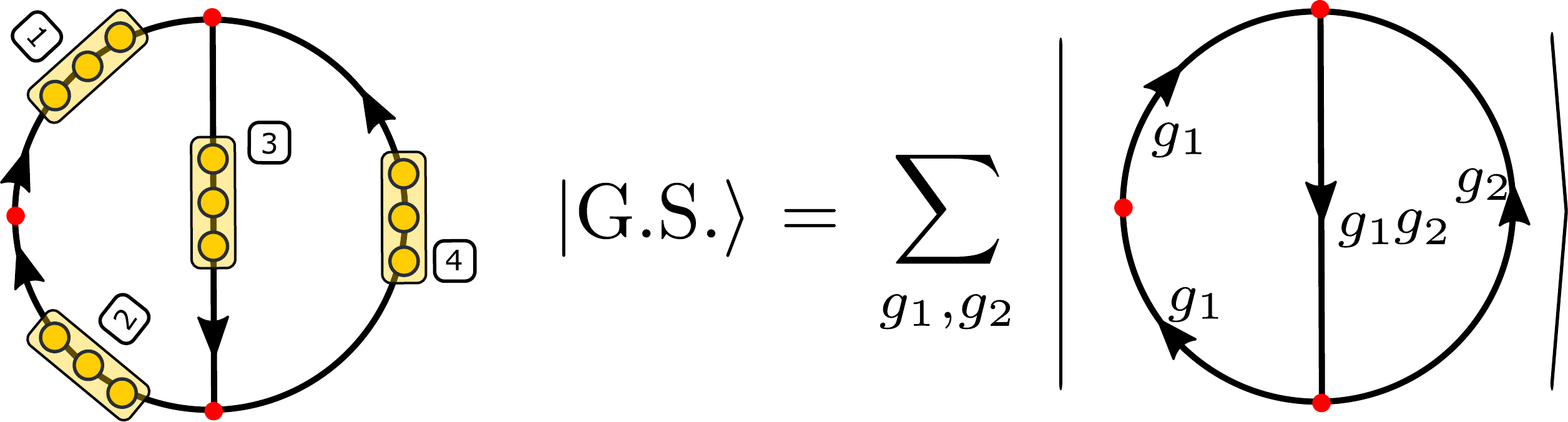}
    \caption{A small two-dimensional graph. Left: Yellow bars denote individual spins associated to edges, which are composed of three qubits each. Edge orientations are needed to define the vertex- and plaquette operators of the corresponding Hamiltonian. Right: The ground state on this small two-dimensional graph. The circuit needed for its preparation is discussed in Section \ref{sec:num:elem}.}
    \label{fig:basketball}
\end{figure}

We also examine fusion on a small two-dimensional graph, as shown in Figure \ref{fig:basketball}.

\subsection{Elemental protocols}
\textbf{Anyon fusion}
is the first and most basic protocol that can demonstrate non-abelianness. 

After preparing the ground state, we apply two ribbon operators. One between sites $s_1$ and $s_2$, and the other between sites $s_2$ and $s_3$.
This creates two pairs of anyons and fuses them on site $s_2$. We then perform a partial charge measurement and flux readout for this site.
This experiment can be performed on both the braiding ladder and the small two-dimensional graph.

We can do this experiment for any anyon types. Here, we will give an example for the case of fusing the pure flux $\Psi_m$ with itself. In contrast to the abelian case, this fusion is not restricted to yield vacuum. Instead we have
$$\Psi_m \times \Psi_m = 0 + \tilde{0} + \Sigma_m + \tilde{\Sigma}_m,$$ where $0$ is the vacuum, $\tilde{0}$ is the abelian flux corresponding to the other element of the centre of $D_4$, $r^2$, and the other two anyons are a pure charge associated with the $\alpha_m$ representation of $D_4$ and the dyon of this pure charge and the abelian flux.

All four of these outcomes can be differentiated by reading out the flux and performing the partial charge measurement using $H_r$ or $H_{mr}$. For concreteness, we choose $H_{mr}$ for which we expect to only observe the measurement outcomes $(1,1)$ and $(1,-1)$, corresponding to no charge and $\alpha_m$, respectively. A discussion of this protocol implemented on a realistic hardware device and a corresponding numerical simulation are shown in Section \ref{sec:num:elem}.

\textbf{Anyon braiding.} The second phenomenon that gives the non-abelian anyons their name is the fact that the image of the braid group, as represented by physically braiding these anyons, is non-abelian. 
This means that the order of interchanges matter, i.e. $\sigma_{12}\sigma_{23} \neq \sigma_{23}\sigma_{12}$, where $\sigma_{i(i+1)}$ is the clockwise interchange of particle $i$ and $i+1$, the generator of the braid group.

\begin{figure}
    \centering
    \includegraphics[width=\linewidth]{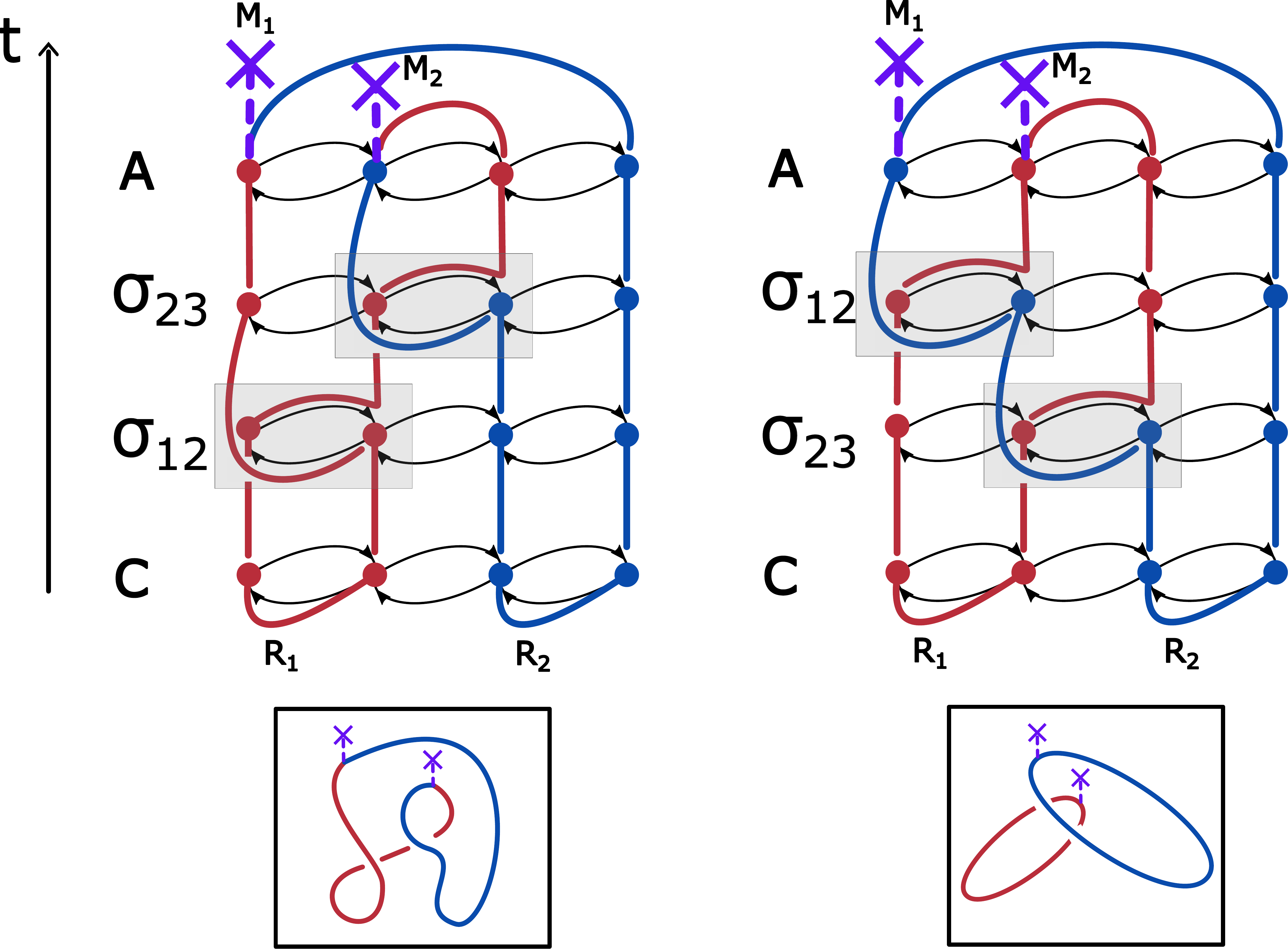}
    \caption{The two braiding protocols, differing only in the order of exchanging the two anyons. C stands for creation, M for measurements.  The protocol on the left will have 4 fusion outcomes, while the protocol on the right can only produce vacuum.}
    \label{fig:flux_braid}
\end{figure}

The braiding procedures we want to implement to show this fact are shown in Figure \ref{fig:flux_braid}. We create two pairs of anyons from the vacuum, perform two interchanges $\sigma_{12} \sigma_{23}$, and then fuse pairwise. Then we repeat the same protocol with the inverted order of the interchanges, i.e., $\sigma_{23} \sigma_{12}$. For concreteness, we again consider pairs of pure fluxes $\Psi_m$.

In the second protocol, we annihilate the pairs that have a fixed fusion channel. Given that they are created from the vacuum they will fuse to the vacuum.
In the first protocol, we annihilate the pairs whose fusion channel is not fixed, hence all four fusion outcomes are expected. We see that the two braidings indeed produce two different states. 

For further discussion of a concrete implementation and numerical results, see Section \ref{sec:num:elem}.

\subsection{Anyon interferometry}\label{subsec:Intef}

In order to measure the relative phase between different braiding processes, we need to devise an interference protocol. This is done by setting up a control qubit $c$, whose state is entangled with different braiding protocols
\begin{equation}
    \ket{\psi}_c\ket{\text{GS}} \rightarrow \ket{0}_c \ket{\Psi_0} + \ket{1}_c \ket{\Psi_1},
\end{equation}
where $\Psi_i$ are the two wave functions of the matter degrees of freedom after two different braiding operations.

If the charge content of the two states is the same $\ket{\Psi_0}$ and $\ket{\Psi_1}$ may only differ by a constant. Hence we can write
\begin{equation}
    \ket{0}_c \ket{\Psi_0} + \ket{1}_c \ket{\Psi_1} = (\ket{0}_c + C_{01}\ket{1}_c)\ket{\Psi_0},
\end{equation}
and by the means of tomography on the control qubit $c$ we can extract the relative constant $C_{01}$.

For a suitable choice of the two braiding protocols, this constant reveals elements of the S- and T-matrix as we will see next.

\textbf{S-matrix elements.}
To measure the S-matrix elements, we create a superposition of two states by conditioning an equal time (closed) ribbon operator shown in blue in Fig.~\ref{fig:S-mat} on the control qubit .

Note that the S-matrix appearing in Figure \ref{fig:S-mat} is normalised
\begin{equation}
    \tilde{S}(a,b) = \frac{\mathcal D}{d_a d_b}S(a,b),
\end{equation}
where $\mathcal D=\sqrt{\sum_i d_i^2}=|G|$ is the total quantum dimension,
which makes $|\tilde{S}(a,b)| \leq 1$. To see that note that~\cite{Kitaev2006} $S_{ab}=\frac{1}{\mathcal D} \sum_c N_{ab}^c \frac{\theta_c}{\theta_a \theta_b} d_c$ and $|\theta_x|=1$. We also know that $\sum_c N_{ab}^c d_c = d_a d_b$, so it follows that $|S_{ab}| \leq \frac{1}{\mathcal D} d_a d_b$.

The conceptually simplest interferometry scheme is shown in Figure~\ref{fig:cond_ex}. Here we apply a ribbon of flavour $a$, then, depending on the state of the controlling qubit, either thread a second ribbon of flavour $b$ around the first ribbon or do nothing. This is very similar to the ideal scheme shown in Figure~\ref{fig:intef_example}.\footnote{Compared to Fig.\ref{fig:S-mat}a, Fig.~\ref{fig:S-mat}b is only missing a measurement checking that applying a closed ribbon $b$ on it's own doesn't introduce any phase, which it doesn't.}

However, in this protocol every single qubit gate of the $b$-ribbon operator becomes a two qubit gate (since the ribbon is conditioned on the control $c$) and every two qubit gate becomes a three qubit gate (unitarily similar to a Toffoli gate). Hence, the number of entangling gates grows quickly with the ribbon length.

\begin{figure}
\begin{equation*}
\Qcircuit @C=0.5em @R=0.7em @!R{
\lstick{c} & \ctrl{3} & \qw& \ctrl{1}& \qw\\
\lstick{m} & \qw & \qw & \targ& \qw\\
\lstick{r} & \qw  & \qw & \qw& \qw\\
\lstick{r^{2}} & \targ  & \qw & \qw& \qw\\
\lstick{a} &  \ctrl{-1} & \qw& \qw& \qw
}%
\qquad \qquad \qquad
\Qcircuit @C=0.5em @R=0.7em @!R{
\lstick{c} & \ctrl{1} & \qw\\
\lstick{m} & \targ & \qw \\
\lstick{r} & \qw  & \qw \\
\lstick{r^{2}} & \targ  & \qw \\
\lstick{a} &  \ctrl{-1} & \qw
}
\end{equation*}
\caption{Controlled multiply circuits of an elementary triangle of a ribbon operator conditioned on a control qubit $c$ (first qubit) acting on a physical edge (middle three qubits) and a ribbon ancilla qubit (last qubit) Left. Implementing a $\Psi_m$-elementary triangle vs vacuum, represented as $0\oplus 0$. Right. Implementing an $\Psi_m$-elementary triangle vs $0\oplus\tilde{0}$.}
\label{fig:flavCond}
\end{figure}
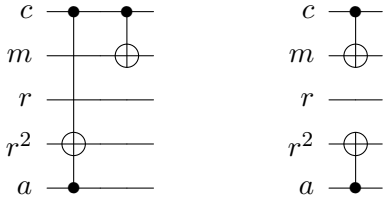

\begin{figure*}
\centering

\begin{subfigure}{0.47\textwidth}
    \includegraphics[width = \linewidth]{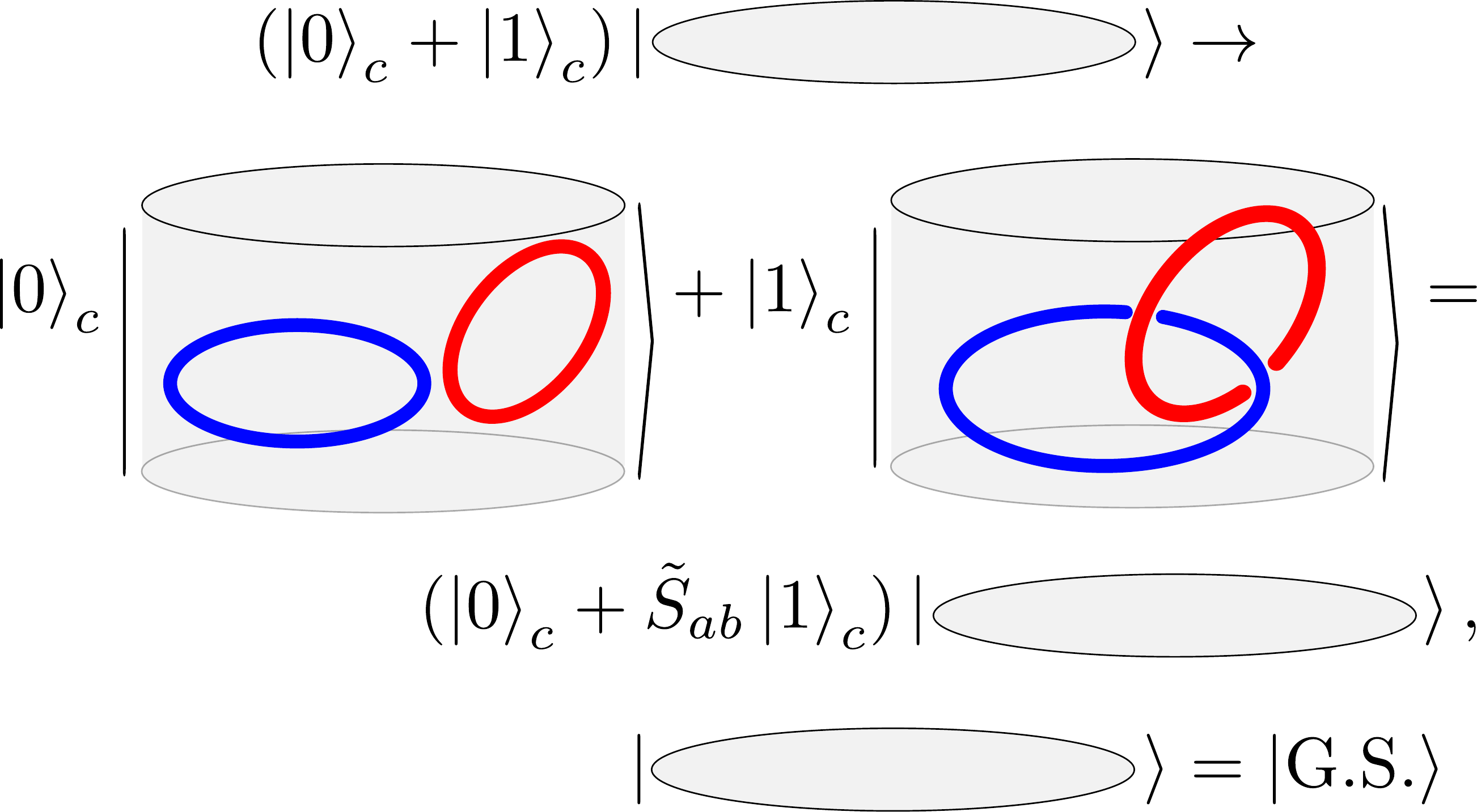}
    \caption{The ideal interferometry scheme for measuring the S-matrix elements. The two different braids are entangled with the controlling qubit $c$.}
    \label{fig:intef_example}
\end{subfigure}\hfill
\begin{subfigure}{0.47\textwidth}
    \includegraphics[width=\linewidth]{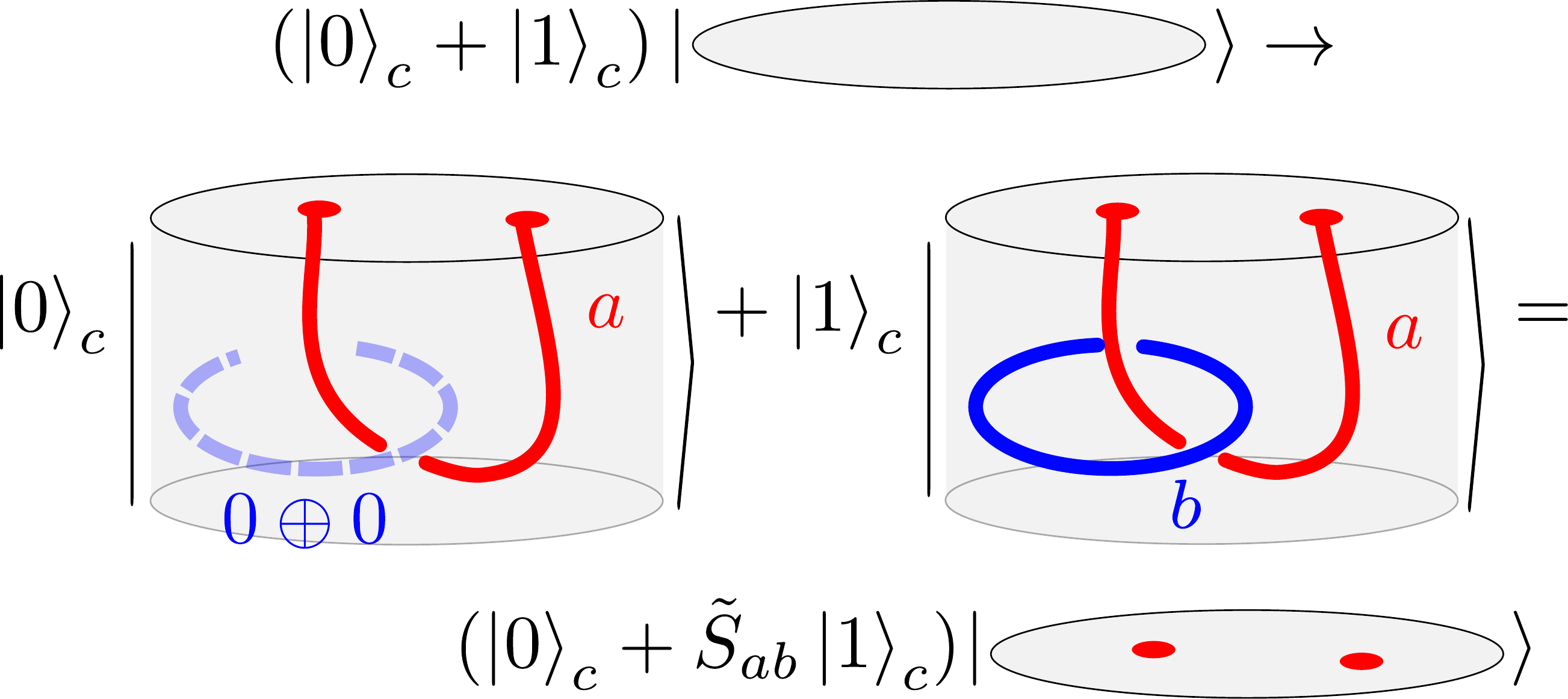}
    \caption{Conditioning the existence of the $b$-ribbon. The $b$-ribbon is only applied, if the state of the control qubit is $\ket{1}_c$. Equivalently, we say that we condition whether to apply $b$ or the reducible representation $0\oplus 0$.}
    \label{fig:cond_ex}
\end{subfigure}

\vspace{15pt}

\begin{subfigure}{0.47\textwidth}
    \includegraphics[width=\linewidth]{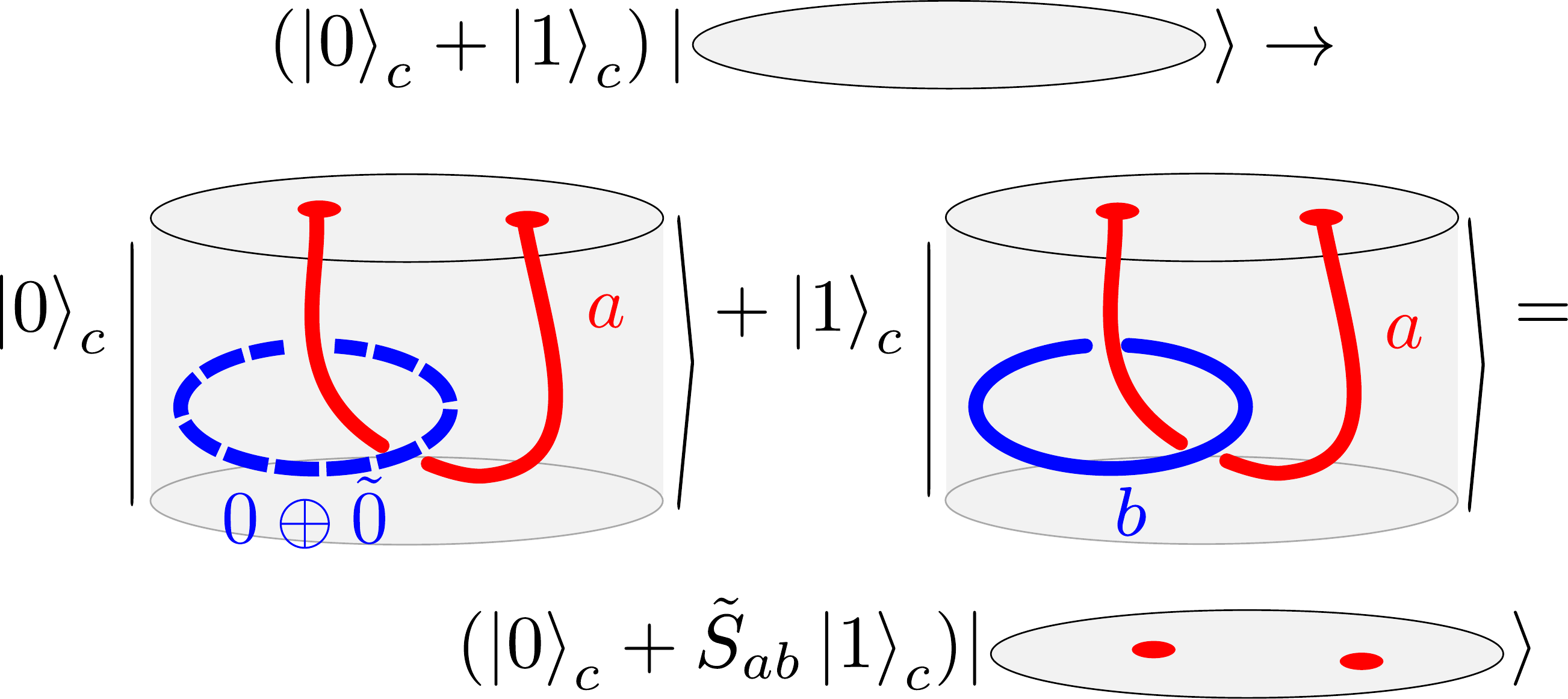}
    \caption{Conditioning the type of the $b$-ribbon. If the state of the control qubit is $\ket{1}_c$, apply ribbon $b$. If it is $\ket{0}_c$, apply the ribbon for the reducible representation $0\oplus\tilde{0}$.}
    \label{fig:cond_flav}
\end{subfigure}

\caption{The interference protocols used for phase sensitive measurement of the (normalised) S-matrix elements  $\tilde{S}(a,b) = \frac{|D_4|}{d_a d_b}S(a,b)$.	}
\label{fig:S-mat}
\end{figure*}

A smarter alternative is to condition the ribbon type instead (see Figure \ref{fig:cond_flav}). In this case we identify where the ribbon operators for the two anyon types differ, and condition only those operations. The circuit for this can be much shorter, see Fig. \ref{fig:flavCond} for an example. In particular, for the case of $D(D_4)$, it turns out, that, if $b$ has a non-trivial flux content, it is easier to compare the linking of anyons $a$ and $b$ to the linking of anyons $a$ and a reducible charge $0 \oplus \tilde 0$. (Note, that this is not a valid anyon label since the representation is reducible.) Thus, we condition whether we apply the ribbon $b$ or the ribbon corresponding to $0 \oplus \tilde 0$. The ribbon operator protocol defined for irreducible representations carries over for reducible ones. Note that the representations $0\oplus 0$ and $0 \oplus \tilde 0$ are two-dimensional and hence distinct from $0$ and $\tilde 0$ respectively.

However, this protocol requires additional knowledge of the theory. Concretely, if we are interested in the S-matrix element $S_{ab}$, we need the additional knowledge of  $S(a, 0)$ and $S(a, \tilde{0})$.
In fact, both can be measured easily by the protocol in Figure~\ref{fig:cond_ex} since anyons $0$ and $\tilde{0}$ are abelian and their ribbon operator only have single qubit gates. A similar protocol was used to measure the $S$-matrix in the case of the toric code~\cite{Satzinger_2021}.

If $\tilde S(a, 0) = \tilde S(a, \tilde{0}) = 1$, we can just read off $\tilde S(a,b)$ after tomographing the control qubit.
In the case of $\tilde S(a, \tilde{0}) = -1$, there is a two qubit gate we need to apply between the controlling qubit and one of the ribbon ancillas before we can simply read off $\tilde S(a,b)$ via tomography.\footnote{See Appendix \ref{app:cirqs}.}
The exact method of tomography will be presented alongside the numerical results in Section \ref{sec:num:intef}.

\begin{figure*}
\centering
\begin{subfigure}{0.47\textwidth}
    \centering
    \includegraphics[width = \linewidth]{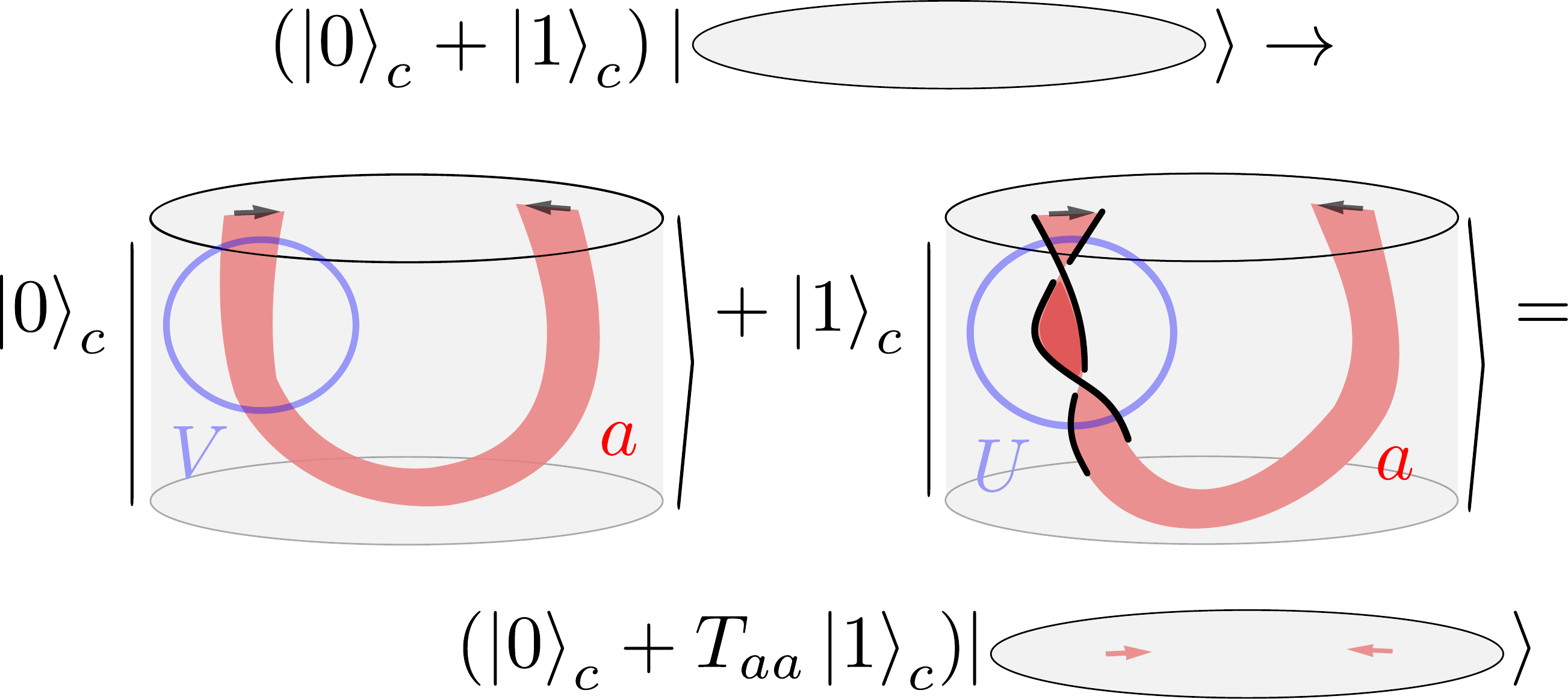}
    \caption{Interference protocol for measuring the phase of the T-matrix elements. Note: $|T_{aa}| = 1$ for all anyons $a$. The inference is done between two paths that differ by only one twist. We have restored the two-dimensionality of the ribbons in our representation in order to showcase the extra twist.}
    \label{fig:Tmat}
\end{subfigure}\hfill
\begin{subfigure}{0.47\textwidth}
\begin{equation*}
\Qcircuit @C=0.5em @R=0.7em @!R{
\lstick{c}&\qw&\ctrl{1}&\qw& \gate{X}&\qw&\ctrl{1}&\qw&\gate{X}&\qw\\
\lstick{\text{anc}}&\qw&\multigate{1}{U}&\qw &\qw&\qw&\multigate{1}{V}&\qw&\qw&\qw\\
\lstick{GF}&\qw&\ghost{U}&\qw&\qw&\qw&\ghost{V}&\qw&\qw&\qw
}
\end{equation*}
\caption{A circuit that, depending on the state of the control qubit $c$, either applies a unitary $U$ or $V$ onto the joint gauge field ($GF$) and ancilla (anc) degrees of freedom.}
\label{fig:condcirq}
\end{subfigure}
\caption{The interferometry scheme to measure the phase difference between two paths alongside with a circuit diagram implementing the difference of the two paths.}
\label{fig:tmatfull}
\end{figure*}

\textbf{T-matrix elements.}
In this section, we will describe the interference protocol for measuring the matrix elements of the diagonal T-matrix, or the spin of the anyons.
Here we note that ribbon operators are in-fact ribbons in space-time, hence they can acquire a twist.
Each twist of anyon $a$ contributes a phase factor to the wave function, $T_{aa} = e^{i\theta_a}$. 

The protocol is illustrated in Figure \ref{fig:Tmat}. We identify where the twisted and untwisted paths of a ribbon operator associated with some anyon $a$ differ and apply a controlled circuit of the form illustrated in Figure \ref{fig:condcirq} accordingly. If for the untwisted (twisted) version, we need to apply the unitary $U$ ($V$), the circuit implementing is straight-forward and shown in Figure \ref{fig:condcirq}.

The endpoint of the two ribbons are on the same site, hence the charge content is the same and we can factor out the gauge field wave function. 
The control qubit is in a pure state (assuming there is no noise) so we can tomograph and read off the twisting phase.

\section{Numerical experiments}\label{sec:num}

In this section, we provide numerical evidence for the feasibility of our proposal on state of the art NISQ devices. We performed simulations using Google's 'cirq\_google' python package on Google's cloud computing platform 'Google Colab'. This package executes the quantum trajectory simulation of the circuit using the Kraus operators obtained from the direct Pauli transfer matrix tomography on various single and two qubit gates on the Sycamore chip~\cite{weber}. This chip comes with two principle constraints. First, we can only perform two-qubit gates between adjacent qubits. Second, there is a limited set of elementary gates that can be implemented.

Knowing the characteristics of the single and two-qubit gates, and single- and multi-qubit readout performances of the chip~\cite{weber} we have chosen a suitable part of the chip for our simulations. A similar setup would be suitable for an actual experiment on the Sycamore chip. However, the optimal allocation of recourses will be chip dependent. The layout we used is shown in Figure \ref{fig:fusion_setup}.

The number of qubits we could simulate classically using this software is limited to 30, which is less than the number of qubits currently available on an actual machine. To see how this comes about, let us note that even though in our protocols there are only a few non-Clifford gates, we can not exploit the advantage of Clifford simulators because we simulate \emph{noisy circuits}.

\subsection{Elemental protocols}\label{sec:num:elem}

In this section, we present the simulation results for the fusion and braiding experiments.

\textbf{Circuit characteristics.} Before we present the results, we report on the qubit layout used for the fusion and braiding experiments, as well as circuit depths achieved, in order to put the noise observed in a useful context.

\emph{Ground state.} As mentioned, we prepare the ground state directly. On the braiding ladder that procedure is depth 2 (see Fig.~\ref{fig:latticeGS}). We prepare the qubits on all the bottom edges in an equal superposition via Hadamard gates and then CNOT the qubits above controlled by the one below. 

On the planar graph in Figure \ref{fig:basketball} this process is more complicated. Repeating the process as for the braiding ladder gives us the following state $\ket{\Psi} = \sum_{g_{12},g_{34}}\ket{g_{12},g_{12}, g_{34}, g_{34}}$. We then apply the full controlled multiply circuit from qubits of the second edge onto the qubits of the third edge. This procedure gives us the state defined in Figure \ref{fig:basketball}.

\begin{figure}
	\centering
	\includegraphics[width=\linewidth]{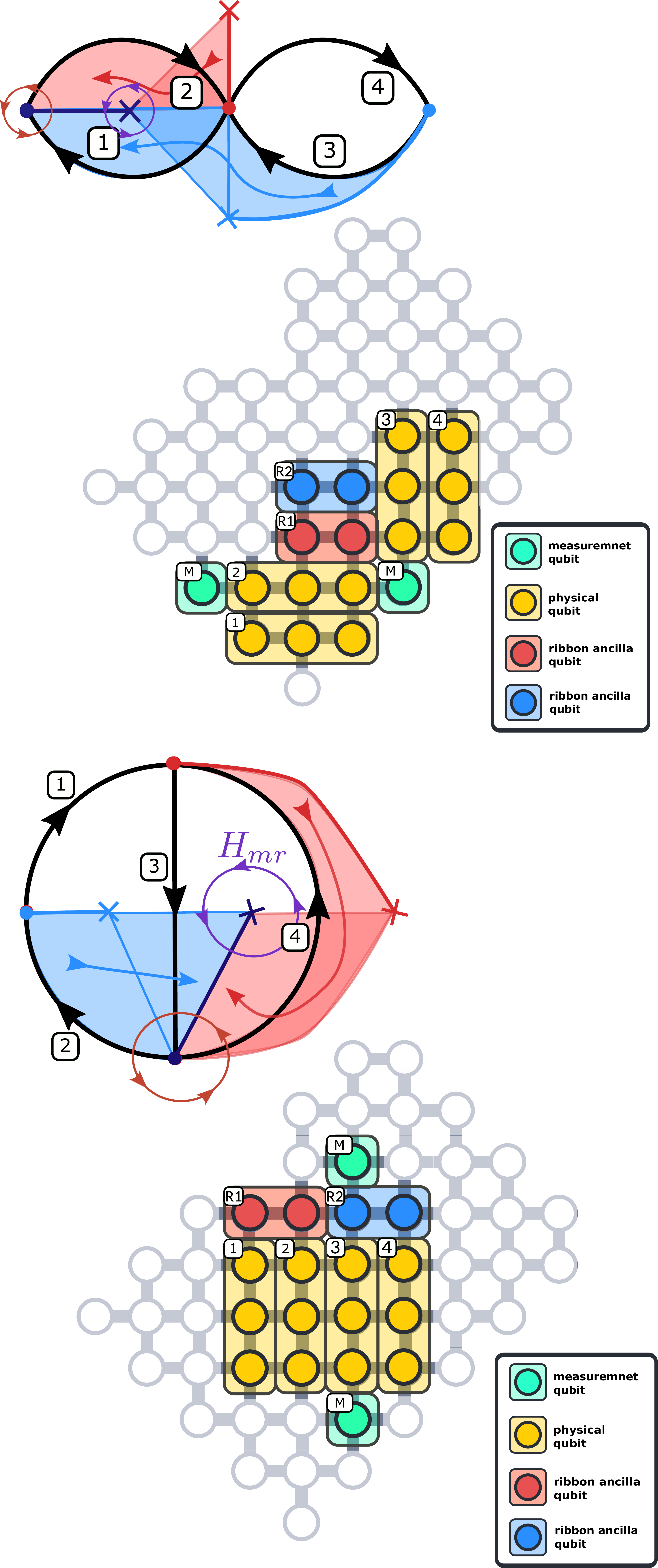}
	\caption{Ribbon operators and qubit layout for the fusion of two $\Psi_m$ anyons on the braiding ladder (top) and a small planar graph (bottom). Note, that the lattice is embedded into a sphere, so the outside plaquette that we labelled twice in the braiding ladder diagram should be identified. Red and blue shadings denote the two ribbon operators, respectively. Purple circles mark the plaquette on which we perform the $H_{mr}$-partial charge measurement. Red circles mark the vertex on which we measure the flux.}
	\label{fig:fusion_setup}
\end{figure}

\emph{Fusion.} 
The exact ribbon operators and the layout of the qubits on the Sycamore chip used in the fusion experiment are shown in Fig.~\ref{fig:fusion_setup}.

Using the circuits listed in Sec.~\ref{subsec:enc} to implement the ribbon operators defined in Sec.~\ref{sec:ribbon_ops} generates a circuit that is not directly implementable on the Sycamore chip due to the constraints mentioned above. We first need to implement swap gates such that all the multi-qubit gates appearing in the original circuit are acting on adjacent qubits. In addition, we need to compile multi-qubit gates into native single- and two-qubit coupler gates. In our case we chose the CZ gate as the coupler (two-qubit) gate. The single qubit gates are unrestricted.  

The additional swaps make up a considerable portion of all coupling gates used in the numerical experiments and hence the positioning of qubits is one of the key factors in minimising the circuit depth.

It is also worth noting that not all anyons are equal in complexity. The $r$-dyon ribbon operator require considerably deeper circuits since the group multiplication controlled by the elements of the $\mathcal{C}_r$ conjugacy class always involves at least one Toffoli gate. Let us recall that the Toffoli gate needs to be compiled into a circuit of at least depth 6 using the CZ as the two qubit coupler, however this neglects any swaps needed to place the qubits acted on by the CZ gates adjacent to one another. Hence, reducing the number of Toffoli gates is the main goal when designing the experiments.

For the fusion of $\Psi_m$-fluxes on the small planar graph, we obtained a device-ready circuit of depth 70. This circuit prepares the ground state, implements the ribbon operators and performs a partial charge measurement. 
On the braiding ladder the same protocol leads to a much shorter circuit of depth 37. This is due to a significantly simpler ground state preparation.

\begin{figure}
	\centering
	\includegraphics[width=\linewidth]{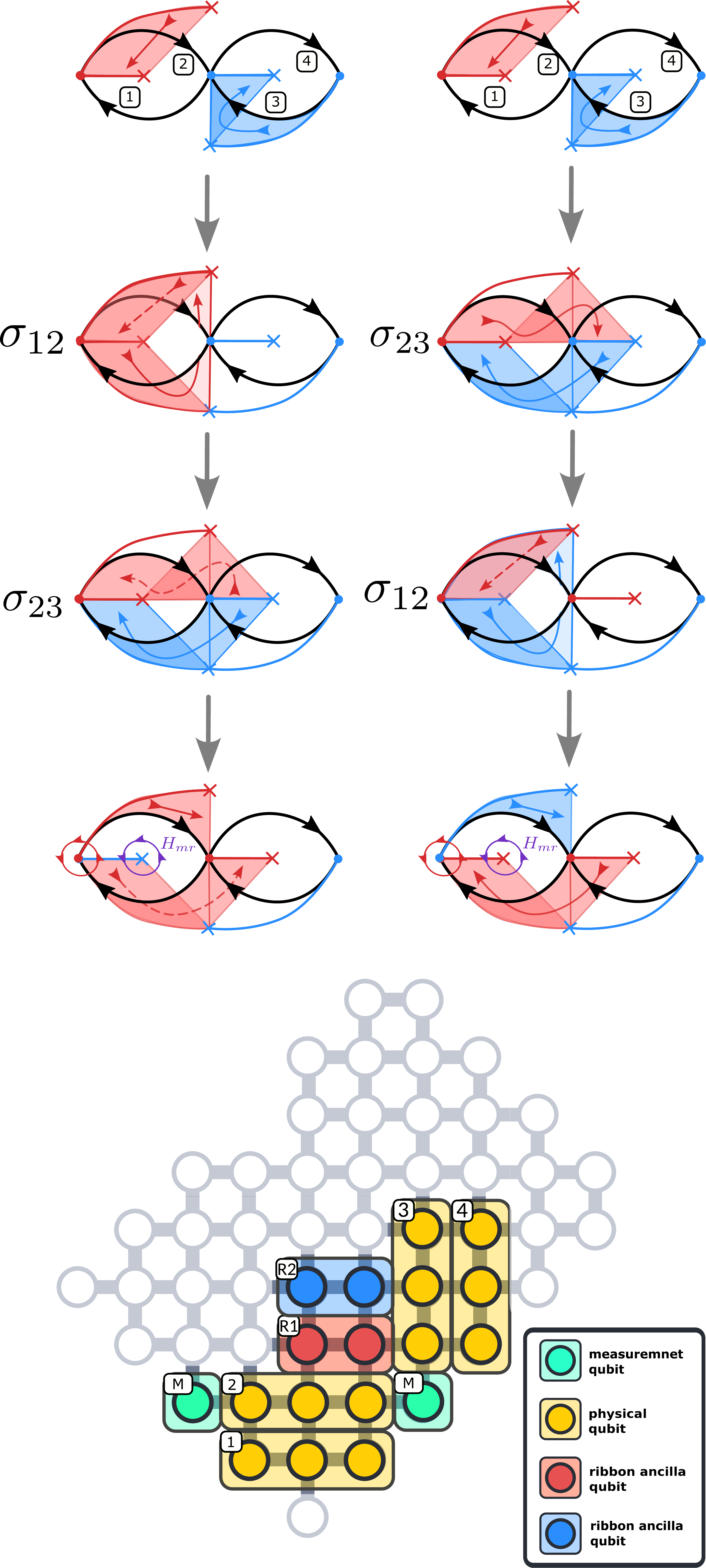}
	\caption{Ribbon operators for the two braiding experiments. On the last step we also draw a circle around the plaquette where we perform the $H_{mr}$-partial charge measurement. Note, that the lattice is embedded into a sphere, so the outside plaquette that we labelled twice for clarity should be identified.}
	\label{fig:braiding_setup}
\end{figure}

\emph{Braiding.} The ribbon operators for the two braiding protocols are shown in Fig.~\ref{fig:braiding_setup}. The qubit layout on the Sycamore chip is the same as for the fusion protocol on the ladder geometry (see Fig.~\ref{fig:fusion_setup}).

The circuit depths achieved for braiding the $\Psi_m$ fluxes are 60 and 68 for the cases of $\sigma_{23}\sigma_{12}$ and $\sigma_{12}\sigma_{23}$, respectively. We performed these experiments on the double braiding ladder. This is due to constraints of the simulation. Adding 6 extra qubits needed for a triple-ladder, dramatically slows down the classical simulation run times to the point of impracticality. On a real quantum device this problem would not occur.

\begin{figure*}
\centering
\begin{subfigure}{0.47\textwidth}
    \includegraphics[width = \linewidth]{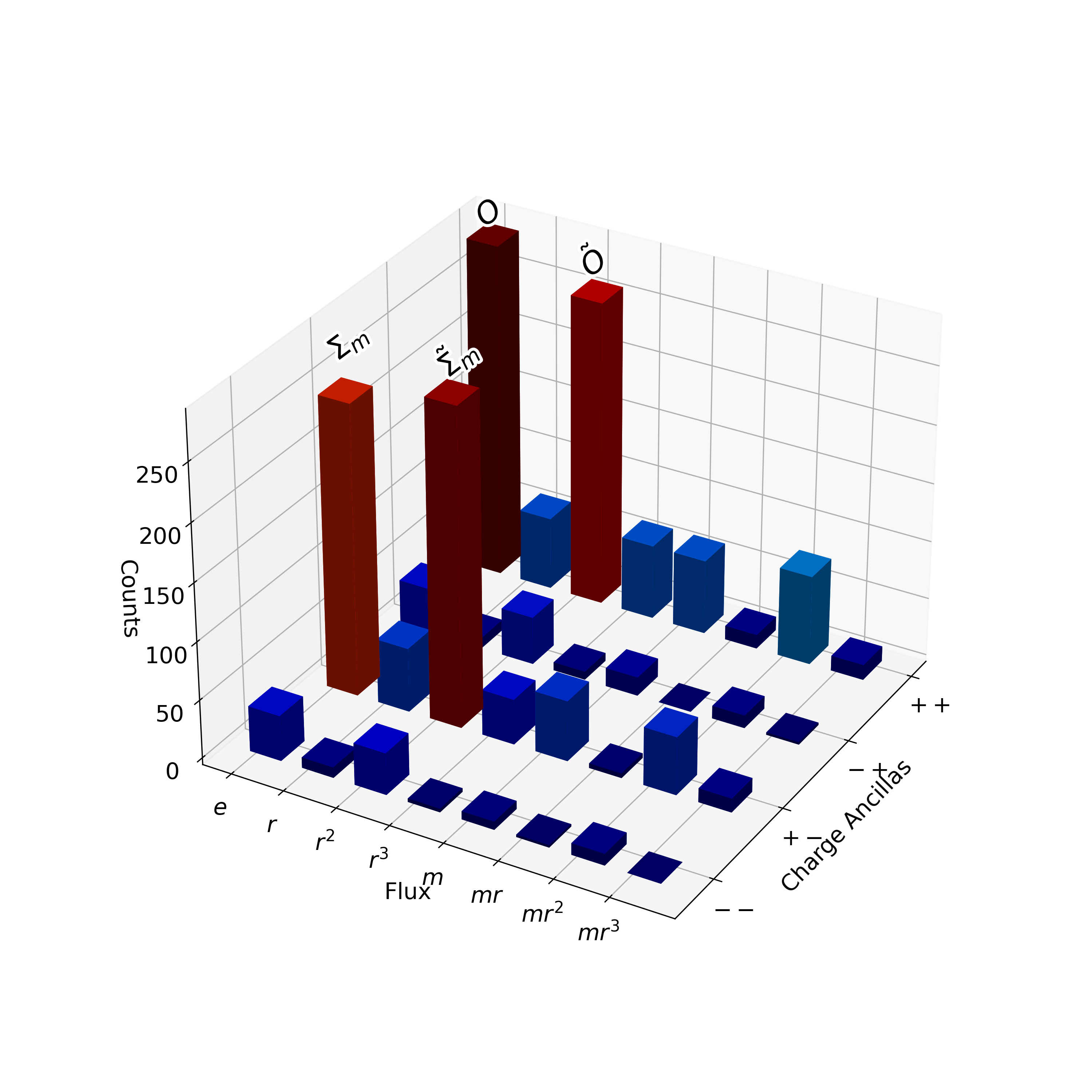}
    \caption{The partial charge-flux measurement after fusing two $\Psi_m$ pure fluxes on the site of fusion. The partial charge measurement was done with $H_{mr}$ subgroup, and the geometry was that of the braiding ladder, Figure \ref{fig:latticeGS}. The theoretical prediction is that the four measurement outcomes $O,\tilde O,\Sigma_m \tilde{\Sigma}_m$ are the only possible outcomes and occur with equal probability.}
    \label{fig:fusion_glass}
\end{subfigure}\hfill
\begin{subfigure}{0.47\textwidth}
    \includegraphics[width=\linewidth]{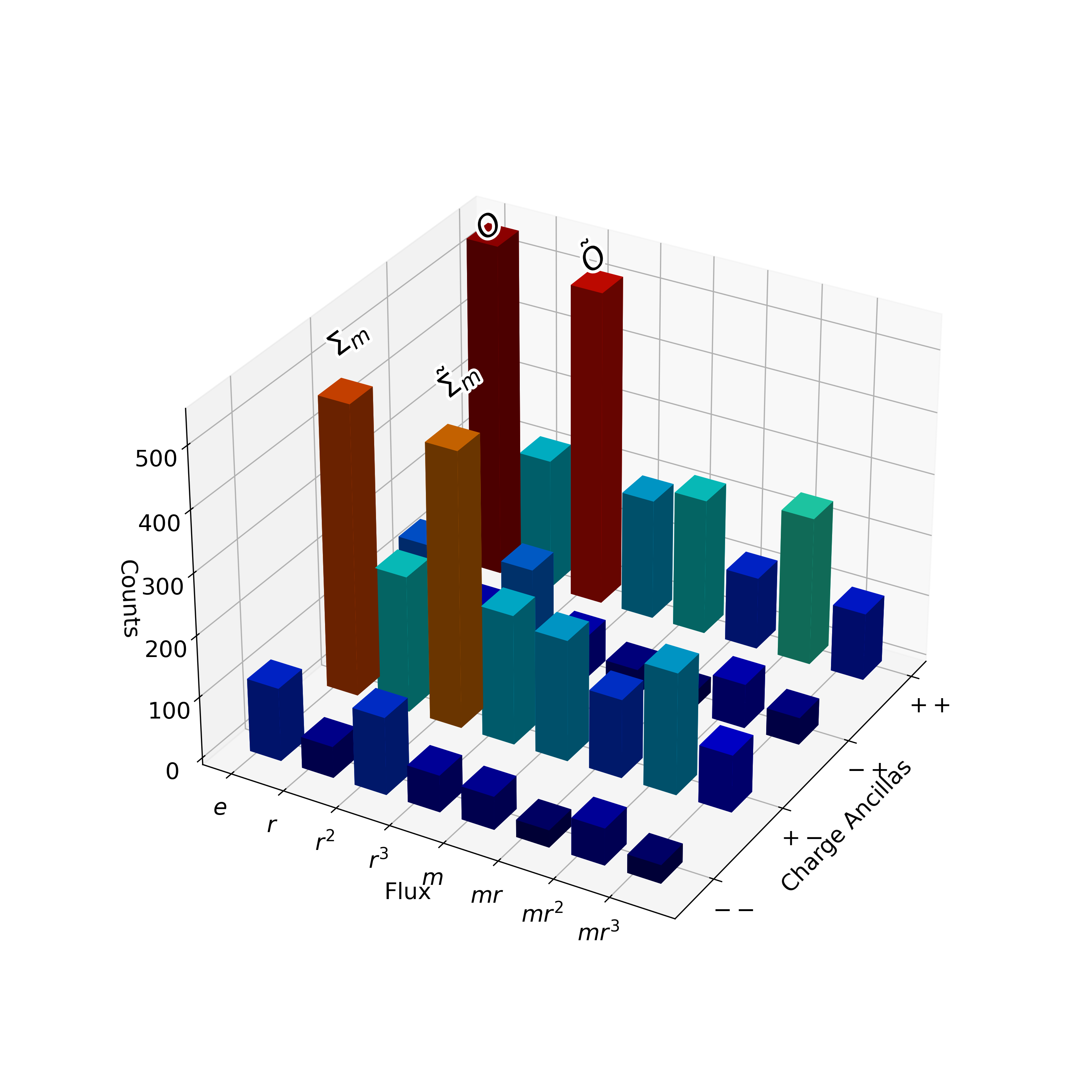}
    \caption{The partial charge-flux measurement after fusing two $\Psi_m$ pure fluxes on the site of fusion. The partial charge was done with $H_{mr}$ subgroup, and the geometry was that of the small two-dimensional graph, Figure \ref{fig:basketball}. The theoretical prediction is that the four measurement outcomes $O,\tilde O,\Sigma_m \tilde{\Sigma}_m$ are the only possible outcomes and occur with equal probability.}
    \label{fig:fusion_basketball}
\end{subfigure}
\vspace{15pt}
\begin{subfigure}{0.47\textwidth}
    \includegraphics[width=\linewidth]{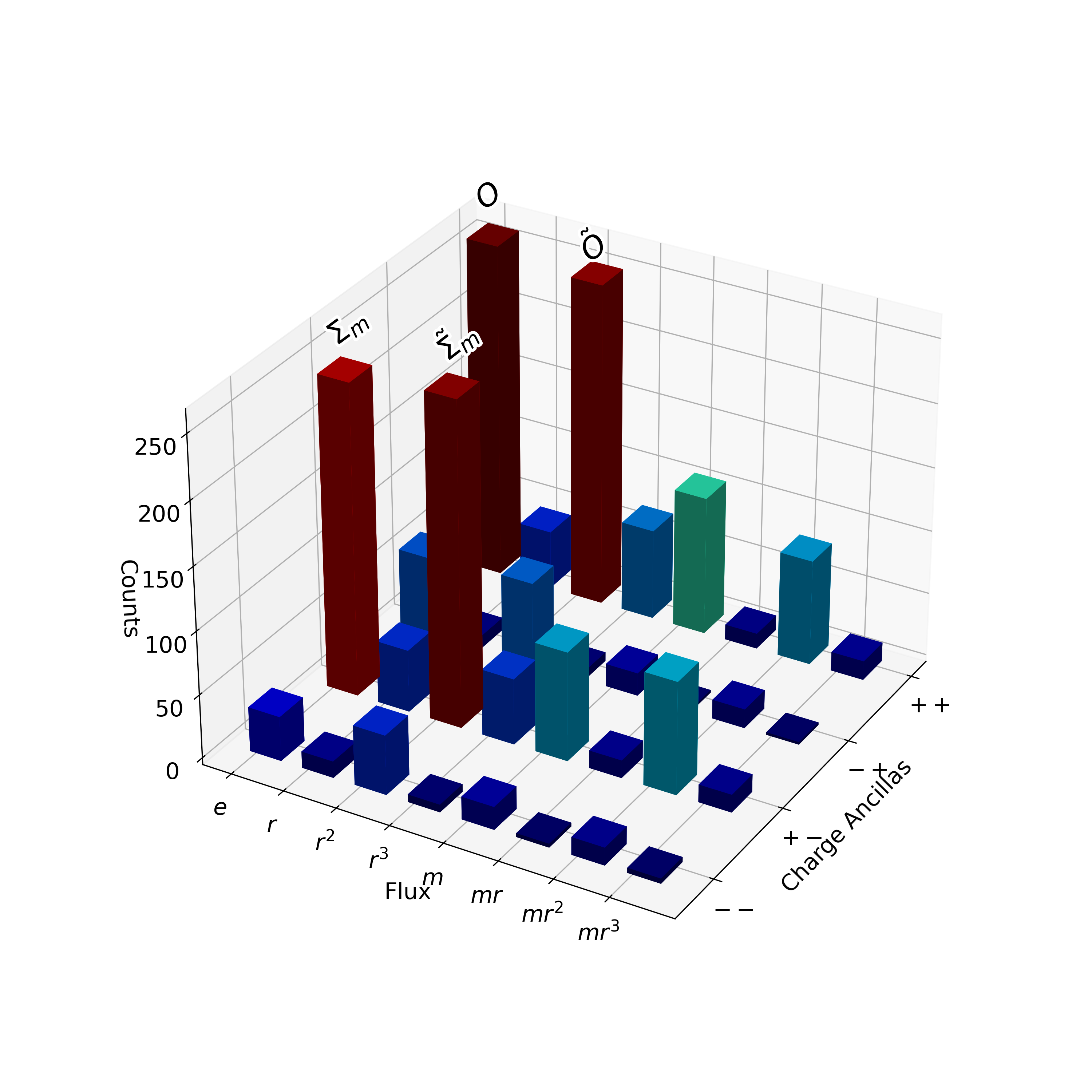}
    \caption{The braiding on a braiding ladder: the partial charge-flux measurement after performing the braiding protocol in $\sigma_{12}\sigma_{23}$ order, Figure \ref{fig:flux_braid} (left), where $\sigma_{ii+1}$ is a braid group generator. The anyons used are the pure fluxes $\Psi_m$. The partial charge measurement was done with respect to $H_{mr}$ subgroup. The theoretical prediction is that the four measurement outcomes $O,\tilde O,\Sigma_m \tilde{\Sigma}_m$ are the only possible outcomes and occur with equal probability.} 
    \label{fig:braid_fuse}
\end{subfigure}\hfill
\begin{subfigure}{0.47\textwidth}
    \includegraphics[width=\linewidth]{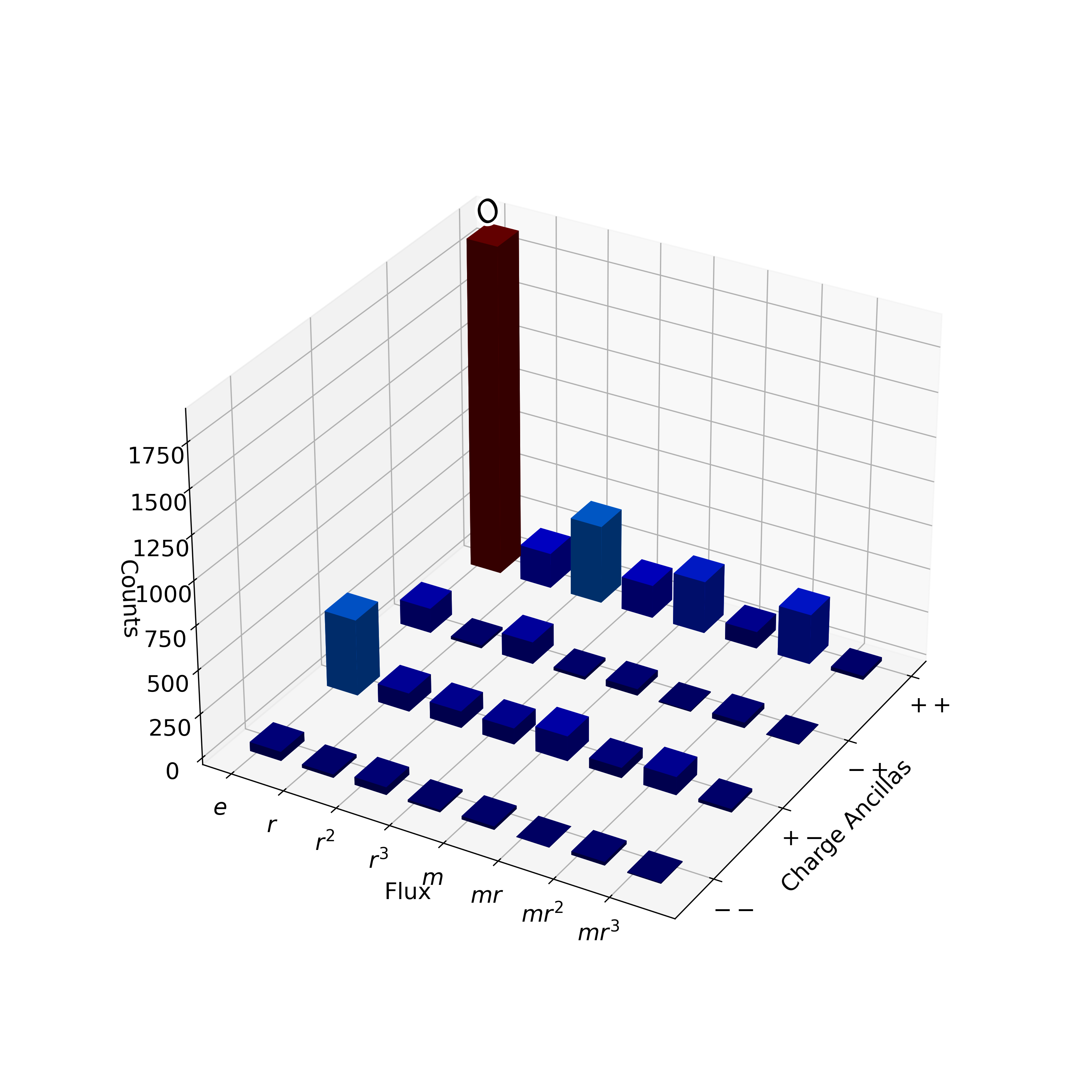}
    \caption{The braiding on a braiding ladder: the partial charge-flux measurement after performing the braiding protocol in $\sigma_{23}\sigma_{12}$ order, Figure \ref{fig:flux_braid} (right), where $\sigma_{ii+1}$ is a braid group generator. The anyons used are the pure fluxes $\Psi_m$. The partial charge measurement was done with respect to $H_{mr}$ subgroup. The theoretical prediction is that the measurement outcome $O$ is the only possible outcome.}
    \label{fig:braid_link}
\end{subfigure}
\caption{The partial topological charge measurements for the fusion and braiding protocols.}
\label{fig:red_charge_res}
\end{figure*}

\textbf{Results.}
The results of the fusion experiments on the ladder and the planar graph, as well as the braiding on the ladder are shown in Figure \ref{fig:red_charge_res}.

\emph{Fusion.} 
On both lattices we measure the charge after fusion via a partial $H_{mr}$-measurement and see the signatures of four fusion outcomes
$$\Psi_m \otimes \Psi_m = 0 \oplus \tilde 0 \oplus \Sigma_m \oplus \tilde{\Sigma}_m,$$
corresponding to measuring no flux or $r^2$ flux combined with no charge ($(1,1)$-outcome) or a non-trivial charge ($(1,-1)$-outcome) identified as $\alpha_m$.
In the case of the small planar graph we see a significantly 
increased background noise due the deeper circuit used in the ground state preparation.

The number of runs for the fusion on the braiding ladder was $16000$. The expected post-selection probability for the projection of the two ribbon ancilla pairs is $1/4^2=0.0625$, while the actual rate of success was $0.113$, due to the circuit noise and measurement readout bias. The four main bins corresponding to the expected topological charges count $270\pm5$ while the largest noise peak is $77$ leaving us with a signal to noise ratio of approximately $3.5$. 

The number of runs for the fusion on the small planar graph was $64000$. The expected post-selection probability for projection of the two ribbon ancilla pairs is again $0.0625$, while the actual number of successes was $0.0826$. The four main bins count $450\pm7$ with the biggest noise peak being $242$ leaving us with a signal to noise ratio of approximately $1.9$.

\emph{Braiding.}
Looking at the results of the charge measurement for the two braiding protocols we clearly see what we expected. The first braiding protocol results in multiple fusion outcomes while for the second braiding protocol the resulting fusion outcome is only vacuum.

The number of runs for both protocols was $16000$. The expected post-selection probabilities for the two protocols are $0.0625$ for $\sigma_{12}\sigma_{23}$ and $0.25$ for $\sigma_{23}\sigma_{12}$, while the actual rates of successes were $0.123$ for $\sigma_{12}\sigma_{23}$ and $0.315$ for $\sigma_{23}\sigma_{12}$, respectively.

In the case of the first braiding we see the four main peaks corresponding to the expected topological charges counting $255\pm 5$ with the biggest peak coming from the noise counting $109$ and resulting in a signal to noise ratio of about $2.3$. In the second case the peak corresponding to the vacuum counts $1890$. The largest peak coming from the noise counts 450 giving a signal to noise ratio of about $4.2$.

\textbf{Supplementary measurements.} In the analysis above we relied on the knowledge of the fusion outcomes to match the observed measurement outcomes of the $H_{mr}$-partial charge measurement with the corresponding charges. 

However, we can still determine what charges the outcomes correspond to even if we do not rely on the knowledge of the fusion rules. This is done by repeating the partial charge measurements for another subgroup, as explained in Section \ref{sec:redchmmt}.

Repeating the protocol with a $H_m$-partial charge measurement we see only two peaks corresponding to the charge measurement outcome $(1,1)$ combined with no or $r^2$ flux. This is due to the fact that all anyons that emerge from the fusion carry either no charge or $\alpha_m$ charge, which both correspond to the measurement outcome $(1,1)$ for the $H_m$ subgroup. Given that we observed the outcomes $(1,1)$ and $(1,-1)$ for $H_{mr}$, we can conclude that the charges present are the trivial and the $\alpha_m$ charge (data shown in Appendix~\ref{app:more_data}).

\subsection{Linking and twist matrices}\label{sec:num:intef}

In this section, we present the results of simulating the interference protocols for measuring the magnitude and the phase of the $S$- and $T$-matrix elements.

\begin{figure}
	\centering
	\includegraphics[width=\linewidth]{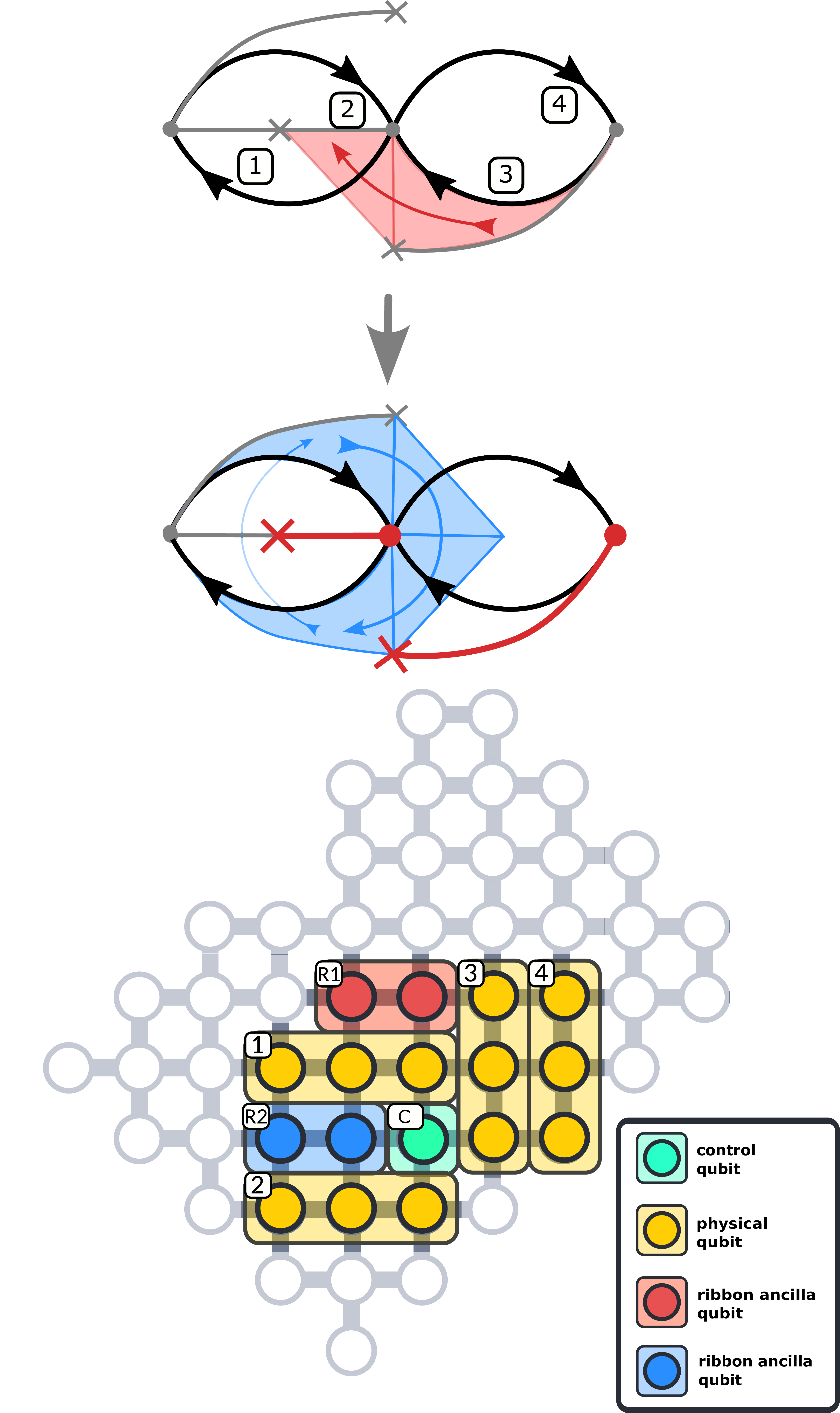}
	\caption{(Top) The ribbon operators applied in the simulation of the $S$-matrix protocol. The existence or the type of the blue (equal-time) ribbon is conditioned on the state of the control qubit. (Bottom) The qubit layout for the interference protocols.}
	\label{fig:intef_setup}
\end{figure}

\textbf{Circuit characteristics.}
Figure \ref{fig:intef_setup} depicts the qubit layout and the exact ribbon operators used for the $S$-matrix protocol.
\begin{figure}
	\centering
	\includegraphics[width=\linewidth]{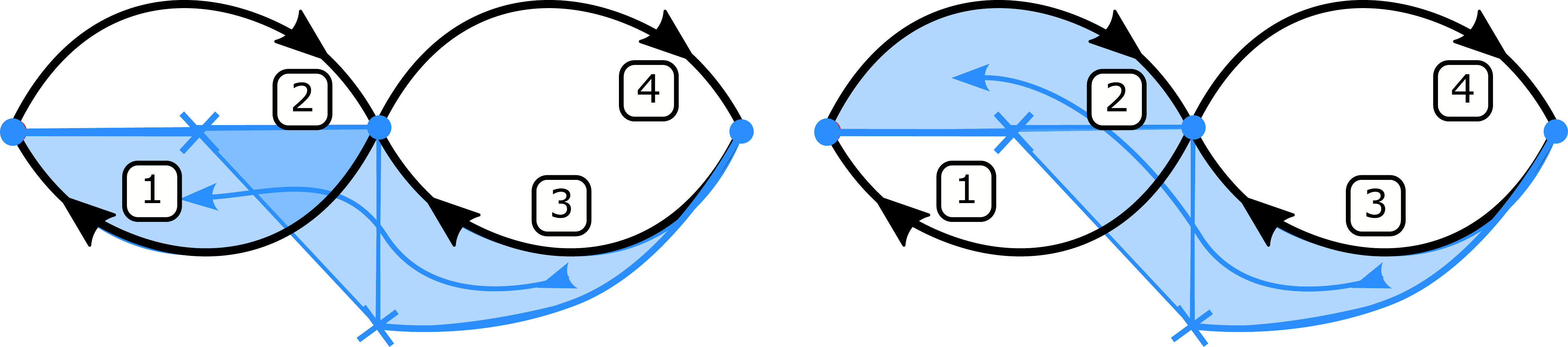}
	\caption{The two ribbon paths between same sites that differ by one twist used in our path interference protocol.}
	\label{fig:T_intef_setup}
\end{figure}
The application or the flavour of the blue (equal-time) ribbon is conditioned on the state of the control qubit.
The layout on the chip is chosen such as to reduce the number of two-qubit gates acting on the control qubit, avoiding an accumulation of errors. This turned out to be more relevant for obtaining reliable estimates than the overall circuit depth. 

For the $T$-matrix experiment we chose the same layout as for the $S$-matrix experiment and the concrete form of the ribbon is shown in Fig.~\ref{fig:T_intef_setup}.

The depths of the circuits for the $S$-matrix protocol heavily depend on the anyons involved.
In the following, we present in detail the circuits and results for the exemplary $S$-matrix elements $\tilde{S}(\Psi_m, \Psi_m)=1$, $\tilde{S}(\Psi_m, \tilde{\Psi}_m)=-1$ and $\tilde{S}(\Psi_m, \Psi_r)$ which are of intermediate depth. For conditioning the ribbon type we have 58 for $S(\Psi_m, \Psi_m)$ and $S(\Psi_m, \tilde{\Psi}_m)$	 and 64  for $S(\Psi_m, \Psi_r)$, while for conditioning the existence of the ribbon we find 84 and 90, respectively. We will return to a more detailed discussion of the circuit depths and results for all other $S$-matrix elements at the end of this section.

For the $T$-matrix elements we chose to measure $T(\Phi_r,\Phi_r)=i$, which together with $T(\tilde{\Phi}_r,\tilde{\Phi}_r)=-i$ is the most non-trivial entry corresponding to the two semions (all other anyons have topological spin $\pm 1$). The circuit depth of 89 for this protocol is larger than for all other $T$-matrix elements and leads to reasonable results. Hence, we did not investigate the other $T$-matrix entries as they seem to be less challenging to measure.

\textbf{Interference and tomography.}
In Section \ref{subsec:Intef}, we proposed an interference scheme that entangles an ancilla qubit with the space-time history of the gauge field excitations in such a way that by the end of the protocol the qubit and the field are disentangled and the qubit is left in a state that depends only on the topological properties of the spacetime history of the anyons -- the $S$- and $T$-matrix elements.

The fact that the qubit is meant to be disentangled from the gauge field and any additional ancillas implies that the qubit is ideally left in a pure state. Hence, it is easy to tomograph and to extract the aforementioned topological properties. The noise in the system alters the situation. For conceptual clarity, we will thus first discuss the ideal case and subsequently comment on the effect of the noise.

The final pure state of the ancilla qubit after the ideal protocol is
\begin{equation}
\begin{split}
    \ket{\psi}_c = \frac{1}{\sqrt{1+|\tilde{S}_{ab}|^2}}(\ket{0}_c + \tilde{S}_{ab}\ket{1}_c),\\
    \rho_c = \frac{1}{1+|\tilde{S}_{ab}|^2}\begin{pmatrix}
    1 & \tilde{S}_{ab}^* \\
    \tilde{S}_{ab} & |\tilde{S}_{ab}|^2
    \end{pmatrix}.
\end{split}
\end{equation}
We write this density matrix in its Pauli basis 
\begin{equation}
    \rho_c = \frac{\mathbb{1} + \vec{r} \cdot \vec{\sigma}}{2},
\end{equation}
where  $r_i = \text{Tr}(\sigma_i\rho_c)$ is the Bloch vector and $\vec{\sigma} = (\sigma_x, \sigma_y, \sigma_z)$ is the vector of Pauli matrices and find
\begin{equation}
    \vec{r} = \left( \frac{2 \text{Re}\tilde{S}_{ab}}{1+|\tilde{S}_{ab}|^2}, \frac{2 \text{Im}\tilde{S}_{ab}}{1+|\tilde{S}_{ab}|^2}, \frac{1 - |\tilde{S}_{ab}|^2}{1+|\tilde{S}_{ab}|^2} \right).\label{eqn:bloch}
\end{equation}
Note that $|\vec{r}| = 1$ as expected for a pure state. The phase and the magnitude of $\tilde S_{ab}$ can be read off from the orientation of the Bloch vector only and the magnitude of $\vec r$ is not needed to determine $\tilde S_{ab}$. The magnitude is strongly affected by the noise in the system as we explain in the following.

The noise in the gates will cause entanglement between the control ancilla qubit $c$ and the gauge field. Hence, when we tomograph the control qubit, we probe a mixed state whose Bloch vector is $|\vec{r}|< 1$. We will assume, that the noise only shortens this Bloch vector, i.e. it acts uniformly across all channels. Under this assumption, we can still read off the S-matrix elements as they only depend on the direction of the Bloch vector. Note, that this assumption is rather strong given that the generic dephasing process usually pulls the Bloch vector towards the $z$-axis. Nevertheless, the assumption is used in order to make the problem tractable without additional machine specific information. In an actual experiment, more elaborate error mitigation techniques and data about the noise bias could be used to replace this simple model.

\emph{Tomography.} In order to determine the orientation of the Bloch vector we measure in a set of different bases. Each basis is parametrised by a vector $\vec{s}_i$ with $\sigma_{s_i} = \vec{s}_i \cdot \vec{\sigma}$ being the associated Hermitian operator.

Given a basis $\vec{s}_i$ and a Bloch vector of a mixed state $\vec{r}$, the quantum mechanical probabilities for the two measurement outcomes are
\begin{equation}\label{eqn:probs}
\begin{split}
p_{QM}(1|\vec{s}_i, \vec{r}) = \frac{1}{2}(1+\vec{r}\cdot\vec{s}_i)\,,\\
p_{QM}(0|\vec{s}_i, \vec{r}) = \frac{1}{2}(1-\vec{r}\cdot\vec{s}_i)\,.
\end{split}
\end{equation}

This probability is, however, modulate by the readout bias
\begin{equation}
    p(b|\vec{s}_i, \vec{r}) = (1-\epsilon_b)p_{QM}(b|\vec{s}_i, \vec{r}) + \epsilon_{\bar{b}} p_{QM}(\bar{b}|\vec{s}_i, \vec{r}), \label{eqn:marg}
\end{equation}
where $\epsilon_b$ is the probability to measure the qubit in state $\bar{b}$ ($\equiv 1-b$) even though it is in the state $b$.

We perform the measurement $N=n_0 + n_1$ times and record the measurement outcomes $(n_0, n_1)$ and define the estimator 
\begin{equation}
    P(\vec{s}_i) = \frac{n_1-n_0}{N},
\end{equation}
which we call \emph{polarisation}.
Introducing $\bar \epsilon = (\epsilon_1+\epsilon_0)/2, \Delta \epsilon = \epsilon_1-\epsilon_0$ and using that 
$$p(b|\vec{s}_i, \vec{r})=\lim_{N\to \infty} \frac{n_b}{N},$$
we find \begin{equation}
\lim_{N \to \infty} P(\vec{s}_i)=(1-2\bar\epsilon)\vec{s}_i \cdot \vec{r} + \Delta\epsilon.\label{eqn:estim}
\end{equation}
 
Performing a sequence of measurements for different bases $\vec{s}_i$ allows us to extract $\vec{r}$, up to a multiplicative constant, i.e., we determine its direction.

For the sake of concreteness we chose the following sets of measurements bases  \begin{enumerate}
    \item \emph{Equatorial Scan.} Fixing the value of the polar angle to $\theta = \pi/2$, we vary the azimuthal angle $\phi \in [0, 2\pi)$. From this scan we extract $\phi_{\text{max}}$ which has the largest polarisation.
    \item \emph{Meridian Scan.} Fixing the value of the azimuthal angle to $\phi = \phi_{\text{max}}$, we scan the polar angle $\theta \in [0, 2\pi)$\footnote{The domain is extended on purpose.} From this scan we extract $\theta_{\text{max}}$ which has the largest polarisation.
\end{enumerate}
The extraction of the relevant angles is done by fitting Eq.~\eqref{eqn:estim}.
The two angles then fix the value of $\tilde{S}_{ab}$ via $$\tilde{S}_{ab} = \sqrt{\frac{1-\cos{\theta_{\text{max}}}}{1+\cos{\theta_{\text{max}}}}}e^{i\phi_{\text{max}}}.$$
The other two parameters, the amplitude and the offset of the polarisation, do not convey any physical information. The offset determines the difference of the effective readout biases, $\Delta\epsilon$, and the amplitude determines the product of the mean readout bias and the length of the Bloch vector, $(1-2\bar{\epsilon})|\vec{r}|$.

In the discussion above, we have neglected the fact that in addition to the measurement of the control qubit, we measure the ribbon ancillas and post-select on their biased measurement outcomes. This causes an increased observed effective readout bias $\Delta \epsilon$, obtained from the fitting procedure described in the last section, which exceeds the $\Delta \epsilon$ estimate from calibration data. A more detailed discussion of this issue is given in Appendix \ref{app:bias}.

For the $T$-matrix protocol the Bloch vector of the control qubit after performing the ideal circuit reads
\begin{equation}
    \vec{r} = \left(  \text{Re}{T}_{aa}, \text{Im}{T}_{aa}, 0 \right).\label{eqn:blochT} 
\end{equation}
We perform the same type of tomography to estimate $\vec r$ and calculate the magnitude and phase of $T_{aa}$ via the equation above.

\emph{Results.}
\begin{figure*}
    \centering
    \begin{subfigure}{\textwidth}
        \centering
        \includegraphics[width=0.7\linewidth]{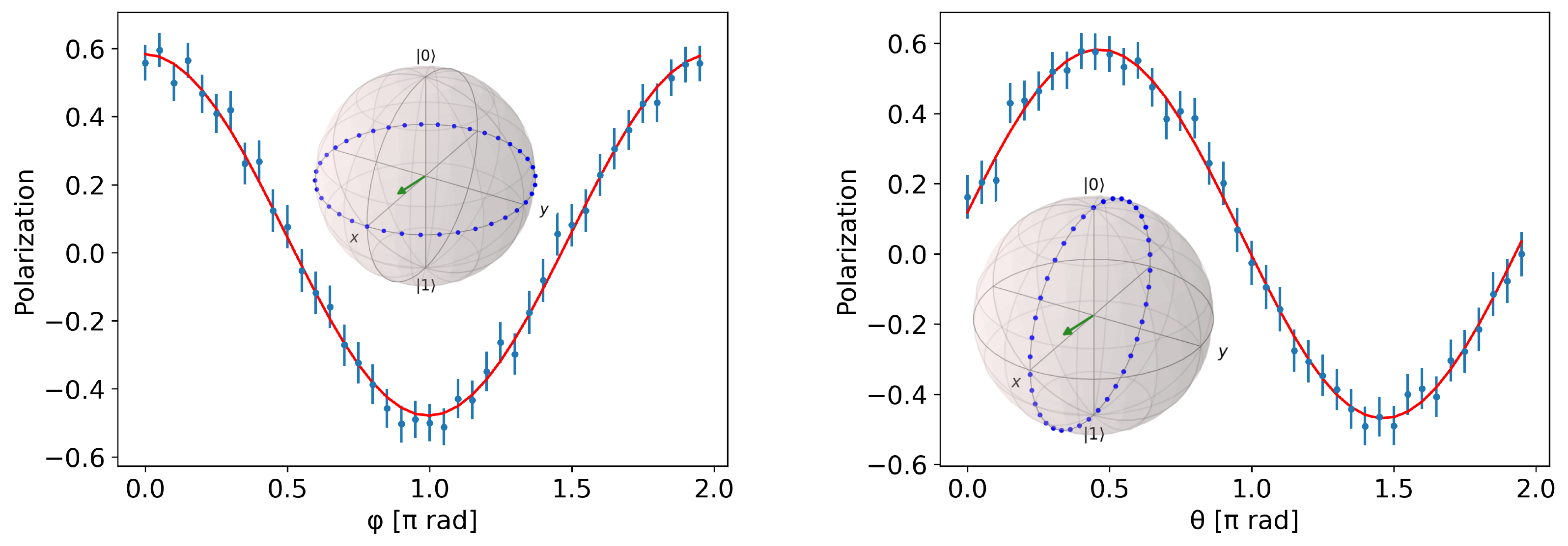}
        \caption{$\tilde{S}(\Psi_m, \Psi_m) = 0.89(1) e^{i\pi 0.004(4)}$ measured, $\tilde S(\Psi_m, \Psi_m) = 1$ predicted.}
        \label{fig:flav_cond_res_plus}
    \end{subfigure}
        \begin{subfigure}{\textwidth}
        \centering
        \includegraphics[width=0.7\linewidth]{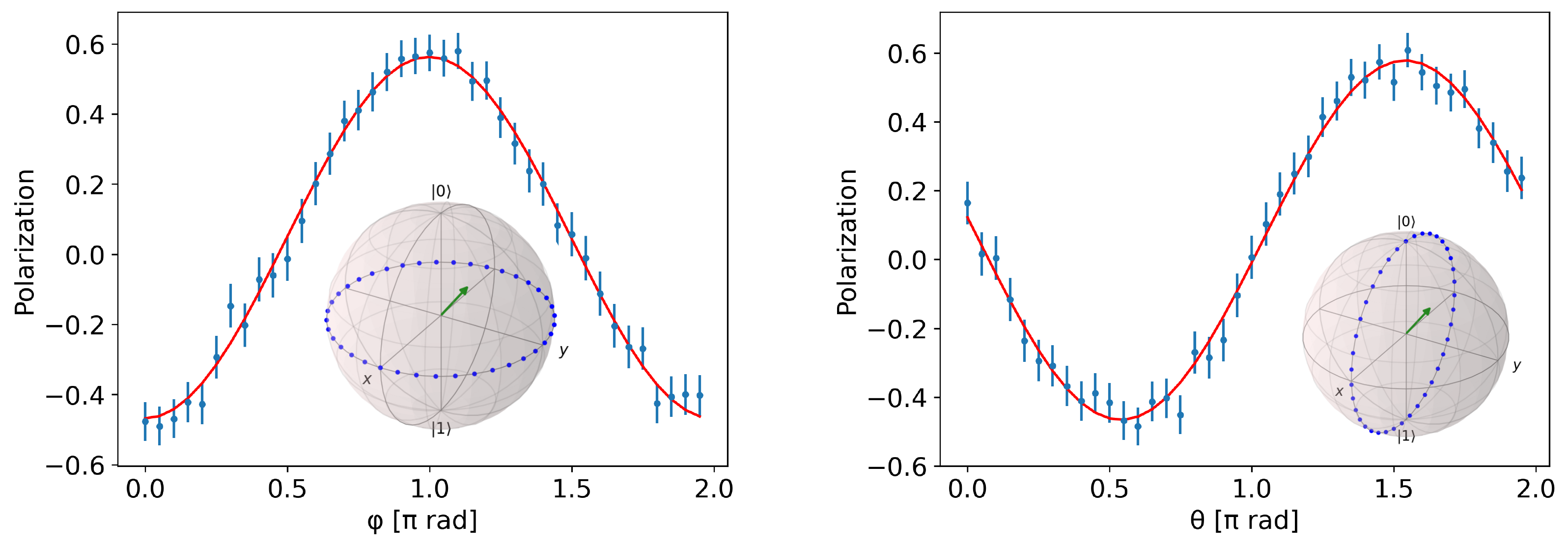}
        \caption{$\tilde{S}(\Psi_m, \tilde{\Psi}_m) = 0.88(1)e^{i\pi 0.998(5)}$ measured, $\tilde S(\Psi_m, \tilde{\Psi}_m) = -1$ predicted.}
        \label{fig:flav_cond_res_minus}
    \end{subfigure}
    \begin{subfigure}{\textwidth}
        \centering
        \includegraphics[width=0.35\linewidth]{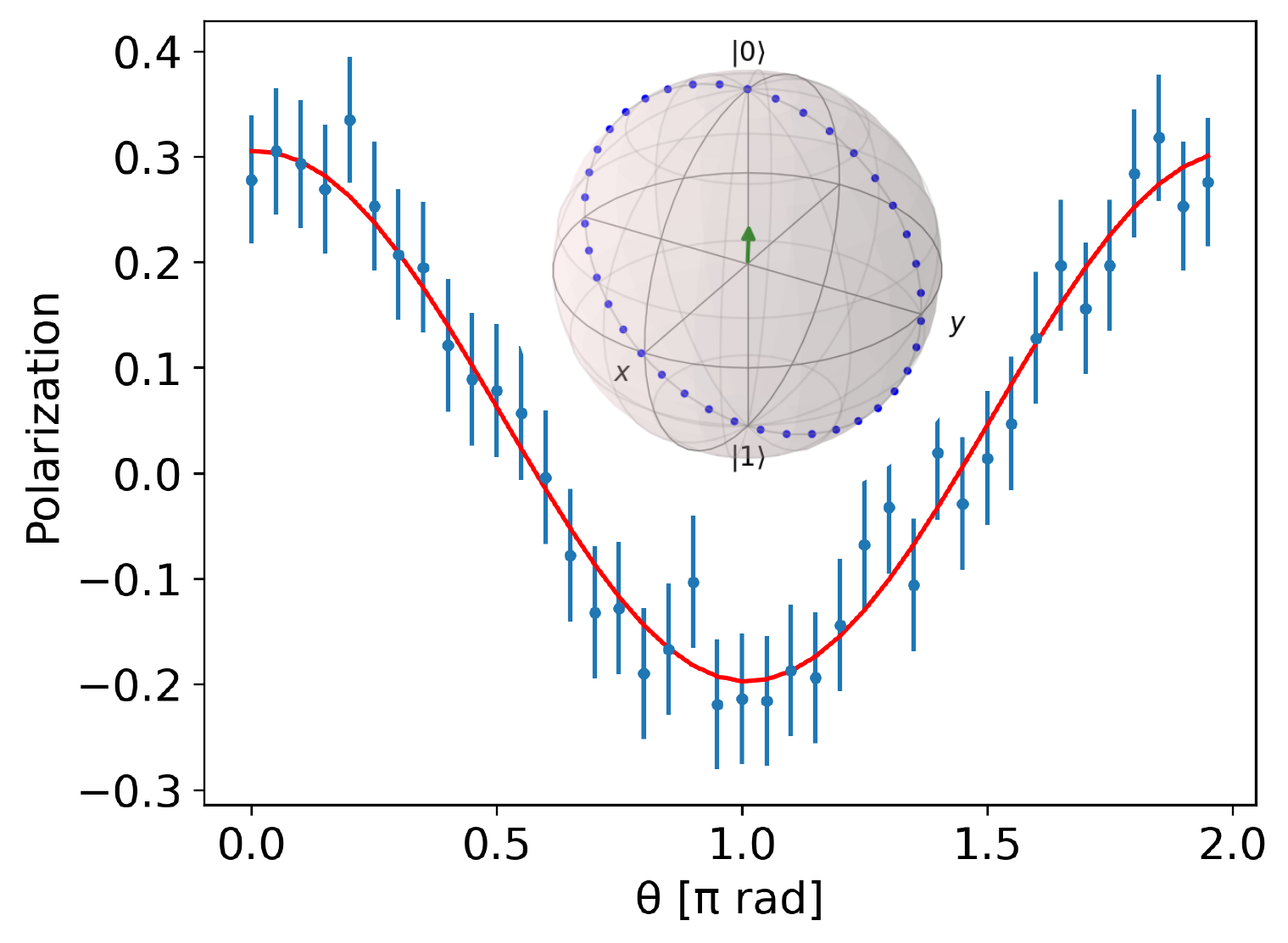}
        \caption{$\tilde{S}(\Psi_m, \Psi_r) = 0.02(2)$ measured, $\tilde S(\Psi_m, \Psi_r) = 0$ predicted. $\phi$-scan omitted, since $\theta=\pi$.}
        \label{fig:flav_cond_res_zero}
    \end{subfigure}
        \begin{subfigure}{\textwidth}
     \centering
        \includegraphics[width=0.7\linewidth]{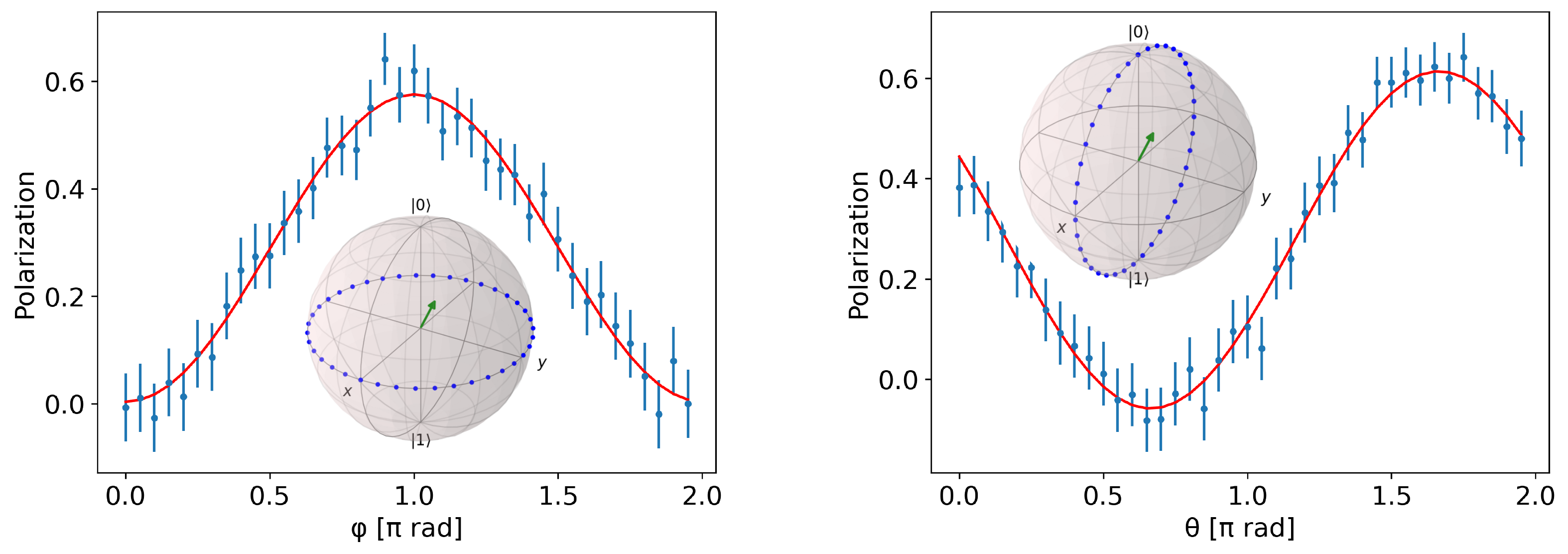}
        \caption{$\tilde{S}(\Psi_m, \tilde{\Psi}_m) = 0.58(1)e^{i\pi 1.002(8)}$ measured, $\tilde{S}(\Psi_m, \tilde{\Psi}_m)=1$ predicted.}
        \label{fig:ex_cond_res_zero}
    \end{subfigure}
    \caption{Numerical results of the control qubit tomography for the S-matrix interference protocol (a)-(c) conditioning the type of the equal time ribbon (cf. Figure \ref{fig:cond_flav}) and (d) conditioning its existence (cf. Figure \ref{fig:cond_ex}). The measurement basis was scanned across two planes, see the Bloch sphere diagram (blue dotted circles). The polarisation $P$ was estimated from these measurements by fitting Eq.~\eqref{eqn:estim}. It yields the Bloch vector of $\rho_c$ (green arrow) and $\tilde S_{ab}$.}
    \label{fig:S_res}
\end{figure*}
\begin{figure*}
    \centering
    \includegraphics[width=\textwidth]{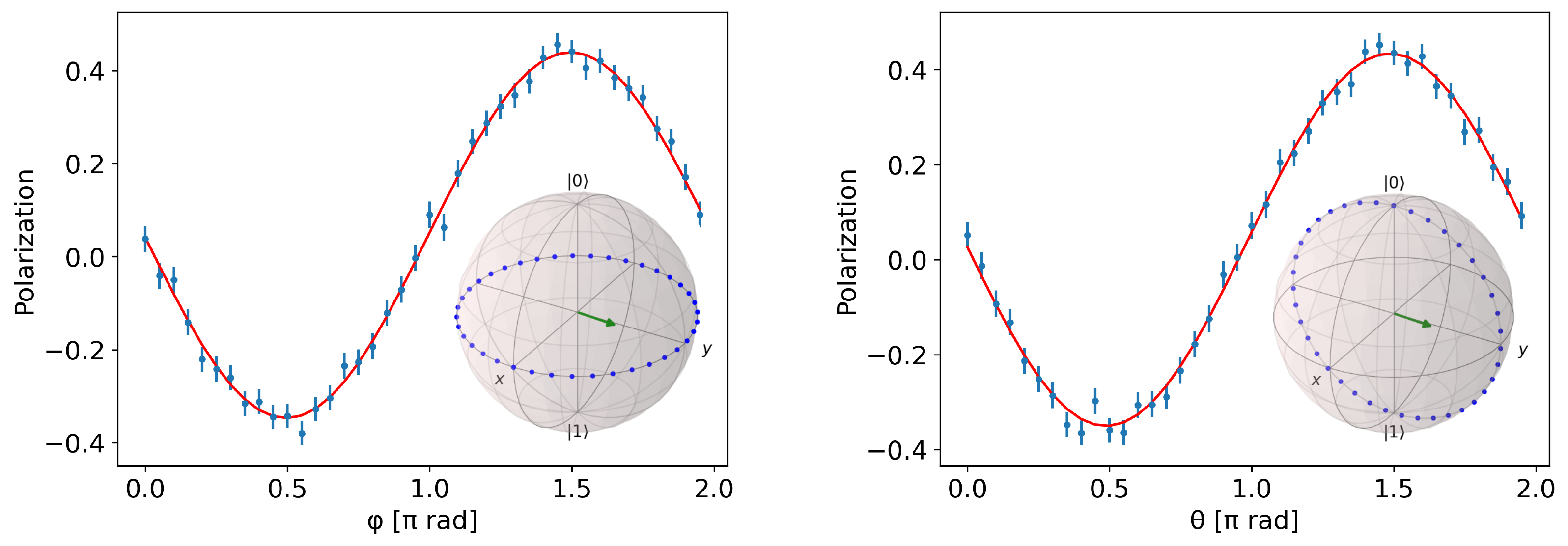}
    \caption{Numerical results of the control qubit tomography for the T-matrix interference protocol where the paths of the ribbon was conditioned (cf. Figure \ref{fig:Tmat}). Tomography performed as for the S-matrix (cf. Fig.\ref{fig:S_res}).  $T(\tilde{\Phi}_r, \tilde{\Phi}_r) = 1.04(1)e^{i\pi 1.496(4)}$ measured, $T(\tilde{\Phi}_r, \tilde{\Phi}_r) = -i$ predicted.}
    \label{fig:t_mat_results}
\end{figure*}
The numerical results for the interference protocols determining $S_{ab}$ with $a=\Psi_m$ and $b=\Psi_m,\tilde \Psi_m, \Psi_r$ are shown in Fig.~\ref{fig:S_res}. The first three panels show the results of the protocol in which the type of the second ribbon is conditioned to be $\Psi_b$ or $0 \oplus \tilde 0$ (cf. Fig.~\ref{fig:cond_flav}), while the last one shows the results for conditioning the existence of the ribbon (cf. Fig.~\ref{fig:cond_ex}). In both cases we note a systematic drift of the Bloch vector towards the $z$-axis. The magnitude of the S-matrix element is dictated by the angle of the Bloch vector with the z-axis and hence this drift leads to an underestimation of $|\tilde S_{ab}|$. The systematic drift is significantly less dramatic in the case of the first protocol, i.e., conditioning on $\Psi_b$ or $0 \oplus \tilde 0$.  Hence, the estimate of the magnitudes of the S-matrix elements is much closer to the theoretical value. 
 This is due to the fact that the circuit implementing this protocol is much shallower due to the simpler forms of the conditional ribbon operators, one Toffoli gate less in the conditional multiplication circuit, cf. Figure \ref{fig:flavCond}. The phases of all measured $\tilde S_{ab}$ are estimated well and agree with the theoretically predicted values.

The result for the T-matrix element $T(\tilde \phi_r, \tilde \phi_r)$ determined by the path conditioning protocol in Figure \ref{fig:Tmat} is shown in Figure \ref{fig:t_mat_results} and agrees with the theoretical value.

\emph{Uncertainty.} In the experiment we used 1000 shots per measurement basis for the S-matrix measurements and 5000 for the T-matrix measurements. See Appendix~\ref{app:DATA_Anl} to see how we estimated the final uncertainties in the measured braiding amplitudes, i.e., $S$- and $T$-matrix elements, as well as for each datapoint in our plots.

We expected a post selection probability of $1/4$ for all T-matrix protocols and S-matrix protocols with measured $|S| \neq 0$. For the case of measured $|S| = 0$ the probability drops to $1/8$. These probabilities were observed in the numerical experiment.

\textbf{Other S-Matrix elements.}
As stated above some anyon ribbons are harder to compile than others. The bottleneck of this anyon interference protocol is the difficulty of the conditioned ribbon. 

We can divide the $S$-matrix protocols into six different difficulty classes measured by the depth of the required circuits. The biggest factor determining those is whether for the anyon in question, $(C, \chi)$ obeys $\chi(r^2) = \chi(e)$. If this is not the case, the representation $A{(g)}$ is faithful. This means that all group elements must be included in the circuit which requires many SWAP gates and increases the circuit depth significantly.

The second factor to consider is the number of Toffoli gates in the controlled multiplication and generalised conjugation circuits. The semions $\Phi_r,\tilde \Phi_r$ are difficult in both regards. 

\begin{figure*}
	\centering
	\includegraphics[width=\textwidth]{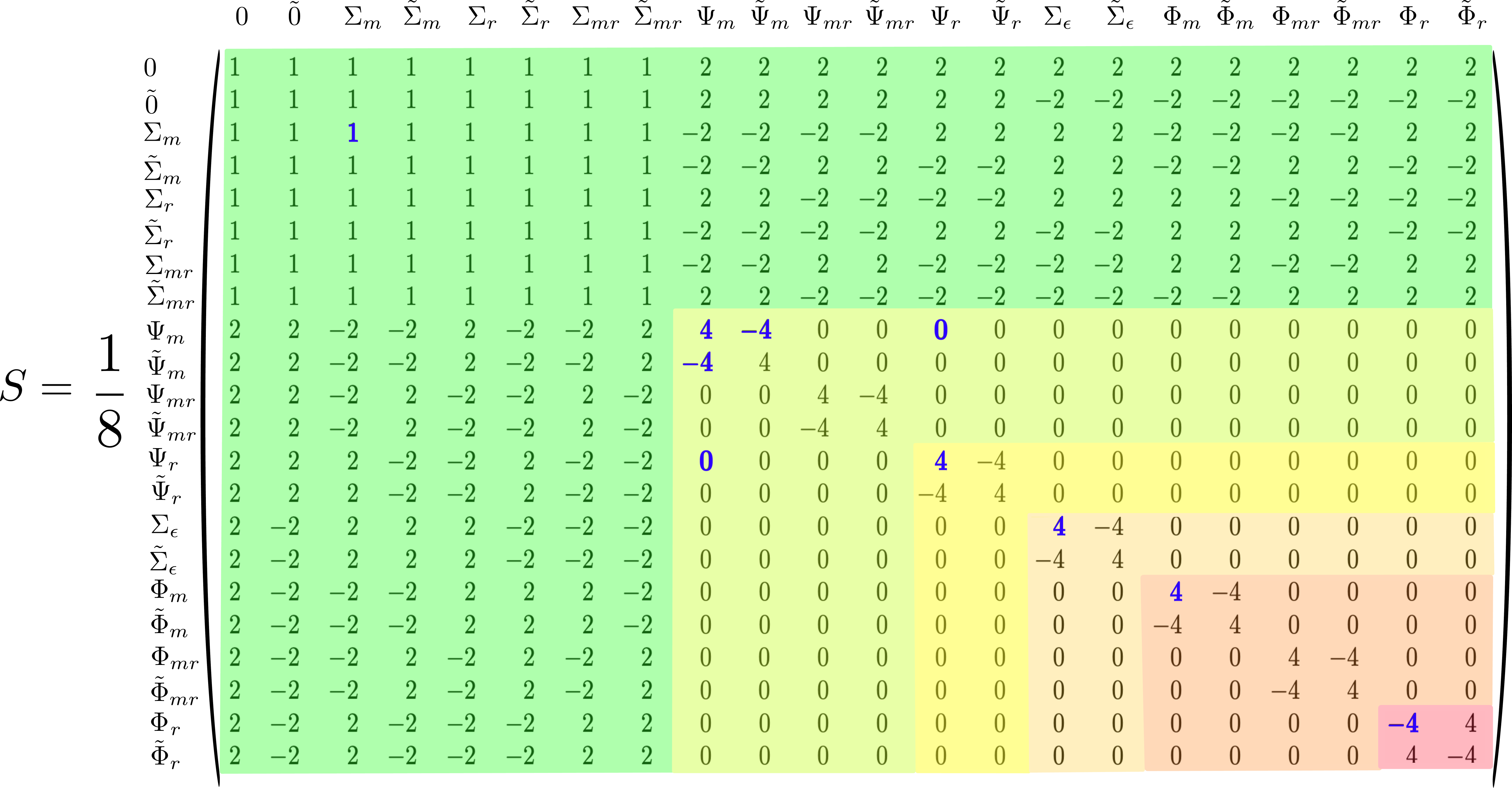}
	\caption{The S-matrix of the $D(D_4)$ theory. The color shading represents the difficulty to observe the values experimentally, where green to red denotes increasing difficulty. For the entries highlighted in bold face and blue we numerically obtained the values by simulating the phase-sentitive measurement protocols.}
	\label{fig:Smat_diff}
\end{figure*}

In Fig.~\ref{fig:Smat_diff} we show the $S$-matrix with all its entries color coded according to the difficulty class. In the list below we present the circuit depths and numerical simulation results for the diagonal S-matrix elements for representative cases of all six difficulty classes. For all protocols, except for 1. and 4., we chose to condition the ribbon $b$ vs $0\oplus \tilde 0$ rather than ribbon $b$ vs $0 \oplus 0$.

\begin{enumerate}
	\item Conditioning an abelian anyon: $\tilde{S}(\Sigma_m, \Sigma_m) = 0.969(6)e^{i\pi 0.004(2)}$, theoretical prediction $\tilde{S}(\Sigma_m, \Sigma_m)=1$. The total compiled circuit depth is 23 with 0 Toffoli gates corresponding to the difficulty class shaded in green in Fig.~\ref{fig:Smat_diff}.
	\item Conditioning a $m$- or $mr$-dyon with $\chi(r^2) = 1$: $\tilde{S}(\Psi_m, \Psi_m) = 0.89(1)e^{i\pi 0.004(4)}$, theoretical prediction $\tilde{S}(\Psi_m, \Psi_m)=1$. The total compiled circuit depth is 58 with 2 Toffoli gates corresponding to the difficulty class shaded in light green in Fig.~\ref{fig:Smat_diff}.
	\item Conditioning a $r$-dyon with $\chi(r^2) = 1$: $\tilde{S}(\Psi_r, \Psi_r) = 0.82(1)e^{i\pi 0.001(4)}$, theoretical prediction $\tilde{S}(\Psi_r, \Psi_r)=1$. The total compiled circuit depth is 77 with 4 Toffoli gates corresponding to the difficulty class shaded in yellow in Fig.~\ref{fig:Smat_diff}.
	\item Conditioning a non-abelian charge: $\tilde{S}(\Sigma_\epsilon, \Sigma_\epsilon) = 0.123(8)e^{i\pi 0.01(2)}$, theoretical prediction $\tilde{S}(\Sigma_\epsilon, \Sigma_\epsilon)=1$. The total compiled circuit depth is 125 with 4 Toffoli gates corresponding to the difficulty class shaded in light orange in Fig.~\ref{fig:Smat_diff}.
	\item Conditioning a $m$- or $mr$-dyon with $\chi(r^2) = -1$: $\tilde{S}(\Phi_m, \Phi_m) = 0.19(2)e^{i\pi 0.04(4)}$, theoretical prediction $\tilde{S}(\Phi_m, \Phi_m)=1$. The total compiled circuit depth is 133 with 4 Toffoli gates corresponding to the difficulty class shaded in orange in Fig.~\ref{fig:Smat_diff}.
	\item Conditioning a $r$-dyon with $\chi(r^2) = -1$ (semions): $\tilde{S}(\Phi_r, \Phi_r) = 0.06(1)e^{i\pi 1.0(3)}$, theoretical prediction $\tilde{S}(\Phi_r, \Phi_r)=-1$. The total compiled circuit depth is 171 with 6 Toffoli gates corresponding to the difficulty class shaded in red in Fig.~\ref{fig:Smat_diff}.
\end{enumerate}
We assume that the first element is conditioned and largely determines the circuit depth while the specific choice of the other, unconditioned ribbon operator only mildly influences the depth. This means that the difficulty is set by the row of the S-matrix and one can use the symmetry of the S-matrix to pick the better of the two interference procedures. For example for measuring $S(\Phi_r, \Sigma_m)$ one would condition the $\Sigma_m$ ribbon and not the semion $\Phi_r$.

We observe that the magnitude of the $S$-matrix in the simulation results decays strongly with the circuit depth. This is due to accumulating errors on the control qubit. We speculate that the main error channel responsible for the drift of the estimated angle $\theta_{\text{max}}$, and hence the braiding amplitude, is dephasing. Dephasing reduces the $r_x$ and $r_y$ components of the Bloch vector, while keeping $r_z$ fixed, so any error in estimating this angle due to any small offset $\delta r_z \neq 0$ will be amplified. On the upside, this error channel does not change the ratio between $r_x$ and $r_y$ and hence the braiding phase estimates are well within the theoretical values.

\section{Extension to $S_3$}\label{sec:other_gauge}

The anyons of $D_4$ lattice gauge theory are not suitable for universal topological quantum computation.
This is due to the simple structure of $D_4$. All of its subgroups are abelian and the group is nilpotent and solvable.

However, the slightly more complex gauge group $S_3$, can be used for universal topological quantum computation, if we allow measurements. Hence, it is interesting to investigate how our schemes carry over to this case. We find that the main tricks we use to achieve a manageable circuit depth for $D_4$, in particular the simplified group multiplication circuits for elements restricted to subgroups or conjugacy classes, would carry over straight forwardly to $S_3$ if the quantum hardware had native \emph{qutrits}. However, for a qubit based hardware implementing $S_3$ is considerably more difficult. In the following, we will discuss the difficulties and increase in circuit depth for all steps in detail. 

\textbf{Encoding and basic circuits.} The first obstacle we face is the fact that the qubit encoding of elements of $S_3$ is not trivial, i.e. some states of the encoding qubits will not be allowed.

We assume the same encoding as for the case of $D_4$, i.e., $\ket{i_1,i_2,i_3} \to \ket{m^{i_1}r^{i_2}r^{2 i_3}}$. However, since $r^3 = e$ in $S_3$ we need to project out the states $\ket{0,1,1}$ and $\ket{1,1,1}$.

\begin{figure}
\begin{gather*}
\Qcircuit @C=0.4em @R=0.7em @!R{
\lstick{m} & \qw & \qw & \qw & \qw & \qw & \qw & \qw & \qw & \qw & \qw  & \ctrl{5} & \qw \\
\lstick{r} & \qw \link{1}{1}& &\link{1}{-1} & \ctrl{5} & \ctrl{5} & \qw    & \qw & \qw\link{1}{1}& &\link{1}{-1} & \qw  & \qw \\
&&&&&&&&&&&&&&\\
\lstick{r^{2}} & \qw \link{-1}{1}& &\link{-1}{-1} & \qw & \qw  & \ctrl{4} & \ctrl{4}  & \qw \link{-1}{1}& &\link{-1}{-1} & \qw  & \qw \\
\\
\lstick{m} & \qw & \ctrl{-3}  & \qw & \qw & \qw  & \qw & \qw & \qw  & \ctrl{-3} & \qw  & \targ & \qw \\
\lstick{r} & \qw & \qw & \qw & \ctrl{1}  & \targ & \targ & \qw & \qw & \qw & \qw & \qw & \qw & \qw \\
\lstick{r^{2}} & \qw & \qw & \qw & \targ  & \qw  & \ctrl{-1} & \targ & \qw & \qw & \qw  & \qw & \qw 
}
\\
\\
\Qcircuit @C=0.4em @R=0.7em @!R{
\lstick{m} & \qw & \ctrl{6} & \qw & \ctrl{4} & \qw & \qw & \qw  & \qw & \qw \\
\lstick{r} & \qw  & \qw & \qw & \qw & \ctrl{4} & \ctrl{4} & \qw & \qw & \qw \\
\lstick{r^{2}} & \qw  & \qw & \qw  & \qw & \qw & \qw & \ctrl{3} & \ctrl{5} & \qw \\
\\
\lstick{m} & \qw & \qw  & \qw & \targ & \qw & \qw & \qw & \qw & \qw \\
\lstick{r} & \qw \link{1}{1}& &\link{1}{-1} & \qw & \ctrl{2} & \targ & \targ & \qw & \qw \\
&&&&&&&&&&&&&&&\\
\lstick{r^{2}} & \qw \link{-1}{1}& &\link{-1}{-1}& \qw & \targ & \qw & \ctrl{-2} & \targ & \qw 
}
\end{gather*}
\caption{The general controlled multiplication circuit for the group $S_3$. The top circuit is the left multiplication, $\ket{g,h} \rightarrow \ket{g, gh}$, and the bottom circuit is the right multiplication, $\ket{g,h} \rightarrow \ket{g, hg}$. The first three qubits encode the first group element and the second three the second group element.}
\label{fig:S3mult}
\end{figure}
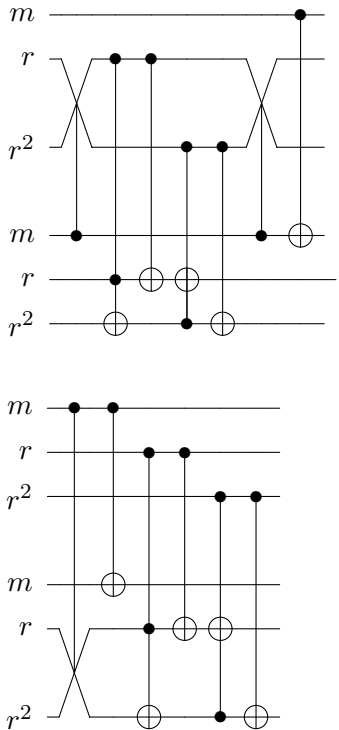

This alone complicates the $\mathbb{Z}_3$ multiplication of $S_3 = \mathbb{Z}_2 \ltimes \mathbb{Z}_3$ significantly. In Figure \ref{fig:S3mult}, we show the general controlled left and right $S_3$-multiplication circuit. For the left multiplication the controlled SWAP gates implement the defining commutation rule between the two generators, $mr = r^2m$, while the middle portion implements the $\mathbb{Z}_3$-multiplication. The circuit contains 8 Toffoli gates, once the SWAPs are decomposed. The circuit for the right multiplication is simpler, containing 6 Toffoli gates.

There are two reasons for this complication. One is the aforementioned unnatural encoding, while the second is the fact that $S_3$ has a trivial center. In the case of $D_4$ the commutation of the two generators amounts to multiplying the expression by $r^2 \in Z(G)$ which is a simple operation.

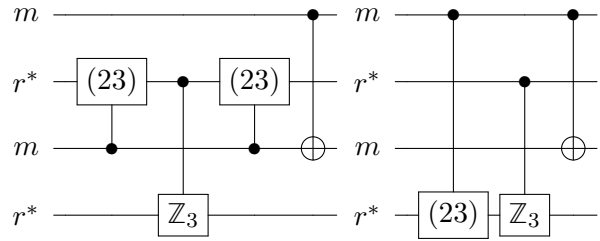
\begin{figure}
\begin{equation*}
\Qcircuit @C=0.4em @R=0.7em @!R{
\lstick{m} & \qw & \qw & \qw & \qw & \ctrl{2}& \qw\\
\lstick{r^*} & \qw & \gate{(23)} & \ctrl{2} & \gate{(23)} & \qw& \qw\\
\lstick{m} & \qw & \ctrl{-1} & \qw & \ctrl{-1}& \targ& \qw\\
\lstick{r^*} & \qw & \qw & \gate{\mathbb{Z}_3}& \qw & \qw& \qw
}
\quad\quad
\Qcircuit @C=0.4em @R=0.7em @!R{
\lstick{m} & \qw & \ctrl{3} & \qw & \ctrl{2} & \qw\\
\lstick{r^*} & \qw & \qw & \ctrl{2} & \qw& \qw\\
\lstick{m} & \qw & \qw & \qw & \targ & \qw\\
\lstick{r^*} & \qw & \gate{(23)} & \gate{\mathbb{Z}_3}& \qw & \qw 
}
\end{equation*}
\caption{The general controlled multiplication circuit for the group $S_3$. The left circuit is the left multiplication, $\ket{g,h} \rightarrow \ket{g, gh}$, and the circuit on the right is the right multiplication, $\ket{g,h} \rightarrow \ket{g, hg}$. The first qubit-qutrit encode the first group element and the second pair the second group element. The gate $C(23)$ is the controlled permutation implementing the commutation relation of the $S_3$ group and the gate $C(\mathbb{Z}_3)$ is the qutrit-qutrit generalisation of the CX gate, i.e. $\mathbb{Z}_3$-addition.}
\label{fig:S3mult_qutrit}
\end{figure}

The situation is simplified significantly if we encode the group elements by a qubit and a qutrit, assuming that we have a full one-qutrit control as one has for qubits in current devices. A possible controlled group multiplication circuit utilising this recourse is shown in Figure~\ref{fig:S3mult_qutrit}.
 
\begin{figure}
\begin{gather*}
\Qcircuit @C=0.4em @R=0.7em @!R{
\lstick{r} & \qw & \targ & \gate{X} & \ctrl{3} & \ctrl{3} & \gate{X} & \ctrl{3} & \ctrl{4} & \targ & \qw \\
\\
\lstick{m} & \qw & \ctrl{-2} & \qw & \qw & \qw & \qw & \qw & \qw & \ctrl{-2} & \qw \\
\lstick{r} & \qw & \qw & \qw & \ctrl{1} & \targ & \qw & \targ & \qw & \qw & \qw \\
\lstick{r^{2}} & \qw & \qw & \qw & \targ & \qw & \qw  & \ctrl{-1} & \targ & \qw & \qw
}
\\
\\
\Qcircuit @C=0.4em @R=0.7em @!R{
\lstick{r}  & \qw & \qw & \qw & \qw & \ctrl{5}  & \ctrl{6} & \qw\link{1}{1}& &\link{1}{-1} & \qw & \qw & \qw \\
&&&&&&&&&&&&&&\\
\lstick{r^{2}} & \qw & \ctrl{3} & \ctrl{3} & \qw & \qw & \qw  & \qw \link{-1}{1}& &\link{-1}{-1} & \qw & \qw & \qw \\
\\
\lstick{m} & \gate{X} & \qw & \qw & \qw & \qw & \qw & \qw & \ctrl{-3} & \qw & \qw & \qw & \qw \\
\lstick{r} & \qw & \ctrl{1} & \targ & \qw & \targ & \qw & \qw & \qw & \qw & \qw & \qw  & \qw \\
\lstick{r^{2}} & \qw & \targ & \qw  & \qw & \ctrl{-1} & \targ & \qw & \qw & \qw & \qw & \qw & \qw  
}
\end{gather*}
\caption{The $S_3$ controlled left multiplication with the domain restricted to a single conjugacy class, $\mathcal{C}_r$ on the top and $\mathcal{C}_m$ on the bottom. Note, that the encoding of $\mathcal{C}_r$ elements is slightly different from the encoding of the general group elements.}
\label{fig:C_S3mult}
\end{figure}
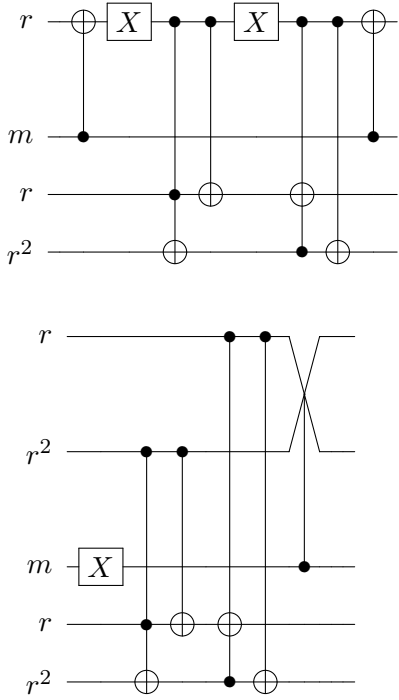

\begin{figure}
\begin{equation*}
\Qcircuit @C=0.4em @R=0.7em @!R{
\lstick{m} & \qw & \ctrl{2} & \qw \\
\\
\lstick{m} & \qw & \targ & \qw \\
\lstick{r} & \qw & \qw & \qw \\
\lstick{r^{2}} & \qw & \qw & \qw 
}\quad\quad
\Qcircuit @C=0.4em @R=0.7em @!R{
\lstick{r}  & \qw & \qw & \qw & \qw & \ctrl{4} & \ctrl{5} & \qw\\
\lstick{r^{2}} & \qw & \ctrl{3} & \ctrl{3} & \qw & \qw & \qw & \qw  \\
\\
\lstick{m} & \qw & \qw & \qw & \qw &  \qw &  \qw &  \qw \\
\lstick{r} & \qw & \ctrl{1} & \targ & \qw & \targ & \qw & \qw\\
\lstick{r^{2}} & \qw & \targ & \qw  & \qw & \ctrl{-1} & \targ & \qw
}\quad\quad
\Qcircuit @C=0.4em @R=0.7em @!R{
\lstick{m} & \qw & \gate{X} & \ctrl{2} & \gate{X} & \qw \\
\lstick{r} & \qw & \qw \link{1}{1}& &\link{1}{-1} & \qw\\
&&&&&\\
\lstick{r^{2}} & \qw & \qw \link{-1}{1}& &\link{-1}{-1} & \qw 
}
\end{equation*}
\caption{(Left) The controlled left multiplication circuit, $\ket{g,h} \rightarrow \ket{g, gh}$, for $g \in H_m$. (Middle) The controlled right multiplication circuit, $\ket{g,h} \rightarrow \ket{g, hg^{-1}}$, for $g \in H_r$. (Right) The inverse circuit, $\ket{g} \rightarrow \ket{g^{-1}}$, needed for reversing arrows.}
\label{fig:H_red_S3}
\end{figure}
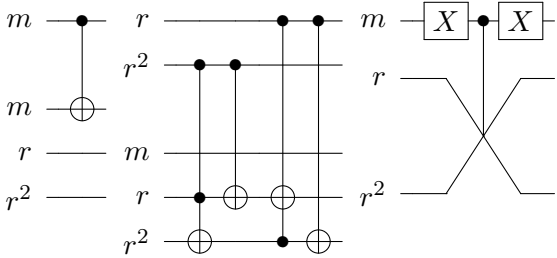

\textbf{Anyon content.} The group $S_3$ has three conjugacy classes, $\mathcal{C}_e = \{e\}$, $\mathcal{C}_r = \{r, r^2\}$ and $\mathcal{C}_m = \{m, mr, mr^2\}$. The two nontrivial conjugacy classes have abelian centres. 
The anyons of this theory alongside their braiding properties have been completely studied in Ref.~\cite{Cui_2015}. In this section, we just note that there are two abelian, four two-dimensional and two three-dimensional anyons.

In our ribbon application protocol we rely on two operations, $\mathcal C$-controlled multiplication and generalised conjugation. When restricted to a conjugacy class the controlled multiplication circuit simplifies significantly for the two-dimensional anyons with $\mathcal{C}_r$ flux. $\mathcal{C}_r$ is naturally encoded by a qubit and the $\mathcal{C}_r$-controlled multiplication circuit has only one Toffoli gate for both left and right multiplication.

For the three dimensional anyons with $\mathcal{C}_m$-flux, we can reduce the number of Toffoli gates to 5 in the case of right and to 2 in the case of left controlled multiplication. The conjugacy class is encoded with two qubits with one disallowed state. Figure \ref{fig:C_S3mult} shows these circuits.

The generalised conjugations are derived from the representation matrices $A{(g)}$ and can be compiled into circuits of similar complexity as the controlled multiplication circuits of restricted domain.

As in the case of the unrestricted controlled multiplication circuit we would see significant reduction in circuit depth, if we had access to well controlled qutrits.

\textbf{Ground state.}
The ground state preparation protocol (without feed-forward) requires setting up the superposition $\sum_{g \in S_3} \ket{g}=\ket{+}\otimes \ket{+_3}$, where $\ket{+_3}=\frac{\ket{00}+\ket{10}+\ket{01}}{{\sqrt{3}} }$. We therefore need a circuit that implements the unitary $U_{+_3}\ket{00}= \ket{+_3}$. This unitary can be chosen to be purely real, i.e., $U_{+_3}\in SO(4)$ such that we can use the minimal gate decomposition from Ref.~\cite{decomp} to write it as $U_{+_3}=M^\dag(A\otimes B)M$, where $M$ is the magic basis rotation that can be implemented with a single CNOT and two single qubit gates  (see Fig.~2 of Ref.~\cite{decomp}). A suitable choice for $A$ and $B$ is $B=\mathbb 1$ and $A=R_z(5\pi/4)R_y(\theta_m) R_z(\pi/4)$, where $R_z$ and $R_y$ are rotations around $z$ and $y$ axis, respectively and $\theta_m=2 \arctan(1/\sqrt{2})$. 

\textbf{Charge measurement.}
Arguably, where we loose out the most is with the partial charge measurement. The group $S_3$ has only few very simple proper subgroups, $H_r = \mathbb{Z}_3^r$, $H_m = \mathbb{Z}_2^m$, $H_{mr^2} = \mathbb{Z}_2^{mr^2}$ and $H_{mr} = \mathbb{Z}_2^{mr}$.

The circuits involved in the partial charge measurement with respect to the subgroups $H_m$ and $H_r$ are shown in Figure~\ref{fig:H_red_S3}. We chose to implement left and right multiplication for the two subgroups respectively because the circuits are significantly shallower. If the orientation of the arrows differs from what we need we can always reverse using the inverse circuit shown in the same figure.

\begin{table}[h]
\centering
\begin{tabular}{|l|lll|}\hline
  $\braket{\chi_{H_r}, \chi}$ & $1$ & $-1$ & $\epsilon$  \\ \hline
$1$ & 3   & 3            & 0                             \\ 
$\omega$ & 0   & 0            & 3                              \\ 
$\bar{\omega}$ & 0   & 0            & 3                          \\ \hline
\end{tabular}

\begin{tabular}{|l|lll|}\hline
  $\braket{\chi_{H_m}, \chi}$ & $1$ & $-1$ & $\epsilon$  \\ \hline
$1$ & 2   & 0            & 2                             \\ 
$-1$ & 0   & 2            & 2                              \\ \hline

\end{tabular}
\caption{Partial orthogonality of the character table for $S_3$ with respect to its two proper subgroups.}
\label{tab:red_ch_S3}
\end{table}

Table \ref{tab:red_ch_S3} shows the partial orthogonality of character tables of $S_3$ and two of its proper subgroups. The fact that given the results of both $H_m$- and $H_r$-partial charge measurements we can uniquely determine the charge still holds for this group as well. 

\textbf{Summary.}
The computational universality of $D(S_3)$ complemented with measurements comes at the expense of significantly longer elementary circuits for our protocols. The elementary circuits are about ten times deeper once compiled into native gates that the Sycamore chip can perform, neglecting the SWAPs needed to place the relevant qubits into appropriate proximity to one another.

This is mainly a consequence of the unnatural way we encode the group elements. Unnatural in the sense that it does not respect the group structure ($\mathbb{Z}_2 \ltimes \mathbb{Z}_3$). A more suitable encoding is achieved by introducing qutrits alongside the full one-qutrit control and ability to condition arbitrary one-qutrit gates. In this encoding all of the elementary circuits become even shallower than in the case of $D_4$. 

However, for any qubit based architecture we reckon that realising $D(S_3)$ is beyond the limits of current NISQ technology.

\section{Conclusions and outlook} \label{sec:outlook}
In this work we explored the feasibility of preparing the ground state of the quantum double model $D(D_4)$, as well as creating, manipulating, and measuring its (non-abelian) anyons on current NISQ technology, in particular Google's Sycamore chip. We demonstrated that by exploiting the structure of the group $D_4$ one can achieve moderate circuit depths for the creation and manipulation of anyons with ribbon operators. We also proposed a partial charge measurement which uniquely determines the anyon content without relying on prohibitively costly full group multiplications. Our numerical results suggest that current NISQ technology is capable of probing the full modular data of $D(D_4)$ on a quasi one-dimensional lattice architecture and that ground state preparation without feed-forward protocols are possible on small two-dimensional lattices.

The qubit layout we used respects the geometry of the physical lattice and can straightforwardly be extended to larger system sizes once more qubits are available. However, with current noise levels for larger two-dimensional lattices, feed-forward protocols would need to be invoked for the ground state preparation.

We note, that in this proposal we did not include error mitigation or noise reducing strategies such as decoupling and expect that applying the latter would further increase the circuit depths for which acceptable signal to noise ratios can be obtained.

We investigated how the circuit simplifications carry over to the group $S_3$ and its double $D(S_3)$ which is more powerful for topological quantum computation in the sense that braiding anyons assisted by measurement allows for universal computations.
We found that for an architecture that features qubits and qutrits, all of our simplifications carry over and the circuit depths become even shorter than the qubit implementation of $D(D_4)$. In contrast, for a qubit-based architecture, the circuit depths we could achieve lie outside of the range currently accessible by roughly one order of magnitude. However, given the steady improvement in quantum technologies, the implementation of $D(S_3)$ might soon be possible and the methods and concrete protocols proposed in this work could be useful for that.

All in all we conclude that our work contributes to the thriving and exciting field of exploring the usability and readiness of current NISQ devices for realising and utilising topological quantum order and anyons. The non-abelian quantum double model $D(D_4)$ presents an important intermediate step between simple abelian systems such as the toric code and more complex ones such as $D(S_3)$ or general string-net models, in particular, the (doubled) Fibonacci model for which braiding anyons alone allows for universal quantum computation. 

\section*{Acknowledgments}
J.J. is partly funded by the Oxford-ShanghaiTech Collaboration Agreement.
C.W. acknowledges support from the European Research Council under the
European Union Horizon 2020 Research and Innovation
Programme via Grant Agreement No. 804213-TMCS and support from the Engineering and Physical Sciences Research Council (EPSRC) via Grant EP/S020527/1. S.S. acknowledges support from the EPSRC via Grants EP/S020527/1 and EP/X030881/1.

Additional data and the code used to run the simulations can be found 
\href{https://drive.google.com/drive/folders/1bPn1ezCVAowkMWX3Pf2Iyu_kSI81uDz5?usp=sharing}{here} and on \href{https://github.com/DaanTimmers/kitaevsim}{GitHub}.

\FloatBarrier
\onecolumn
\appendix

\section{Ribbon types}\label{app:ribs}

In this appendix, we present all variants of the elementary triangles that constitute the ribbon operators introduced in Section \ref{sec:ribbon_ops}. Apart from the main distinction into type I and type II triangles which correspond to controlled group multiplication and controlled generalised conjugation, respectively, there are different variants of the concrete operation which depend on how exactly we couple the ancilla qudit with the gauge field degrees of freedom.

\begin{figure}
\centering
\caption*{$U_{CM}\ket{c', i'}_{a_b}\ket{c, i}_{a_f}\ket{g_i}_{\text{phys}} = $}
\vspace{20pt}
	\begin{tabular}{llll}
 \includegraphics[width=0.12\linewidth]{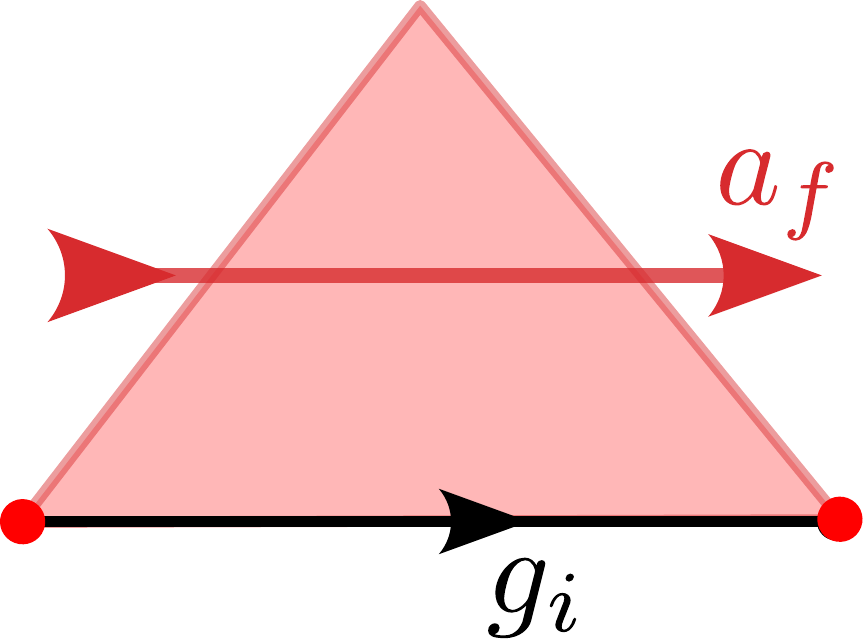} &   $\ket{c', i'}_{a_b}\ket{c, i}_{a_f}\ket{cg_i}_{\text{phys}}$ &  \includegraphics[width=0.12\linewidth]{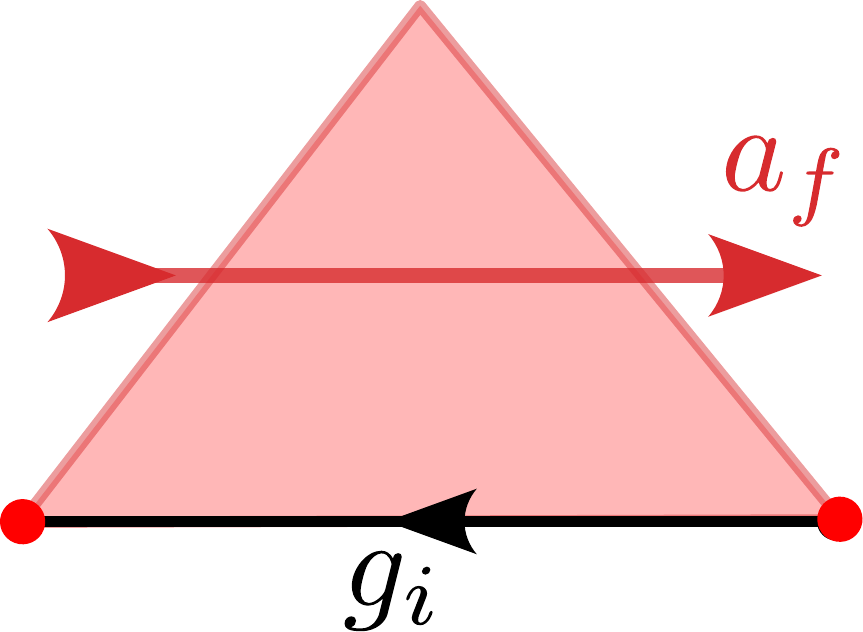}           &       $\ket{c', i'}_{a_b}\ket{c, i}_{a_f}\ket{g_i c^{-1}}_{\text{phys}}$                  \\ 
\includegraphics[width=0.12\linewidth]{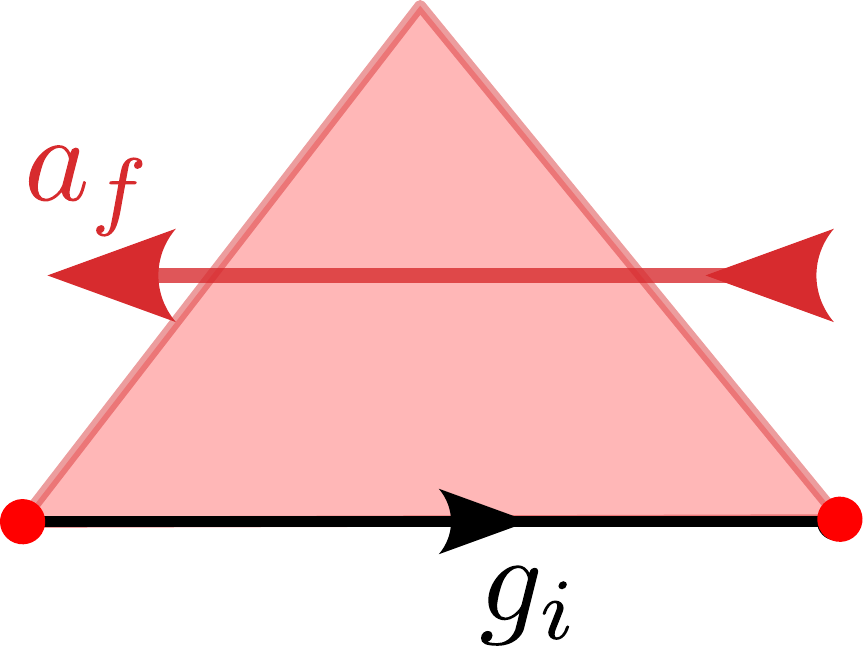} &   $\ket{c', i'}_{a_b}\ket{c, i}_{a_f}\ket{c^{-1}g_i}_{\text{phys}}$ &  \includegraphics[width=0.12\linewidth]{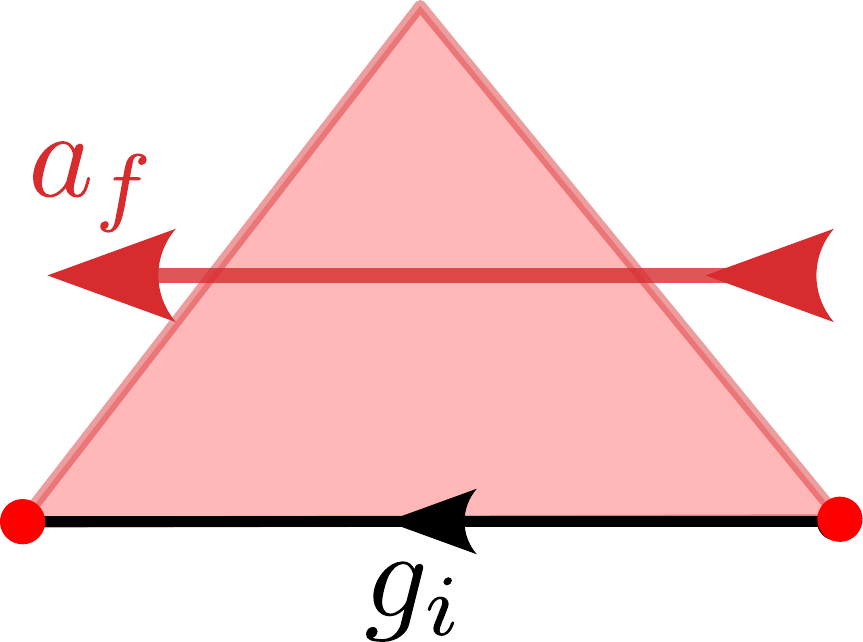}           &       $\ket{c', i'}_{a_b}\ket{c, i}_{a_f}\ket{g_i c}_{\text{phys}}$                  \\  
 \includegraphics[width=0.12\linewidth]{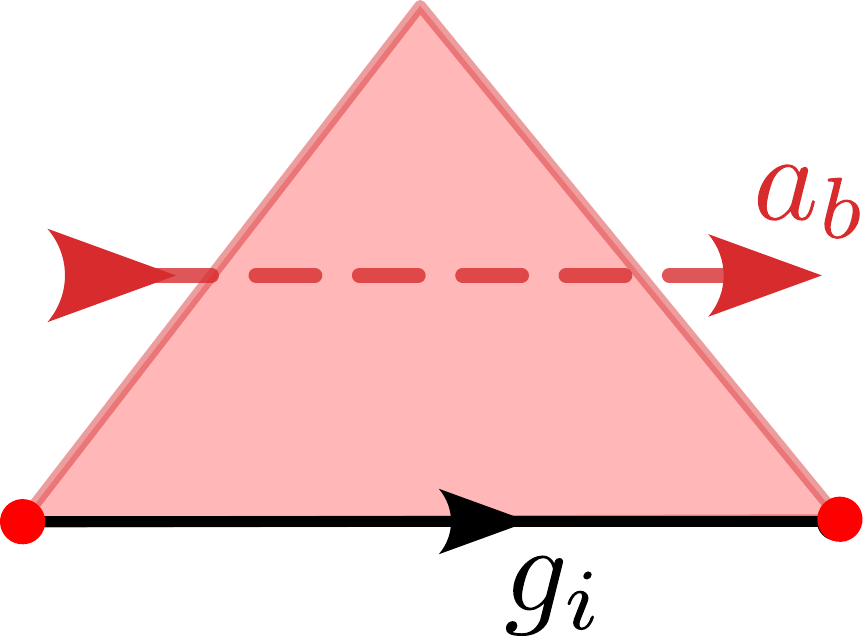} &   $\ket{c', i'}_{a_b}\ket{c, i}_{a_f}\ket{c'g_i}_{\text{phys}}$ &  \includegraphics[width=0.12\linewidth]{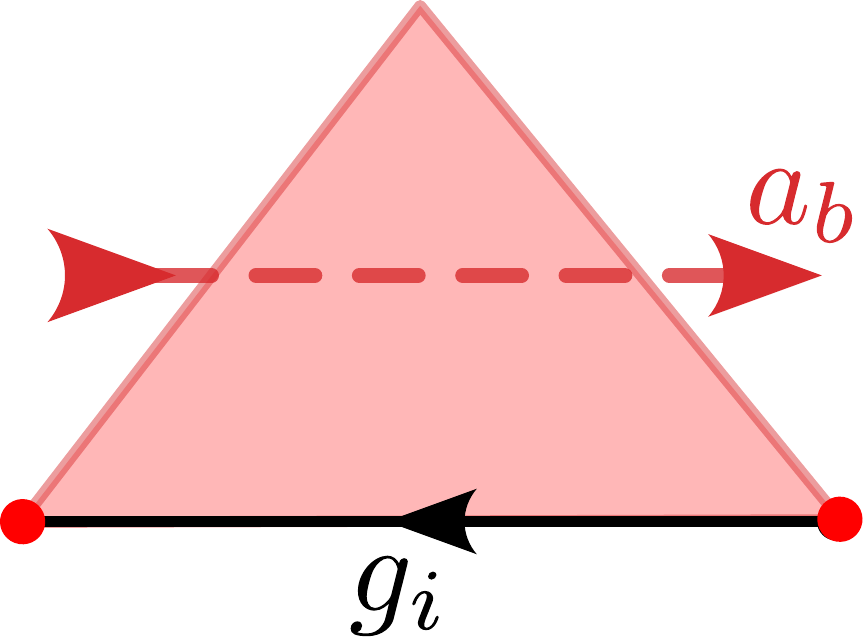}           &       $\ket{c', i'}_{a_b}\ket{c, i}_{a_f}\ket{g_i c'^{-1}}_{\text{phys}}$                  \\ 
\includegraphics[width=0.12\linewidth]{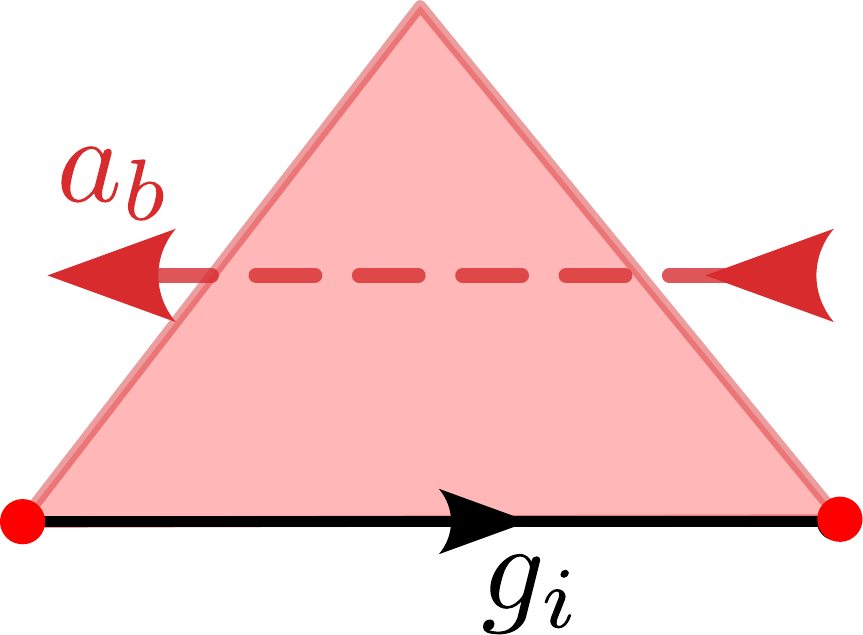} &   $\ket{c', i'}_{a_b}\ket{c, i}_{a_f}\ket{c'^{-1}g_i}_{\text{phys}}$ &  \includegraphics[width=0.12\linewidth]{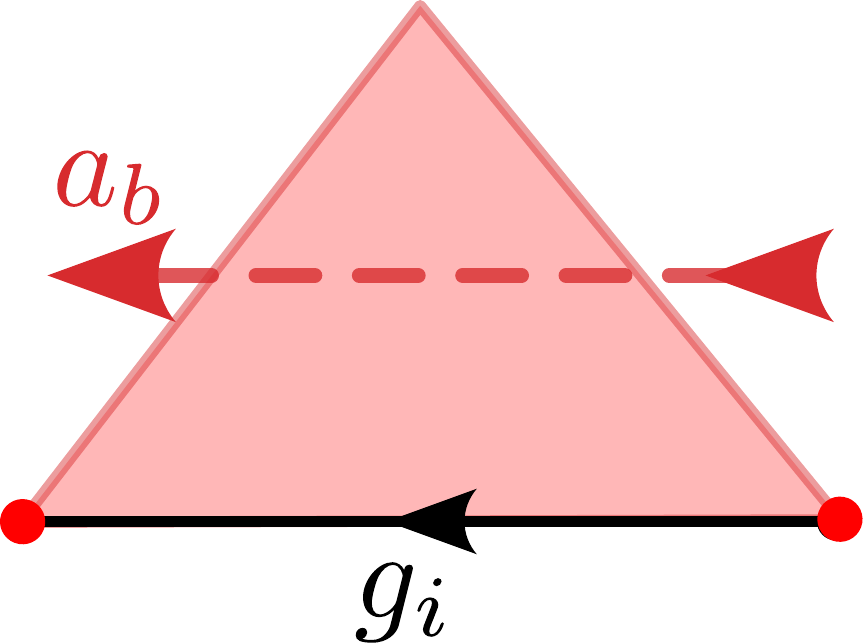}           &       $\ket{c', i'}_{a_b}\ket{c, i}_{a_f}\ket{g_i c'}_{\text{phys}}$                  \\  
\end{tabular}\vspace{60pt}

\caption*{$U_{GC}\ket{\alpha'}_{a_b} \ket{\alpha}_{a_f}\ket{h_i}_{\text{phys}} = $}
\vspace{20pt}
	\begin{tabular}{llll}
 \includegraphics[width=0.12\linewidth]{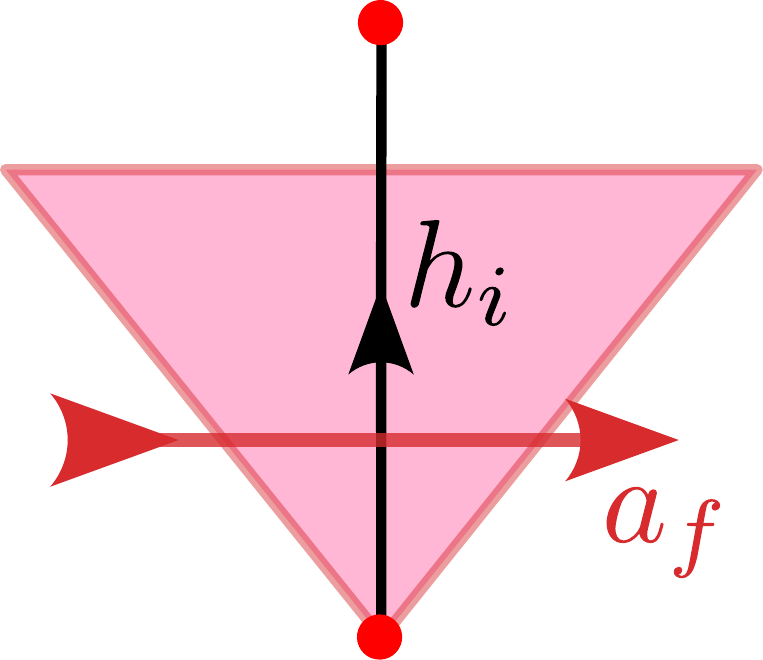} &   $\ket{\alpha'}_{a_b}A^T(h_i)\ket{\alpha}_{a_f}\ket{h_i}_{\text{phys}}$ &  \includegraphics[width=0.12\linewidth]{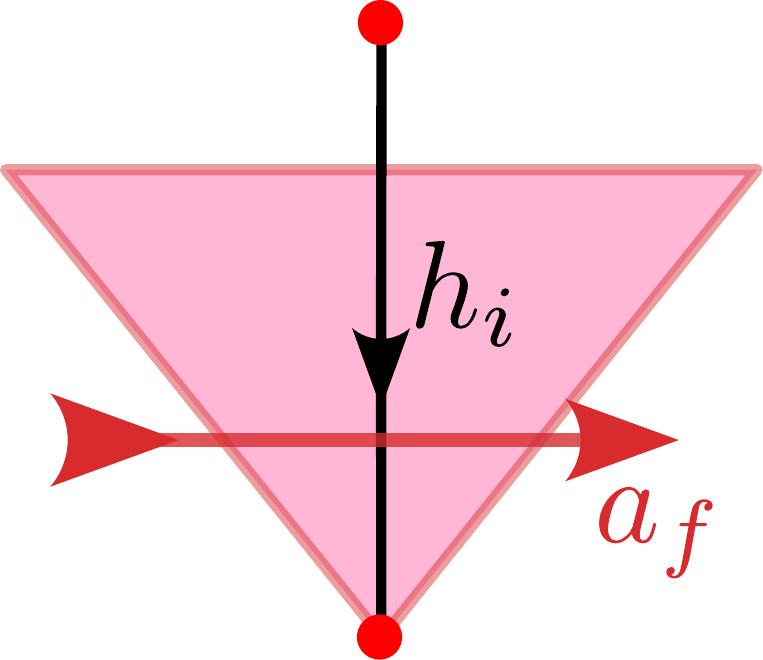}           &       $\ket{\alpha'}_{a_b}A^T(h_i^{-1})\ket{\alpha}_{a_f}\ket{h_i}_{\text{phys}}$                  \\ 
\includegraphics[width=0.12\linewidth]{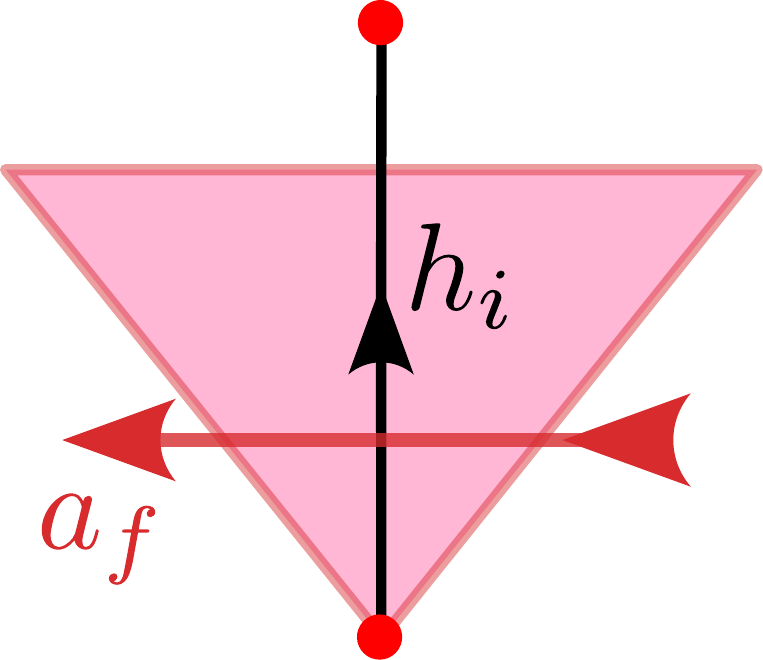} &   $\ket{\alpha'}_{a_b}A^T(h_i^{-1})\ket{\alpha}_{a_f}\ket{h_i}_{\text{phys}}$ &  \includegraphics[width=0.12\linewidth]{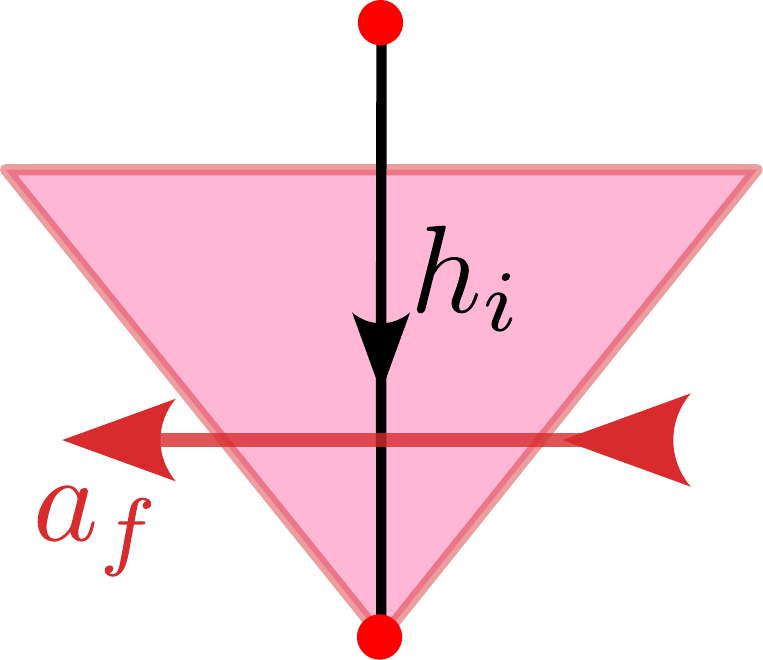}           &       $\ket{\alpha'}_{a_b}A^T(h_i)\ket{\alpha}_{a_f}\ket{h_i}_{\text{phys}}$                  \\  
 \includegraphics[width=0.12\linewidth]{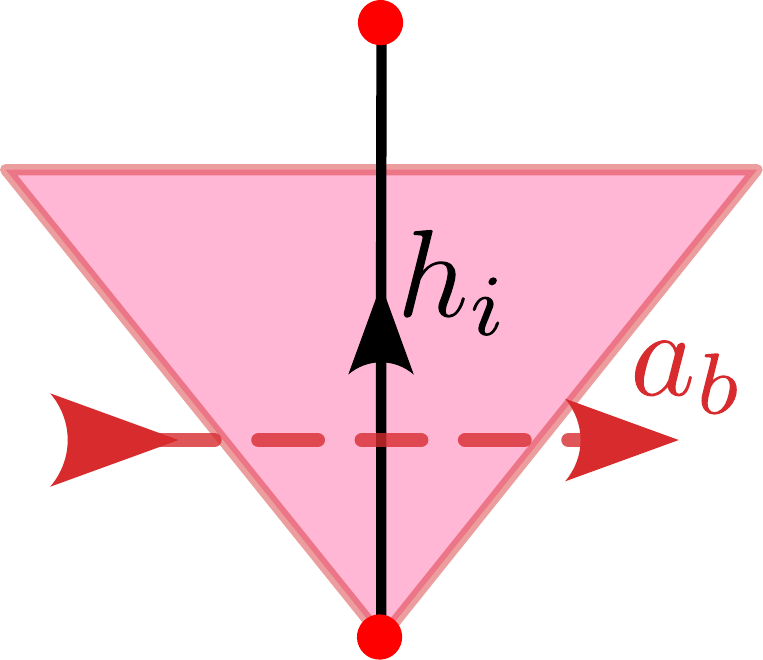} &   $A(h_i)\ket{\alpha'}_{a_b}\ket{\alpha}_{a_f}\ket{h_i}_{\text{phys}}$  &  \includegraphics[width=0.12\linewidth]{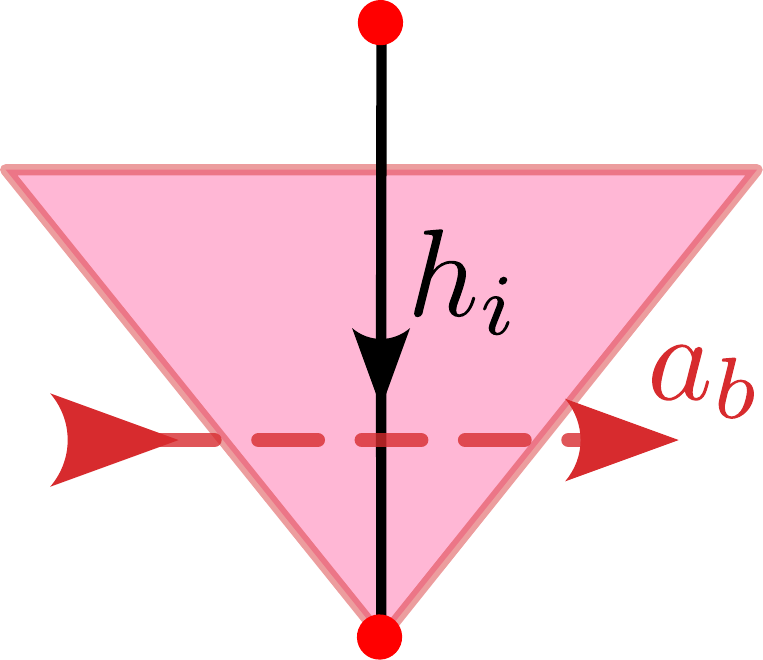}           &       $A(h_i^{-1})\ket{\alpha'}_{a_b}\ket{\alpha}_{a_f}\ket{h_i}_{\text{phys}}$               \\ 
\includegraphics[width=0.12\linewidth]{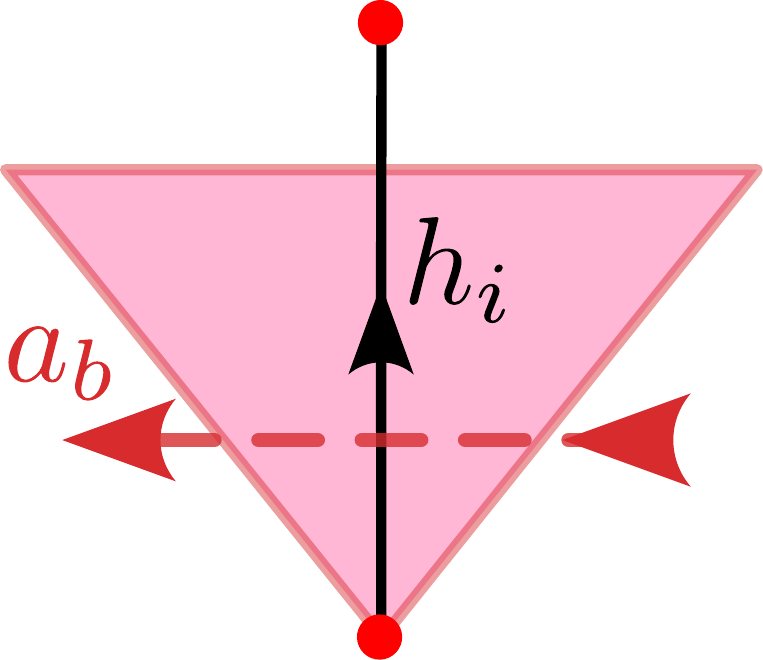} &   $A(h_i^{-1})\ket{\alpha'}_{a_b}\ket{\alpha}_{a_f}\ket{h_i}_{\text{phys}}$ &  \includegraphics[width=0.12\linewidth]{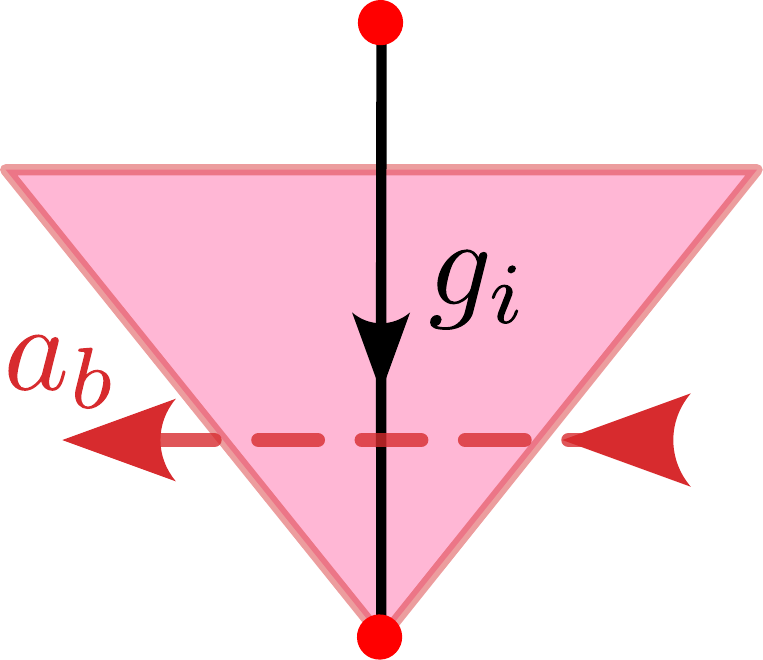}           &       $A(h_i)\ket{\alpha'}_{a_b}\ket{\alpha}_{a_f}\ket{h_i}_{\text{phys}}$                  \\  
\end{tabular}\vspace{20pt}
\caption{The 16 variants of the elementary triangles of type I (top) and type II (bottom).}
\label{fig:al_trigs}
\end{figure}

In the main text we have chose one particular case to keep the description of the algorithm readable. In this section, we provide an exhaustive list of all sixteen variants (8 per type). All triangles can be freely rotated and we chose to rotate them such that the lattice edge of type I triangles is the bottom edge, and such that type II triangles stand on their tip. The triangles are then further distinguished by whether the triangle is appended to the front end or the back end of the ribbon, i.e., involving the ancilla qubit $a_f$ or $a_b$, by the orientation of the lattice edge and by the direction of extension (being aligned or anti-aligned with the lattice edge orientation). All operators are shown explicitly in Figure \ref{fig:al_trigs}.

The controlled group multiplication $U_{CM}$ of type I triangles is given by --  depending on these specifications -- right or left multiplication with the group element $c$ encoded by the forward or backward ancilla $a_f$ or $a_b$ or its inverse $c^{-1}$. 

For type II triangles we need to apply different variants of generalised conjugation $U_{GC}$. In the main text in Section~\ref{sec:ribbon_ops} we defined the generalised conjugation as $U_{GC}^{(C, \chi)}: \ket{c,i} \ket{h}_\text{phys}\rightarrow \ket{h c h^{-1}}\Gamma^\chi_c(h)\ket{i} \ket{h}_\text{phys}$, where $\Gamma^\chi_c(g)$ are matrices defined by the representation of the algebra $(C, \chi)$ spanned by basis vectors $\ket{c,i}$. In this form it is obvious why we call this operation generalised conjugation, however, this operator is more conveniently defined by the $(C, \chi)$-representation matrices $A(h)$ defined in Section \ref{sec:anyon}. To make the connection between the two, we unify the two indices of the $(C, \chi)$-representation into one multi index $\ket{c, i} \equiv \ket{\mu}$ and identify $\ket{c,i} \to \ket{hch^{-1}} \Gamma_c^\chi(h) \ket{i}$ with $\ket{\mu} \to A^{(C,\chi)}(h) \ket{\mu}$. The superscript $(C,\chi)$ is dropped, if no confusion arises. The different type II triangle variants then implement $A(h)$ or $A(h^{-1})$ or their transpose to the ancilla qubits $a_f$ or $a_b$.

\section{Representation theory of $D(D_4)$}\label{app:reps}

In this appendix, we cover the representation theory, viz. the anyon content, of the quantum double algebra of $D_4$. Tab.~\ref{tab:anyons} below shows the naming convention of the anyons and Tab.~\ref{tab:reps} lists the corresponding representation matrices used for the generalised conjugation.

The (abelian) fusion algebra of the excitations of this theory is highly symmetrical and can be broken down into a set of rules listed below. For writing down these rules we introduce the function $\lfloor \rfloor: G \rightarrow \{e, m , r, mr\} \equiv G/Z(G)$ that divides out $r^2$ and use the notation $\Sigma_e \equiv 0$.
\begin{enumerate}
	\item Fusion with the vacuum: $0 \otimes a = a$. 
	\item Fusion with the trivial flux $\tilde{0}$:\begin{enumerate}
		\item $\tilde{0} \otimes \tilde{0} = 0$,
		\item $\tilde{0} \otimes \Sigma_x = \tilde\Sigma_x$, $x \in \{m, r, mr, \epsilon\}$, 
		\item $\tilde{0} \otimes \Psi_x = \Psi_x$ and $\tilde{0} \otimes \Phi_x = \tilde\Phi_x$,  $x \in \{m, r, mr\}$.
	\end{enumerate}
	\item Fusion with abelian charges $\Sigma_x$, $x \in\{m, r, mr\}$:\begin{enumerate}
		\item $\Sigma_x \otimes \Sigma_y = \Sigma_{\lfloor xy \rfloor}$, $ y \in \{m, r, mr\}$, 
		\item $\Sigma_x \otimes \Sigma_\epsilon = \Sigma_\epsilon$,
		\item $\Sigma_x \otimes \Psi_y = \delta_{xy}\Psi_y \oplus(1-\delta_{xy})\tilde\Psi_y$,
		\item $\Sigma_x \otimes \Phi_y = \delta_{xy}\Phi_y \oplus(1-\delta_{xy})\tilde\Phi_y$,	
		\end{enumerate}
	\item Fusion with non-abelian charge $\Sigma_\epsilon$:\begin{enumerate}
		\item $\Sigma_\epsilon \otimes \Sigma_\epsilon = 0 \oplus \Sigma_m \oplus \Sigma_r \oplus \Sigma_{mr}$,
		\item $\Sigma_\epsilon \otimes \Psi_x = \Phi_x \oplus \tilde\Phi_x$ and $\Sigma_\epsilon \otimes \Phi_x = \Psi_x \oplus \tilde\Psi_x$, for $x \in \{m , r, mr\}$. 
	\end{enumerate}
	\item Fusion with dyons of nontrivial flux $x \in \{m, r, mr\}$:\begin{enumerate}
		\item $\Psi_x \otimes \Psi_x = \tilde\Psi_x \otimes \tilde\Psi_x = 0 \oplus \tilde 0 \oplus \Sigma_x \oplus \tilde\Sigma_x$ and $\Psi_x \otimes \tilde\Psi_x = \bigoplus_{y\in \{m, r, mr\}}(1-\delta_{xy})\Sigma_y \oplus \tilde \Sigma_y$,
		\item $\Psi_x \otimes \Phi_x = \Sigma_\epsilon \oplus \tilde\Sigma_\epsilon$,
		\item $\Phi_x \otimes \Phi_x = 0 \oplus \Sigma_x \oplus (\bigoplus_{y \in \{m, r, mr\}}(1-\delta_{xy})\tilde\Sigma_y)$,
		\item $\Psi_x \otimes \Psi_y = \Psi_{\lfloor xy \rfloor} \oplus \tilde\Psi_{\lfloor xy \rfloor}$ and $\Psi_x \otimes \Phi_y = \Phi_{\lfloor xy \rfloor} \oplus \tilde\Phi_{\lfloor xy \rfloor}$, for $x \neq y$ and $y \in \{m, r, mr\}$,
		\item $\Phi_x \otimes \Phi_y  = \Psi_{\lfloor xy \rfloor} \oplus \tilde\Psi_{\lfloor xy \rfloor}$, for $x \neq y$ and $y \in \{m, r, mr\}$.
\end{enumerate}
\end{enumerate}
In the fusion rules stated so far, we have omitted several rules where one or both of the anyons carry a tilde symbol. Note, that while for the pure charges, the tilde symbol is meaningful and distinguishes between the two conjugacy classes defined for the two elements of the centralizer $C_e$ anr $C_{r^2}$, the convention of naming the four dyons for the other conjugacy classes as $\Psi,\tilde \Psi,\Phi,\tilde \Phi$ is somewhat arbitrary (cf. Table \ref{tab:anyons}). Nevertheless, the additional fusion rules involving the tilde anyons can be derived from the ones above by (multiple) application of a simple rule. To state this rule, we note that for any anyon $X$ the anyon $\tilde X$ can be read off from Table \ref{tab:anyons}. For convenience, we also define $\tilde{\tilde X}=X$. The rule can then be given as follows. Pick one anyon $X$ on the LHS ($\otimes$-side) and replace $X$ by $\tilde X$. For all elements $X_i$ in the direct sum $\oplus_i X_i$ on the RHS, replace $X_i$ by $\tilde X_i$.	
 
Note that this prescription is only suitable to derive the rules not explicitly stated above. In particular, it is \emph{not} a symmetry (or a $\mathbb{Z}_2$-grading) of the algebra and is not true for pairs involving $\Psi_x$ and $\tilde\Psi_x$, for $x \in \{m, r, mr\}$, as can be seen from 5(a) above.

\begin{table}
\centering
\begin{subtable}{\textwidth}
\raggedright
\begin{tabular}{l||lllll|lllll}

$D(D_4)$  & $O$     & $\Sigma_{r}$ & $\Sigma_{mr}$ & $\Sigma_{m}$ & $\Sigma_{\epsilon}$ & $\tilde{O}$ & $\tilde{\Sigma}_{r}$ & $\tilde{\Sigma}_{mr}$ & $\tilde{\Sigma}_{m}$ & $\tilde{\Sigma}_{\epsilon}$ 
\\ \hline
$\mathcal C$    & $\mathcal C_e$ &       &       &       &              &   $\mathcal C_{r^2}$     &           &       &           &                \\ $\chi$ & $1$ & $\alpha_{r}$ & $\alpha_{mr}$ & $\alpha_{m}$ & $\epsilon$          & $1$     & $\alpha_{r}$         & $\alpha_{mr}$         & $\alpha_{m}$         & $\epsilon$                  \\ 
\hline 
T & 1 & 1 & 1 & 1 & 1 & 1 & 1 & 1 & 1 & -1\\
dim $d$ & 1 & 1 & 1 & 1 & 2 & 1 & 1 & 1 & 1& 2
\end{tabular}
\end{subtable}

\vspace{0.5cm}
\medskip
\begin{subtable}{\textwidth}
\begin{tabular}{l||llll|llll}
$D(D_4)$  & $\Psi_{m}$ & $\tilde{\Psi}_{m}$ & $\Phi_{m}$ & $\tilde{\Phi}_{m}$ & $\Psi_{mr}$ & $\tilde{\Psi}_{mr}$ & $\Phi_{mr}$ & $\tilde{\Phi}_{mr}$ \\ \hline
$\mathcal C$    & $\mathcal C_m$    &          &    &            & $\mathcal C_{mr}$    &           &   &           \\ 
$\chi$ & $(1,1)$    & $(1,-1)$              & $(-1,1)$    & $(-1,-1)$              &$(1,1)$    & $(1,-1)$              & $(-1,1)$    & $(-1,-1)$             \\ \hline
T  & 1 & -1 & 1 & -1 & 1 & -1 & 1 & -1 \\
dim $d$ & 2 & 2 & 2 & 2 & 2 & 2 & 2 & 2 
\end{tabular}
\end{subtable}
\medskip

\vspace{0.5cm}
\begin{subtable}{\textwidth}
\raggedright
\begin{tabular}{l||rrrr}
$D(D_4)$  & $\Psi_{r}$ &  $\Phi_{r}$ & $\tilde{\Psi}_{r}$ & $\tilde{\Phi}_{r}$ \\ \hline
$\mathcal C$    & $\mathcal C_r$    &            &    &             \\ 
$\chi$ & $1$    & $ i$              & $-1$     & $- i$   \\           
\hline
T  & 1 & $i$ & $-1$ & $-i$ \\
dim $d$ & 2 & 2 & 2 & 2
\end{tabular}
\end{subtable} 
\caption{  Anyon content of $D(D_4)$ defined by flux-charge pairs $(\mathcal C,\chi)$, listing the topological spin as given by the (diagonal) $T$-matrix entry and the quantum dimension $d$. }\label{tab:anyons}
\end{table}

\begin{table}[]
    \centering
    \begin{tabular}{|l||l|l|l|l|l|l|l|l|}
    \hline
  $A(g)$ & $e$ &   $r$ & $r^2$ & $r^3$ &  $m$ & $mr$ & $mr^2$ & $mr^3$ \\
\hline\hline
  $O$ & 1 &   1 &  1 &   1 &  1 &  1 &   1 &    1 \\\hline
 $\tilde{O}$ & 1 &   1 &  1 &   1 &  1 &  1 &   1 &    1 \\\hline
  $\Sigma_{r}$ & 1 &   1 &  1 &   1 & -1 & -1 &  -1 &   -1 \\\hline
 $\Sigma_{mr}$ & 1 &  -1 &  1 &  -1 & -1 &  1 &  -1 &    1 \\\hline
  $\Sigma_{m}$ & 1 &  -1 &  1 &  -1 &  1 & -1 &   1 &   -1 \\\hline
 $\tilde{\Sigma}_{r}$ & 1 &   1 &  1 &   1 & -1 & -1 &  -1 &   -1 \\\hline
$\tilde{\Sigma}_{mr}$ & 1 &  -1 &  1 &  -1 & -1 &  1 &  -1 &    1 \\\hline
 $\tilde{\Sigma}_{m}$ & 1 &  -1 &  1 &  -1 &  1 & -1 &   1 &   -1 \\\hline
  $\Sigma_{\epsilon}$ &$\mathbb{1}$&$-i\sigma_y$ &$-\mathbb{1}$& $i\sigma_y$ & $\sigma_z$ &$-\sigma_x$ & $-\sigma_z$ &   $\sigma_x$ \\\hline
 $\tilde{\Sigma}_{\epsilon}$ &$\mathbb{1}$&$-i\sigma_y$ &$-\mathbb{1}$& $i\sigma_y$ & $\sigma_z$ &$-\sigma_x$ & $-\sigma_z$ &   $\sigma_x$ \\\hline
$\Psi_{r}$ &$\mathbb{1}$&  $\mathbb{1}$& $\mathbb{1}$&  $\mathbb{1}$& $\sigma_x$ & $\sigma_x$ &  $\sigma_x$ &   $\sigma_x$ \\\hline
$\Phi_{r}$ &$\mathbb{1}$& $i\sigma_z$ &$-\mathbb{1}$&$-i\sigma_z$ & $\sigma_x$ & $\sigma_y$ & $-\sigma_x$ &  $-\sigma_y$ \\\hline
$\tilde{\Psi}_{r}$ &$\mathbb{1}$& $-\mathbb{1}$& $\mathbb{1}$& $-\mathbb{1}$& $\sigma_x$ &$-\sigma_x$ &  $\sigma_x$ &  $-\sigma_x$ \\\hline
$\tilde{\Phi}_{r}$ &$\mathbb{1}$&$-i\sigma_z$ &$-\mathbb{1}$& $i\sigma_z$ & $\sigma_x$ &$-\sigma_y$ & $-\sigma_x$ &   $\sigma_y$ \\\hline
 $\Psi_{m}$ &$\mathbb{1}$&  $\sigma_x$ & $\mathbb{1}$&  $\sigma_x$ & $\mathbb{1}$& $\sigma_x$ &  $\mathbb{1}$&   $\sigma_x$ \\\hline
 $\tilde{\Psi}_{m}$ &$\mathbb{1}$&  $\sigma_x$ & $\mathbb{1}$&  $\sigma_x$ &$-\mathbb{1}$&$-\sigma_x$ & $-\mathbb{1}$&  $-\sigma_x$ \\\hline
 $\Phi_{m}$ &$\mathbb{1}$&$i\sigma_y$ &$-\mathbb{1}$& $-i\sigma_y$ & $\sigma_z$ & $\sigma_x$ & $-\sigma_z$ &  $-\sigma_x$ \\\hline
 $\tilde{\Phi}_{m}$ &$\mathbb{1}$&$i\sigma_y$ &$-\mathbb{1}$& $-i\sigma_y$ &$-\sigma_z$ &$-\sigma_x$ &  $\sigma_z$ &   $\sigma_x$ \\\hline
$\Psi_{mr}$ &$\mathbb{1}$&  $\sigma_x$ & $\mathbb{1}$&  $\sigma_x$ & $\sigma_x$ & $\mathbb{1}$&  $\sigma_x$ &   $\mathbb{1}$\\\hline
$\tilde{\Psi}_{mr}$ &$\mathbb{1}$&  $\sigma_x$ & $\mathbb{1}$&  $\sigma_x$ &$-\sigma_x$ &$-\mathbb{1}$& $-\sigma_x$ &  $-\mathbb{1}$\\\hline
$\Phi_{mr}$ &$\mathbb{1}$&$i\sigma_y$ &$-\mathbb{1}$& $-i\sigma_y$ &$-\sigma_x$ & $\sigma_z$ &  $\sigma_x$ &  $-\sigma_z$ \\\hline
$\tilde{\Phi}_{mr}$ &$\mathbb{1}$&$i\sigma_y$ &$-\mathbb{1}$& $-i\sigma_y$ & $\sigma_x$ &$-\sigma_z$ & $-\sigma_x$ &   $\sigma_z$ \\\hline
\end{tabular}
    \caption{$A_g$ matrices for every representation of $D(D_4)$}
    \label{tab:reps}
\end{table}

\section{Elementary circuits for the case of $D(D_4)$}\label{app:cirqs}

In Appendix~\ref{app:ribs}, we have presented all the elementary operations associated with the application of the ribbon operators and in Section~\ref{sec:qm_double}, the other elementary operations used for charge measurements and ground state preparation.
In Section~\ref{sec:D4_double} we have spelled out some of the above mentioned operations for the actual group element-to-qubit encoding used in our simulations of the protocols. In this Appendix, we present the concrete circuit elements for all relevant operations.

\textbf{Controlled multiplication.} 
There are four kinds of controlled multiplications that appear in all of our elementary protocols
\begin{equation}\label{eq:circs}
U_{CM}^{(1)}: \ket{g,h} \rightarrow \ket{g, gh}, \quad
U_{CM}^{(2)}: \ket{g,h} \rightarrow \ket{g, hg}, \quad
U_{CM}^{(3)}: \ket{g,h} \rightarrow \ket{g, g^{-1}h}, \quad
U_{CM}^{(4)}: \ket{g,h} \rightarrow \ket{g, hg^{-1}}.	
\end{equation}

\begin{figure}
\begin{gather*}
\bold{G}:\qquad
\Qcircuit @C=0.4em @R=0.7em @!R{
\lstick{m} & \qw & \ctrl{3} & \qw & \qw & \qw & \qw\\
\lstick{r} & \ctrl{2} & \qw & \ctrl{3} & \ctrl{3} & \qw & \qw\\
\lstick{r^{2}} & \qw  & \qw & \qw & \qw & \ctrl{3} & \qw\\
\lstick{\quad m} &  \ctrl{2} & \targ & \qw & \qw & \qw & \qw\\
\lstick{r} & \qw & \qw & \ctrl{1} & \targ & \qw & \qw\\
\lstick{r^{2}} & \targ & \qw & \targ & \qw & \targ & \qw\\
\qquad\qquad\qquad\quad U_{CM}^{(1)}
}\qquad\qquad
\Qcircuit @C=0.4em @R=0.7em @!R{
\lstick{m} & \ctrl{4} & \ctrl{3} & \qw & \qw & \qw & \qw\\
\lstick{r} & \qw & \qw & \ctrl{3} & \ctrl{3} & \qw & \qw\\
\lstick{r^{2}} & \qw  & \qw & \qw & \qw & \ctrl{3} & \qw\\
\lstick{\quad m} &  \qw & \targ & \qw & \qw & \qw & \qw\\
\lstick{r} & \ctrl{1} & \qw & \ctrl{1} & \targ & \qw & \qw\\
\lstick{r^{2}} & \targ & \qw & \targ & \qw & \targ & \qw\\
\qquad\qquad\qquad\quad U_{CM}^{(2)}
}\qquad\qquad
\Qcircuit @C=0.4em @R=0.7em @!R{
\lstick{m} & \qw & \ctrl{3} & \qw & \qw & \targ & \ctrl{1} & \targ & \qw \\
\lstick{r} & \ctrl{2} & \qw & \ctrl{3} & \ctrl{3}  & \qw & \ctrl{4} & \qw & \qw \\
\lstick{r^{2}} & \qw  & \qw & \qw & \qw & \ctrl{3}  & \qw & \qw & \qw \\
\lstick{\quad m} &  \ctrl{2} & \targ & \qw & \qw & \qw & \qw & \qw & \qw \\
\lstick{r} & \qw & \qw & \ctrl{1} & \targ & \qw & \qw & \qw & \qw \\
\lstick{r^{2}} & \targ & \qw & \targ & \qw & \targ  & \targ & \qw & \qw\\
\qquad\qquad\qquad\qquad U_{CM}^{(3)}
}\qquad\qquad
\Qcircuit @C=0.4em @R=0.7em @!R{
\lstick{m} & \qw & \ctrl{3} & \qw & \qw & \targ & \ctrl{1} & \targ & \qw \\
\lstick{r} & \ctrl{2} & \qw & \ctrl{3} & \ctrl{3}  & \qw & \ctrl{4} & \qw & \qw \\
\lstick{r^{2}} & \qw  & \qw & \qw & \qw & \ctrl{3}  & \qw & \qw & \qw \\
\lstick{\quad m} &  \ctrl{2} & \targ & \qw & \qw & \qw & \qw & \qw & \qw \\
\lstick{r} & \qw & \qw & \ctrl{1} & \targ & \qw & \qw & \qw & \qw \\
\lstick{r^{2}} & \targ & \qw & \targ & \qw & \targ  & \targ & \qw & \qw\\
\qquad\qquad\qquad\qquad U_{CM}^{(4)}
}\\
\\
\\ \bold{H_m}:\qquad
\Qcircuit @C=0.4em @R=0.7em @!R{
\lstick{m} & \ctrl{2} &  \qw & \qw\\
\lstick{r^{2}} & \qw  & \ctrl{3} & \qw\\
\lstick{\quad m} &  \targ & \qw & \qw \\
\lstick{r} & \qw & \qw & \qw\\
\lstick{r^{2}} & \qw & \targ & \qw\\
\qquad U_{CM}^{(1)} = U_{CM}^{(3)}
}\qquad\qquad
\Qcircuit @C=0.4em @R=0.7em @!R{
\lstick{m} & \ctrl{2} &  \qw & \qw & \ctrl{3} & \qw\\
\lstick{r^{2}} & \qw  & \ctrl{3} & \qw & \qw & \qw\\
\lstick{\quad m} &  \targ & \qw & \qw & \qw & \qw\\
\lstick{r} & \qw & \qw & \qw & \ctrl{1} & \qw\\
\lstick{r^{2}} & \qw & \targ & \qw & \targ & \qw\\
\qquad\qquad U_{CM}^{(2)} = U_{CM}^{(4)}
}\qquad\bold{H_{mr}}:\qquad\quad
\Qcircuit @C=0.4em @R=0.7em @!R{
\lstick{mr} & \ctrl{2} &  \ctrl{2} & \ctrl{3} & \ctrl{3} & \qw & \qw\\
\lstick{r^{2}} & \qw  & \qw & \qw & \qw & \ctrl{3} 7 \qw\\
\lstick{\quad m} &  \ctrl{2} & \targ & \qw & \qw & \qw & \qw\\
\lstick{r} & \qw & \qw & \ctrl{1} & \targ & \qw & \qw\\
\lstick{r^{2}} & \targ & \qw & \targ & \qw & \targ & \qw\\
\qquad\qquad\qquad U_{CM}^{(1)} = U_{CM}^{(3)}
}\qquad\qquad
\Qcircuit @C=0.4em @R=0.7em @!R{
\lstick{mr} & \ctrl{2} &  \ctrl{3} & \qw & \qw\\
\lstick{r^{2}} & \qw  & \qw & \ctrl{3} & \qw\\
\lstick{\quad m} &  \targ & \qw & \qw & \qw \\
\lstick{r} & \qw & \targ & \qw & \qw\\
\lstick{r^{2}} & \qw & \qw & \targ & \qw\\
\qquad\quad U_{CM}^{(2)} = U_{CM}^{(4)}
}\\
\\
\\ \bold{H_r}:\qquad
\Qcircuit @C=0.4em @R=0.7em @!R{
\lstick{r} & \ctrl{2}  & \ctrl{3} & \ctrl{3} & \qw & \qw\\
\lstick{r^{2}} & \qw  & \qw & \qw & \ctrl{3} & \qw\\
\lstick{\quad m} &  \ctrl{2} & \qw & \qw & \qw & \qw\\
\lstick{r} & \qw & \ctrl{1} & \targ & \qw & \qw\\
\lstick{r^{2}} & \targ & \targ & \qw & \targ & \qw\\
\qquad\qquad U_{CM}^{(1)}
}\qquad\qquad
\Qcircuit @C=0.4em @R=0.7em @!R{
\lstick{r} & \ctrl{3} &  \ctrl{3} & \qw & \qw\\
\lstick{r^{2}} & \qw  & \qw & \ctrl{3} & \qw\\
\lstick{\quad m} &  \qw & \qw & \qw & \qw \\
\lstick{r} & \ctrl{1} & \targ & \qw & \qw\\
\lstick{r^{2}} & \targ & \qw & \targ & \qw\\
\qquad\qquad U_{CM}^{(2)}
}\qquad\qquad
\Qcircuit @C=0.4em @R=0.7em @!R{
\lstick{r} & \ctrl{2}  & \ctrl{3} & \ctrl{3} & \qw & \ctrl{4} & \qw\\
\lstick{r^{2}} & \qw  & \qw & \qw & \ctrl{3} & \qw & \qw\\
\lstick{\quad m} &  \ctrl{2} & \qw & \qw & \qw & \qw & \qw\\
\lstick{r} & \qw & \ctrl{1} & \targ & \qw & \qw & \qw\\
\lstick{r^{2}} & \targ & \targ & \qw & \targ & \targ & \qw\\
\qquad\qquad\qquad\quad U_{CM}^{(3)}
}\qquad\qquad
\Qcircuit @C=0.4em @R=0.7em @!R{
\lstick{r} & \ctrl{3} &  \ctrl{3} & \qw & \ctrl{4} & \qw\\
\lstick{r^{2}} & \qw  & \qw & \ctrl{3} & \qw & \qw
\\
\lstick{\quad m} &  \qw & \qw & \qw & \qw & \qw\\
\lstick{r} & \ctrl{1} & \targ & \qw & \qw & \qw\\
\lstick{r^{2}} & \targ & \qw & \targ & \targ & \qw\\
\qquad\qquad\quad U_{CM}^{(4)}
}\\
\\
\\ \bold{C_m}:\qquad
\Qcircuit @C=0.4em @R=0.7em @!R{
\lstick{a} & \qw  & \ctrl{3} & \qw\\
\lstick{\quad m} &  \targ & \qw & \qw \\
\lstick{r} & \qw & \qw & \qw\\
\lstick{r^{2}} & \qw & \targ & \qw\\
\qquad\quad U_{CM}^{(1)} = U_{CM}^{(3)}
}\qquad\qquad
\Qcircuit @C=0.4em @R=0.7em @!R{
\lstick{a} & \qw  & \ctrl{3} & \qw & \qw & \qw\\
\lstick{\quad m} &  \targ & \qw & \qw & \qw & \qw\\
\lstick{r} & \qw & \qw & \qw & \ctrl{1} & \qw\\
\lstick{r^{2}} & \qw & \targ & \qw & \targ & \qw\\
\qquad \qquad U_{CM}^{(2)} = U_{CM}^{(4)} 
}\qquad \bold{C_{mr}}: \qquad
\Qcircuit @C=0.4em @R=0.7em @!R{
\lstick{a} & \qw  & \qw & \qw & \qw & \ctrl{3} 7 \qw\\
\lstick{\quad m} &  \ctrl{2} & \targ & \qw & \qw & \qw & \qw\\
\lstick{r} & \qw & \qw & \ctrl{1} & \targ & \qw & \qw\\
\lstick{r^{2}} & \targ & \qw & \targ & \qw & \targ & \qw\\
\qquad\qquad\qquad U_{CM}^{(1)} = U_{CM}^{(3)}
}\qquad\qquad
\Qcircuit @C=0.4em @R=0.7em @!R{
\lstick{a} & \qw  & \qw & \ctrl{3} & \qw\\
\lstick{\quad m} &  \targ & \qw & \qw & \qw \\
\lstick{r} & \qw & \targ & \qw & \qw\\
\lstick{r^{2}} & \qw & \qw & \targ & \qw\\
\qquad \qquad U_{CM}^{(2)} = U_{CM}^{(4)} 
}\\
\\
\\ \bold{C_r}:\qquad
\Qcircuit @C=0.4em @R=0.7em @!R{
\lstick{a} & \qw  & \qw & \qw & \ctrl{3} & \qw\\
\lstick{\quad m} &  \ctrl{2} & \qw & \qw & \qw & \qw\\
\lstick{r} & \qw & \ctrl{1} & \targ & \qw & \qw\\
\lstick{r^{2}} & \targ & \targ & \qw & \targ & \qw\\
\qquad\qquad\qquad U_{CM}^{(1)}
}\qquad\qquad
\Qcircuit @C=0.4em @R=0.7em @!R{
\lstick{a} & \qw  & \qw & \ctrl{3} & \qw\\
\lstick{\quad m} &  \qw & \qw & \qw & \qw \\
\lstick{r} & \ctrl{1} & \targ & \qw & \qw\\
\lstick{r^{2}} & \targ & \qw & \targ & \qw\\
\qquad\qquad U_{CM}^{(2)}
}\qquad\qquad
\Qcircuit @C=0.4em @R=0.7em @!R{
\lstick{a} & \qw  & \qw & \qw & \ctrl{3} & \qw & \qw\\
\lstick{\quad m} &  \ctrl{2} & \qw & \qw & \qw & \qw & \qw\\
\lstick{r} & \qw & \ctrl{1} & \targ & \qw & \qw & \qw\\
\lstick{r^{2}} & \targ & \targ & \qw & \targ & \targ & \qw\\
\qquad\qquad\qquad U_{CM}^{(3)}
}\qquad\qquad
\Qcircuit @C=0.4em @R=0.7em @!R{
\lstick{a} & \qw  & \qw & \ctrl{3} & \qw & \qw\\
\lstick{\quad m} &  \qw & \qw & \qw & \qw & \qw\\
\lstick{r} & \ctrl{1} & \targ & \qw & \qw & \qw\\
\lstick{r^{2}} & \targ & \qw & \targ & \targ & \qw\\
\qquad\qquad\qquad U_{CM}^{(4)}
}
\end{gather*}

    \caption{The circuits implementing the controlled group multiplications $U_{CM}$ defined in Eq.~\eqref{eq:circs}
for $g \in G$ (unrestricted), $g$ restricted to subgroups $H$ and $g$ restricted to conjugacy classes $C$.}
    \label{fig:CM}
\end{figure}
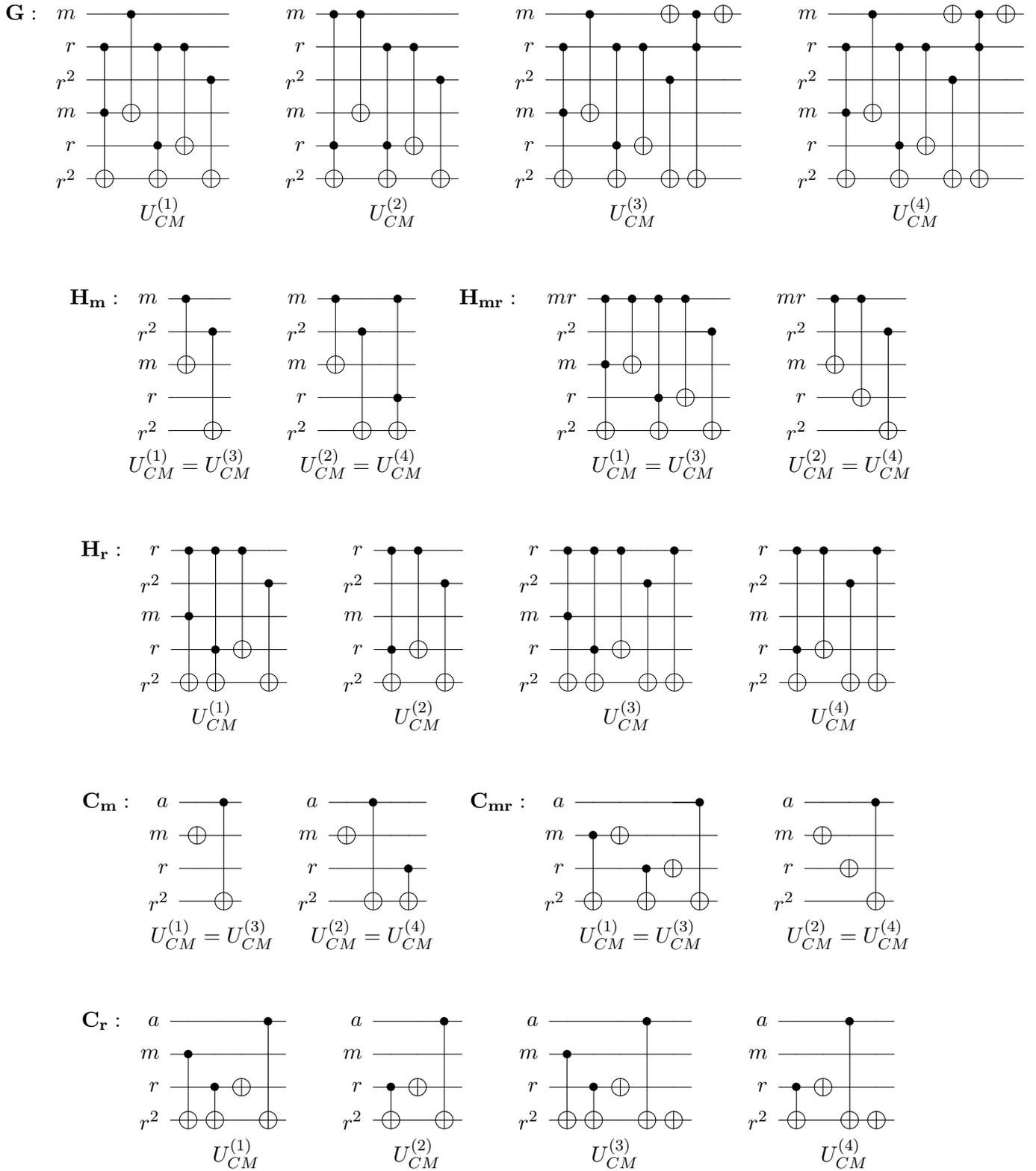

Depending on the context in which it appears, i.e., weather the controlled multiplication is a part of the ground state preparation, the partial charge measurement or  the ribbon operator application, the controlling group element $g$ is unrestricted $g \in G$ or is restricted to one of the subgroups $\{H_m, H_r, H_{mr}\}$ or one of the conjugacy classes $\{C_m, C_r, C_{mr}\}$. As mentioned in the main text, the circuits for the latter two cases are drastically simplified compared to the unrestricted case. The circuits for all cases above are shown in Figure~\ref{fig:CM}.

\begin{figure}
\begin{gather*}
\Qcircuit @C=0.4em @R=0.7em @!R{
\lstick{\quad m} &  \ctrl{3} & \qw & \qw & \qw\\
\lstick{r} & \qw & \ctrl{2} & \qw & \qw\\
\lstick{r^{2}} & \qw & \qw & \ctrl{1} & \qw \\
\lstick{a_i} & \gate{A^T(m)} & \gate{A^T(r)} & \gate{A^T(r^2)}  & \qw
}\qquad\qquad
\Qcircuit @C=0.4em @R=0.7em @!R{
\lstick{\quad m} &  \qw & \qw & \ctrl{3} & \qw\\
\lstick{r} & \qw & \ctrl{2} & \qw & \qw\\
\lstick{r^{2}} & \ctrl{1} & \qw & \qw & \qw \\
\lstick{a_i} & \gate{A^*(r^2)} & \gate{A^*(r)} & \gate{A^*(m)}  & \qw
}
\\
\\
\Qcircuit @C=0.4em @R=0.7em @!R{
\lstick{\quad m} &  \ctrl{3} & \qw & \qw & \qw\\
\lstick{r} & \qw & \ctrl{2} & \qw & \qw\\
\lstick{r^{2}} & \qw & \qw & \ctrl{1} & \qw \\
\lstick{a_i} & \gate{A^\dag(m)} & \gate{A^\dag(r)} & \gate{A^\dag(r^2)}  & \qw
}\qquad\qquad
\Qcircuit @C=0.4em @R=0.7em @!R{
\lstick{\quad m} &  \qw & \qw & \ctrl{3} & \qw\\
\lstick{r} & \qw & \ctrl{2} & \qw & \qw\\
\lstick{r^{2}} & \ctrl{1} & \qw & \qw & \qw \\
\lstick{a_i} & \gate{A(r^2)} & \gate{A(r)} & \gate{A(m)}  & \qw
}
\end{gather*}

    \caption{The four variants ($A^T(g)$, $A^T(g^{-1})$, $A(g^{-1})$ and $A(g)$, left to right top to bottom, respectively) of the generalised conjugation circuits. Depending on the label $(C,\chi)$ of the ribbon operator the appropriate single qubit unitaries from Table~\ref{tab:reps} are inserted.}
    \label{fig:GConj_scheme}
\end{figure}
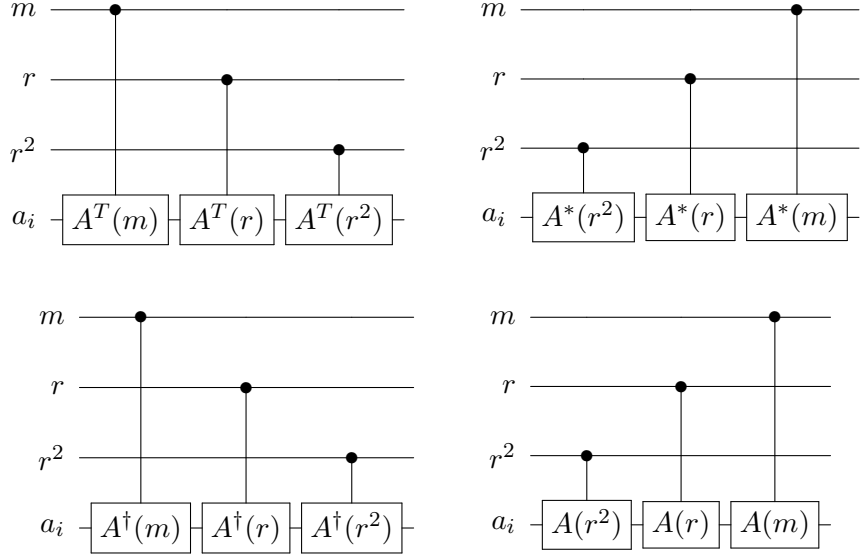

\textbf{Generalised conjugation.}
Another building block of our circuits is the generalised conjugation, $U_{GC}\ket{g}\ket{\alpha}_{a_i} = \ket{g}A^T(g)\ket{\alpha}_{a_i}$. The $A$-matrices for $D(D_4)$, are shown in Appendix~\ref{app:reps}. In Appendix~\ref{app:ribs} we see that we need four variants of the map: $A(g)$, $A(g^{- 1})$, $A^T(g)$, $A^T(g^{-1})$. Figure~\ref{fig:GConj_scheme} shows the corresponding circuits which are to be supplemented by the appropriate unitaries from Table~\ref{tab:reps}.

\section{Partial charge measurement -- additional data}\label{app:more_data}
\begin{figure}
    \centering
    \begin{subfigure}{0.49\linewidth}
        \centering
        \includegraphics[width=\linewidth]{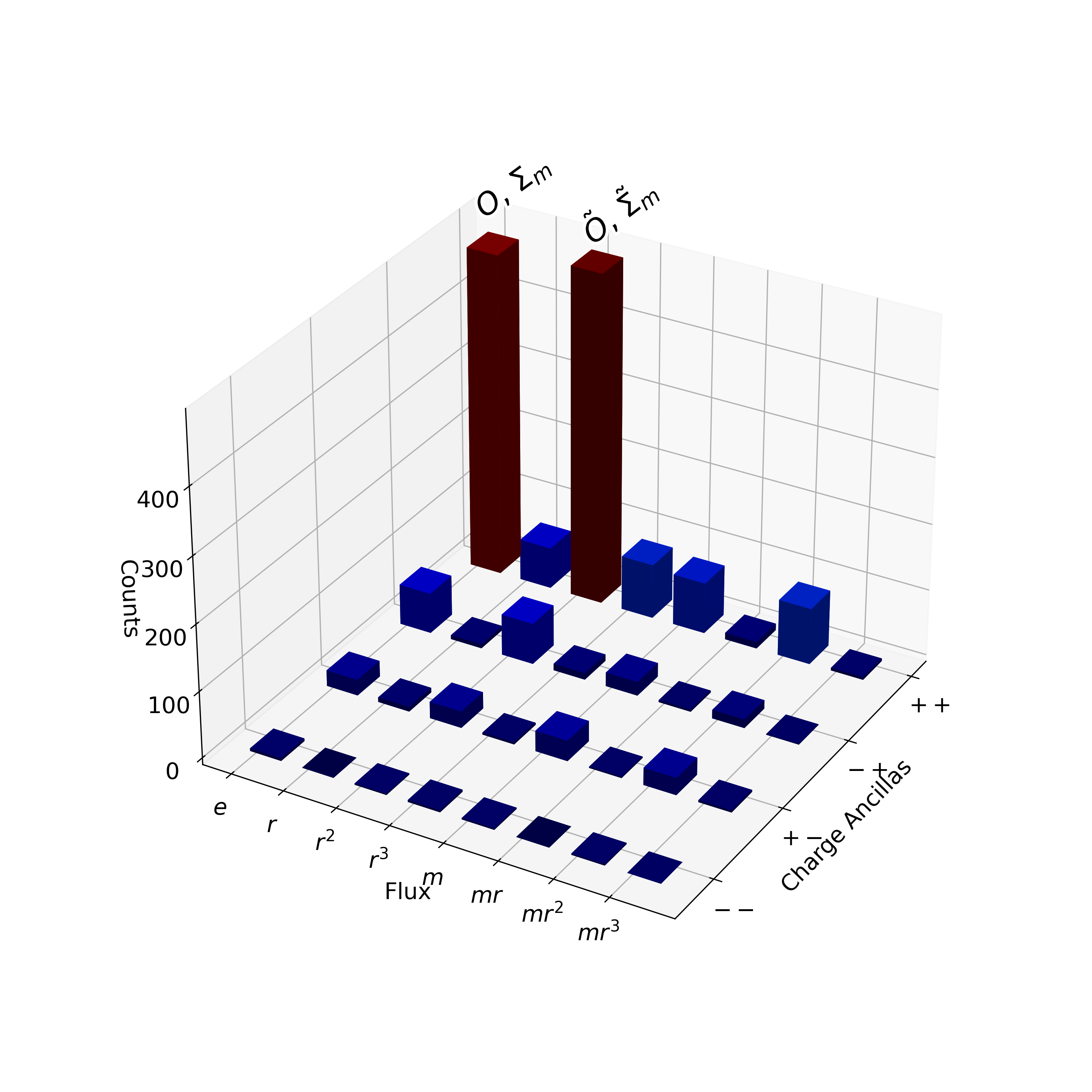}
        \caption{$H_m$-partial charge measurement}
        \label{fig:glasses_Hm}
    \end{subfigure}
    \begin{subfigure}{0.49\linewidth}
        \centering
        \includegraphics[width=\linewidth]{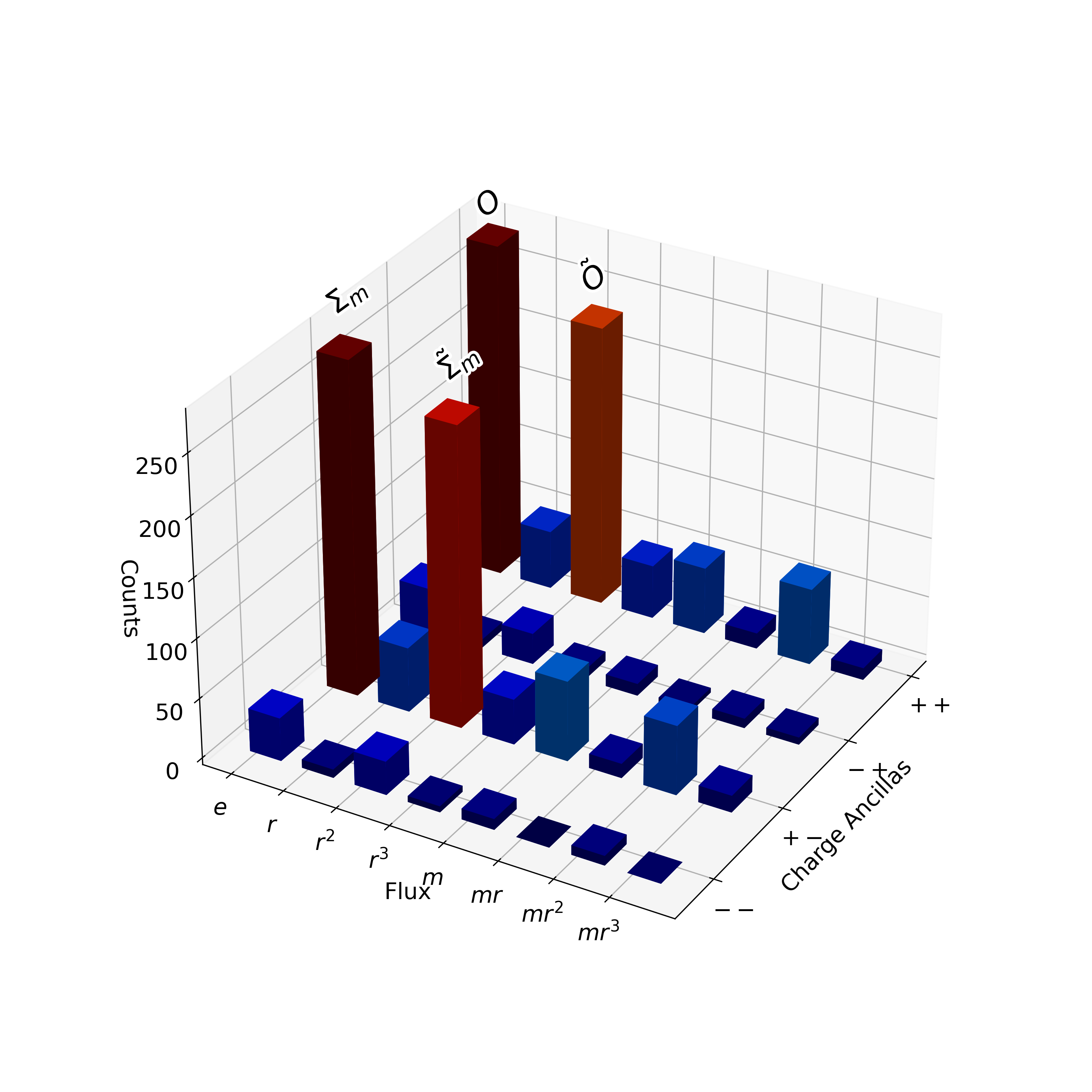}
        \caption{$H_{mr}$-partial charge measurement}
        \label{fig:glasses_Hmr}
    \end{subfigure}
    \begin{subfigure}{0.5\linewidth}
        \centering
        \includegraphics[width=\linewidth]{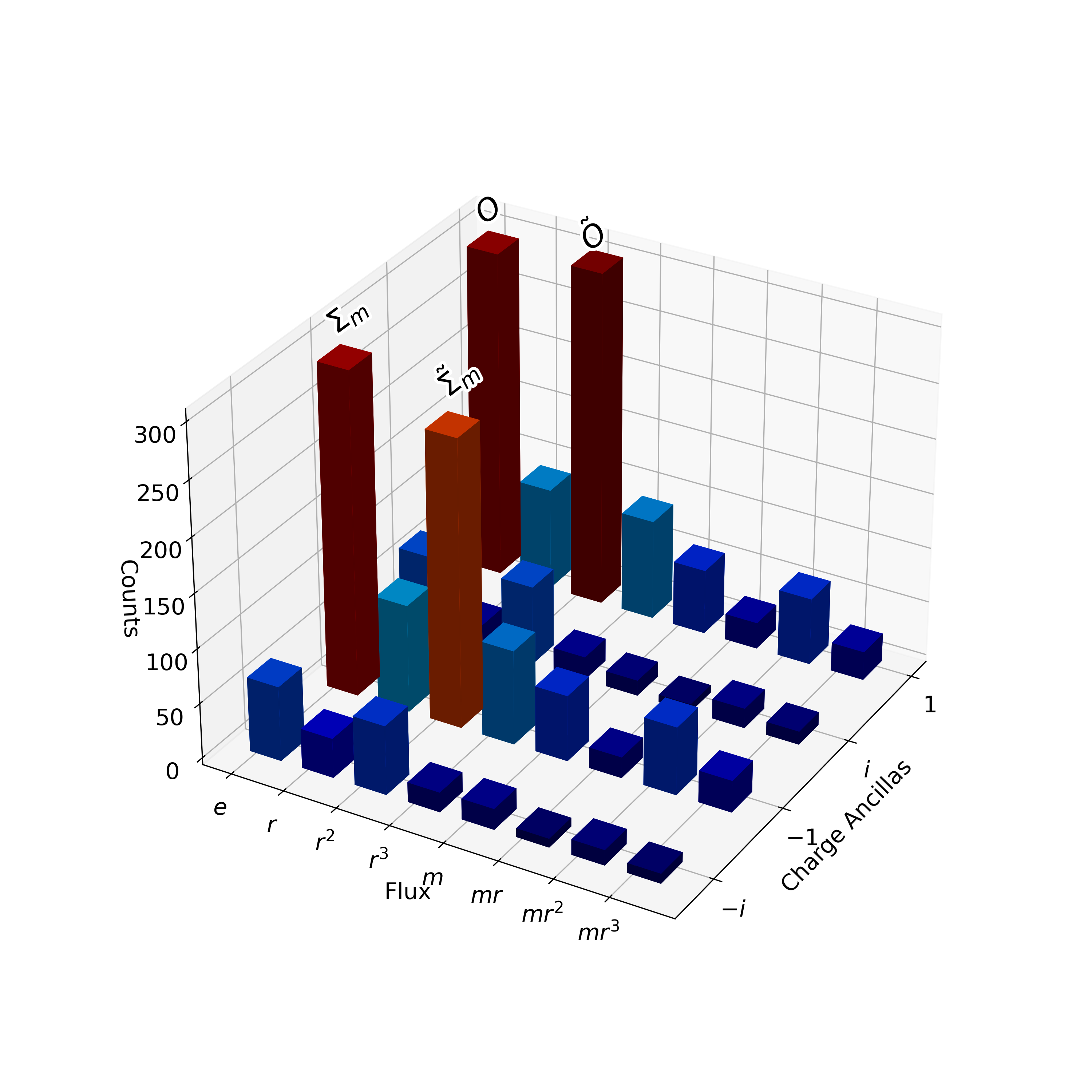}
        \caption{$H_r$-partial charge measurement}
        \label{fig:glasses_Hr}
    \end{subfigure}
    \caption{The results of the partial charge measurements for different four-element subgroups of $D_4$ at the end of the fusion protocol performed on the braiding ladder.}
    \label{fig:glasses_all_subs}
\end{figure}

\begin{figure}
    \centering
    \begin{subfigure}{0.49\linewidth}
        \centering
        \includegraphics[width=\linewidth]{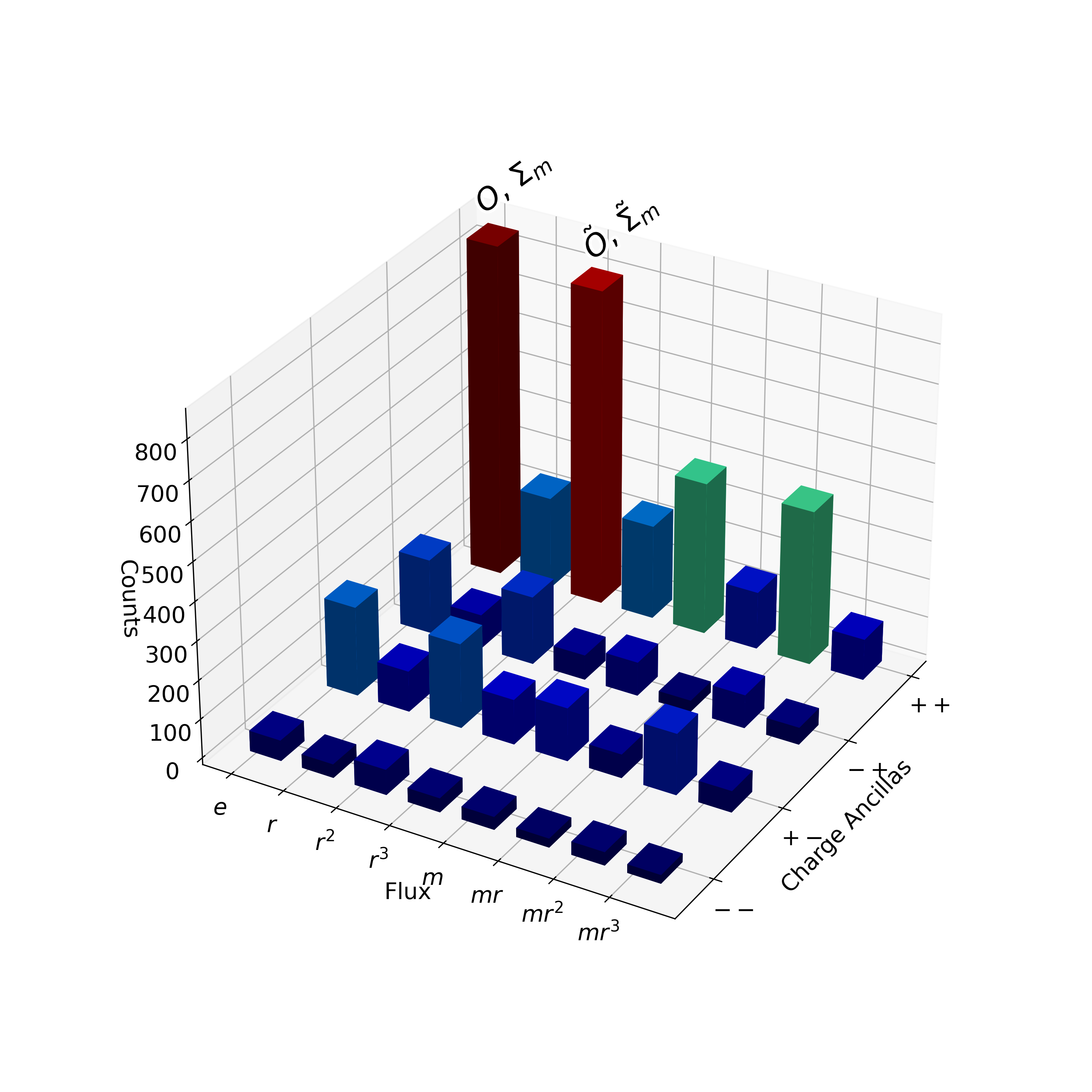}
        \caption{$H_m$-reduced charge measurement}
        \label{fig:basket_Hm}
    \end{subfigure}
    \begin{subfigure}{0.49\linewidth}
        \centering
        \includegraphics[width=\linewidth]{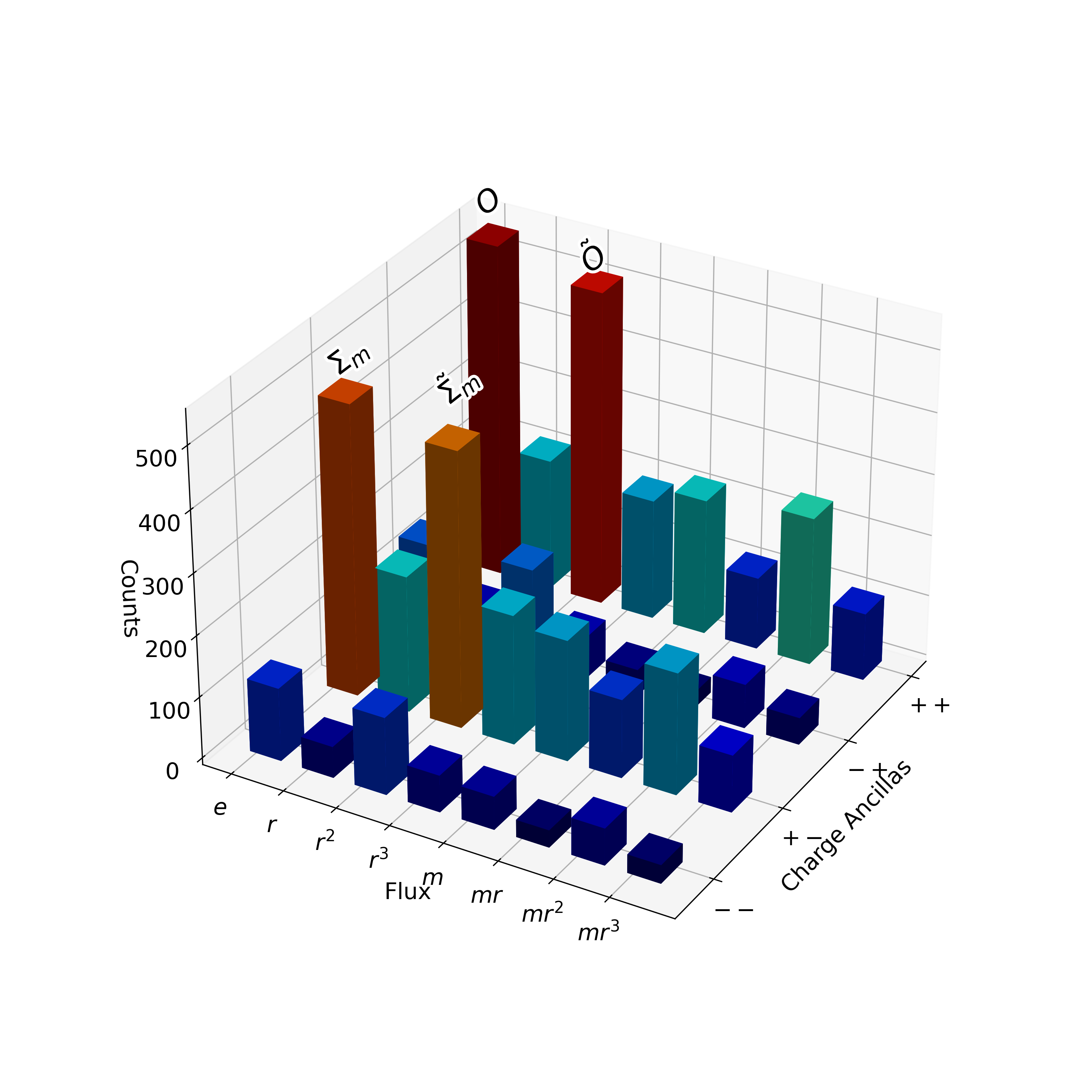}
        \caption{$H_{mr}$-reduced charge measurement}
        \label{fig:basket_Hmr}
    \end{subfigure}
    \begin{subfigure}{0.5\linewidth}
        \centering
        \includegraphics[width=\linewidth]{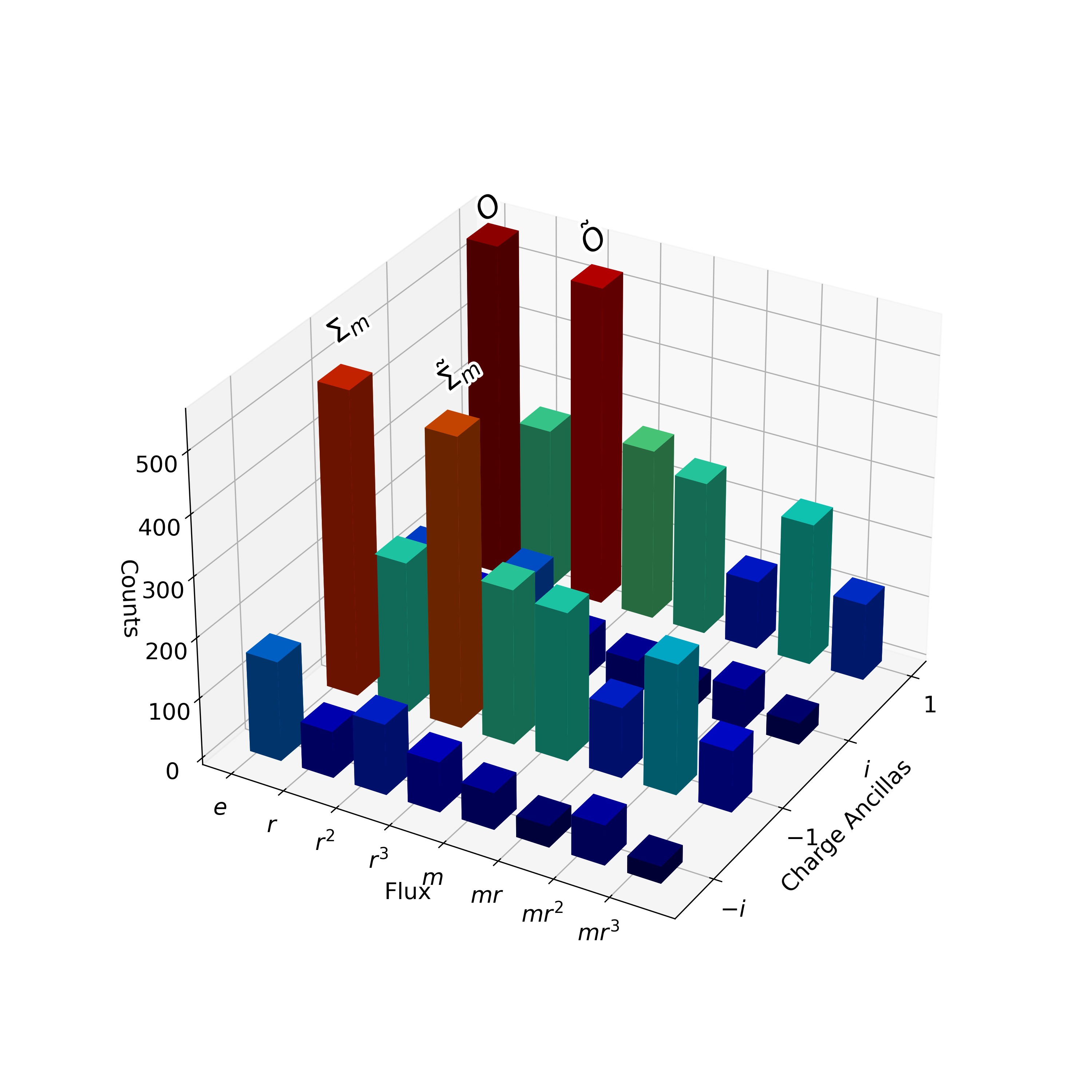}
        \caption{$H_r$-reduced charge measurement}
        \label{fig:basket_Hr}
    \end{subfigure}
    \caption{The results of the partial charge measurements for different four-element subgroups of $D_4$ at the end of the fusion protocol performed on the small planar graph.}
    \label{fig:basket_all_subs}
\end{figure}

In this appendix, we show the results of the partial charge measurements with respect to all four-element subgroups of $D_4$ in the anyon fusion experiment. Fig.~\ref{fig:glasses_all_subs} shows the data for the experiment performed on the braiding ladder and Fig.~\ref{fig:basket_all_subs} shows the data for the experiment performed on the small planar graph.
If one does not want to rely on the knowledge of the fusion algebra to label the peaks in the histograms shown in Figure~\ref{fig:red_charge_res} in the main text, these results are necessary and sufficient as argued in Section~\ref{sec:redchmmt} to uniquely determine the charge labels.

\section{Uncertainty estimation and measurement bias}
In this appendix we provide more details on the measurement uncertainty and the effect of the measurement bias relevant in the $S$-and $T$-matrix protocols in Section \ref{sec:num:intef}.
\subsection{Polarisation uncertainty} \label{app:DATA_Anl}In the following, we discuss the uncertainty estimations for the polarisation $P$ -- the central quantity of interest in the interference protocols in Section \ref{sec:num:intef}, and the resulting uncertainty for the amplitude and phase of the $S$-and $T$-matrix elements. The polarisation was determined for different measurement bases $\vec b$, but here we focus on a single fixed measurement basis first. To estimate $P$, we performed $N$ measurements and recorded the number of measurement outcomes $n_s$, $s\in \{0,1\}$. From these results we estimated the polarisation $P = \frac{p_0 - p_1}{p_0 + p_1}$ via $p_s=n_s/N$. The probability to find the outcome $s$ in $n_s$ of the $N$ shots is given by the binomial distribution, $p(n_s) = {N \choose n_s} p_s^{n_s} p_{\bar{s}}^{N-n_s}$, where $\bar{s} = 1-s$. Thus,
the unbiased estimator for the polarisation is $\hat P = \frac{n_0 - n_1}{n_0 + n_1}$, with $$\text{Mean}(\hat P) =P\text{ and }\text{Var}(\hat{P}) = N\sigma_{P}^2 = P(1-P),$$ where $\sigma_{P}=\sqrt{P(1-P)/N}$ is the error in determining this mean. This procedure of uncertainty estimation was used for all measurement bases and yields the error bars in Figures~\ref{fig:S_res} and ~\ref{fig:t_mat_results}.

To obtain the amplitude and phase of the $S$- and $T$-matrix entries from the interference protocols we used that the measurement basis dependent polarisation $P(\vec b)$ is related to the Bloch vector $\vec r$ of the single qubit density matrix and the read-out biases of the machine via
$$P(\vec{b}) = (1-2\bar\epsilon)\vec{r}\cdot\vec{b} + \Delta\epsilon.$$ 
Given the estimates for the function $P(\vec{b})$ we extracted the relevant quantities $(\vec{r}/|\vec{r}|, \Delta\epsilon, |(1-2\bar\epsilon)\vec{r}|)$ by fitting. To this end, we chose two particular sets of bases referred to as $\phi$-scan and $\theta$-scan. For the $\phi$-scan we fitted the function $f(\phi) = A\cos(\phi - \phi_{\text{max}})+B$, to extract the angle which maximises the polarisation $\phi_{\text{max}}$. Then, with fixeed $\phi = \phi_{\text{max}}$, we performed a $\theta$-scan and fitted to it the more informative function $g(\theta) = |(1-2\bar\epsilon)\vec{r}|\cos(\theta - \theta_{\text{max}}) + \Delta\epsilon$. The two angles $\phi_\text{max}$ and $\theta_\text{max}$ then determine the direction of the Bloch vector and via that, the $S$- or $T$-matrix entry denoted $\mathcal A$ via  
$$\mathcal{A} = \left(\sqrt{\frac{1-\cos{\theta_\text{max}}}{1+\cos{\theta_\text{max}}}}\pm \frac{\sigma_{\theta_\text{max}}}{|1 +\cos{\theta_\text{max}|}}\right)e^{i(\phi_\text{max}\pm \sigma_{\phi_\text{max}})}.$$
Here, the angles and their uncertainty are obtained by the usual $\chi^2$-fitting method.

\subsection{Measurement bias and post selection}\label{app:bias} We investigate the effect of post selecting on biased measurement outcomes for the paradigmatic example of two qubits. We consider a general two-qubit density matrix
\begin{equation}
	\rho=\frac{1}{4} r_{\alpha \beta} \sigma_\alpha \tau_\beta,\quad r_{00}=1,\quad \alpha,\beta=0,\ldots,4\;,
\end{equation} that represents the state after applying a noisy circuit. We tomograph the first qubit while post-selecting on the measurement outcome $0$ in the $z$-basis of the second qubit. The bias to measure $a$ while the outcome should be $\bar a$ is $\epsilon_a$ and assumed to be the same for both qubits. 

After measuring $0$ we obtain with probability $\epsilon_0$  the false positive state
\begin{equation}
	\tilde \rho_{01}=\frac{r_{\alpha0}-r_{\alpha3}}{4} \sigma_\alpha \ket{1}\bra{1}
\end{equation}
and with probability $1-\epsilon_1$ the true state
\begin{equation}
	\tilde \rho_{00}=\frac{r_{\alpha0}+r_{\alpha3}}{4} \sigma_\alpha \ket{0}\bra{0}\;.
\end{equation}
We now perform a measurement of the first qubit in the $\vec s$-basis, which amounts to a full tomography of the first qubit, if we choose several different bases $\vec s$. The probability to measure $0$ is given again by a combination of true and false positive outcomes and reads
\begin{equation}
	p_0=\epsilon_0 \operatorname{tr} \tilde \rho S_1  +(1-\epsilon_1) \operatorname{tr} \tilde \rho S_0  , \quad \tilde \rho=\epsilon_0 \tilde \rho_{01}+(1-\epsilon_1) \tilde \rho_{00} , \quad S_0=\frac{1+\vec S \cdot \vec \sigma}{2}, \;\quad S_1=\frac{1-\vec S \cdot \vec \sigma}{2}\;.
	\end{equation} 
	Similarly,
	\begin{equation}
		p_1=\epsilon_1 \operatorname{tr} \tilde \rho S_0 + (1-\epsilon_0)\operatorname{tr} \tilde \rho S_1
	\end{equation}
	More explicitly, we introduce $s^\pm_\alpha$ with $s_0=\pm 1, s^\pm_i=\vec s_i$ and $r_\alpha^\pm=r_{\alpha0}\pm r_{\alpha 3})$ and obtain
\begin{equation}
\begin{aligned}
p_0=&	\frac{1}{4} [\epsilon_0s^+_\alpha (\epsilon_0r^-_{\alpha}+(1-\epsilon_1)r^+_{\alpha})  
-(1-\epsilon_1) s^-_\alpha(\epsilon_0r^-_{\alpha}+(1-\epsilon_1)r^+_{\alpha})  ]\;.
\end{aligned}
\end{equation}
Likewise the probability to measure $1$ is
\begin{equation}
\begin{aligned}
p_1=&	\frac{1}{4} [-\epsilon_1 s^-_\alpha(\epsilon_0r^-_{\alpha}+(1-\epsilon_1)r^+_{\alpha})
+(1-\epsilon_0) s^+_\alpha (\epsilon_0r^-_{\alpha}+(1-\epsilon_1)r^+_{\alpha})   ]\;.
\end{aligned}
\end{equation}
This yields the polarisation 
\begin{equation}
	P=p_0-p_1= a+ \vec s \cdot \vec b \;,
\end{equation}
where $a,\vec b$ are polynomials in $\epsilon_{0/1}$ and $r_\alpha^\pm$, in particular
\begin{equation}
\begin{aligned}
	a&=(\epsilon_0-\epsilon_1) \frac{r^+_0}{2} +\epsilon_0^2 \frac{r^-_0}{2} -\epsilon_0 \epsilon_1 \frac{r^+_0+r^-_0}{2} +\epsilon_1^2 \frac{r_0^+}{2}\;,\\
	b_j&=\frac{r^+_j}{2}+ \epsilon_0 \frac{r^-_j-r^+_j}{2} -\epsilon_1 r^+_j -\epsilon_0^2 \frac{r^-_j}{2} +\epsilon_0 \epsilon_1 \frac{r^+_j-r^-_j}{2} + \epsilon_1^2 \frac{r_j^+}{2}
	\end{aligned}\;.
\end{equation}
We see that to leading order in $\epsilon$, the offset is proportional to $\Delta \epsilon=\epsilon_1-\epsilon_0$, while the vector $\vec b$ is dominated by terms $\mathcal O(1)$.

For a general setting, where we have $n$ qubits, post-select on the biased measurement outcomes of $k$ qubits and tomograph one qubit, the polarisation will have the same structure, just that $a$ and $b_j$ are higher polynomials in $\epsilon$, with the coefficients depending on the full density matrix. The polarisation can be written as 
\begin{equation}
	P=\vec s \cdot (\vec b^\text{no bias} + \tilde{\epsilon}_- \vec b^\text{bias}) + \tilde \epsilon_+ \;,
\end{equation}
where $\tilde \epsilon_\pm$ are effective errors that depend on $\rho$. In our experiment we are only interested in the direction of the vector $\vec b$ which contains the information of the $S$- or $T$-matrix entry. The offset is purely due to read-out errors and is discarded. In the tomography, we determine $\vec b^\text{observed}=\vec b^\text{no bias} + \tilde{\epsilon}_- \vec b^\text{bias}$ and do not try to distinguish between the two contributions, since we are dealing with an unknown density matrix. 


\bibliographystyle{quantum}
\bibliography{bibliography}

\end{document}